\documentstyle[epsfig,conf_cg]{article}

\newcommand{\be}{\begin{equation}}
\newcommand{\ee}{\end{equation}}
\newcommand{\bea}{\begin{eqnarray}}
\newcommand{\eea}{\end{eqnarray}}
\newcommand{\lb}{\label}

\begin{document}

\heading{Black Holes: Classical Properties, Thermodynamics, 
\\
and Heuristic Quantization
\\
}
\par\medskip\noindent
\author{Jacob D. Bekenstein$^1$}
\address{Racah Institute of Physics, Hebrew University of Jerusalem,
\\
Givat Ram, 91904 Jerusalem, ISRAEL}

\begin{abstract}
I start with a discussion of the no-hair principle. The  hairy black
hole solutions of recent vintage do not deprive it of value because they are
often unstable.  Generic properties of spherical static black holes with
nonvacuum exteriors are derived.  These form the basis for the discussion of
the new no scalar hair theorems.  I discuss the generic phenomenon of
superradiance for its own sake, as well as background for black hole
superradiance.  First I go into uniform linear motion superradiance with some
examples.  I then discuss Kerr black hole superradiance in connection with a
general rotational superradiance theory with possible applications in the
laboratory.  Adiabatic invariants have played a weighty role in theoretical
physics.  I explain why the horizon area of a nearly stationary black hole
can be regarded as an adiabatic invariant, and support this by examples as
well as a general discussion of perturbations of the horizon.  The horizon
area's adiabatic invariance suggests that its quantum counterpart is
quantized in multiples of a basic unit.  Consideration of the quantum analog
of the Christodoulou reversible processes provides support for this idea. 
Area quantization provides a definite discrete black hole mass spectrum. 
Black hole spectroscopy follows: the Hawking semiclassical spectrum is
replaced by a spectrum of nearly uniformly spaced lines whose envelope may be
roughly Planckian.  I estimate the lines' natural broadening.  To check on
the possibility of line splitting, I present a simple algebra involving,
among other operators, the black hole observables.  Under simple assumptions
it also leads to the uniformly spaced area spectrum.
\\
In these lectures I take units for which $c=1$.  Occasionally, where
mentioned explicitly, I also set $G=1$, but always display $\hbar$.
\end{abstract}
%
\section{No scalar hair theorems} \lb{no hair}
\setcounter{equation}{0}
\renewcommand{\theequation}{1.\arabic{equation}}

Almost thirty years ago Wheeler enunciated the Israel-Carter conjecture,
today colloquially known as ``black holes have no hair''~\cite{Wheeler}. 
This influential conjecture has long been regarded as a theorem by large
sectors of the gravity-particle physics community.   But by the early 1990's
solutions for stationary black holes with exterior nonabelian gauge or
skyrmion fields~\cite{VolkovM,Bizon1,Droz,Lee,Greene} had led many workers to
regard the conjecture as having fallen by the wayside.   By now things have
settled down to a new paradigm not very different from Wheeler's original one.

\subsection{Early days of `no-hair'}  \lb{updown}

By 1965 the charged Kerr-Newman black hole metric was known.  Inspired by
Israel's uniqueness theorems for the Schwarzschild and Reissner-Nordstr\"om
black holes~\cite{Israel}, and by Carter's~\cite{Carter} and
Wald's~\cite{Wald} uniqueness theorems for the Kerr black hole,  Wheeler
anticipated that ``collapse leads to  a black hole endowed with mass and
charge and angular momentum, but, so far as we can now judge, no other free
parameters" by which he meant that collapse ends with a Kerr-Newman black
hole.  Wheeler stressed that other `quantum numbers' such as baryon number or
strangeness can have no place in the external observer's description of a
black hole.

What is so special about mass, electric charge and angular momentum ? They
are all conserved quantities subject to a Gauss type law. One can thus
determine these properties of a black hole by measurements from afar. 
Obviously this reasoning has to be completed by including magnetic (monopole)
charge as a fourth parameter because it also is conserved in Einstein-Maxwell
theory, it also submits to a Gauss type law, and duality of the theory
permits Kerr-Newman like solutions with magnetic charge alongside (or instead
of) electric charge. In the updated version of Wheeler's conjecture, the
forbidden ``hair'' is any field not of gravitational or electromagnetic
nature associated with a black hole.

But why is the issue of hair interesting ?   Black holes are in a real sense
gravitational solitons; they play in gravity theory the role atoms played in
the nescent quantum theory of matter and chemistry.  Black hole mass and
charge are analogous to atomic mass and atomic number.  Thus if black holes
could have other parameters, such `hairy' black holes would be analogous to
excited atoms and radicals, the stuff of exotic chemistry.   By contrast, the
absence of a large number of hair parameters would support the conception of
simple black hole exteriors, a situation which is natural for the formulation
of black hole entropy as the measure of the vast number of hidden degrees of
freedom of a black hole.  Indeed, historically, the no-hair conjecture
inspired the formulation of black hole thermodynamics (for the early history
see review~\cite{BekPT}), which has in the interim become a pillar of gravity
theory.

Originally ``no-hair theorems'' meant theorems like Israel's
or Carter's~\cite{Israel,Carter} on the uniqueness of the Kerr-Newman family
within the Einstein-Maxwell theory or like Chase's~\cite{Chase} on its
uniqueness within the Einstein-massless scalar field theory.  Wheeler's
conjecture that baryon and like numbers cannot be specified for a black hole
set off a longstanding trend in the search for new no-hair theorems.  Thus
Hartle~\cite{Hartle} as well as Teitelboim~\cite{Teitelboim} proved that the
nonelectromagnetic force between two ``baryons'' or ``leptons'' resulting
from exchange of various force carriers would vanish if one of the particles
was allowed to approach a black hole horizon.  I developed an alternative and
very simple approach~\cite{Beknohair} to show that classical massive scalar
or vector fields cannot be supported at all by a stationary black hole
exterior, making it impossible to infer any information about their sources
in the black hole interior.

In modernized form this goes as follows.  Start with the action 
\be S_\psi=-{\scriptstyle 1\over \scriptstyle 2}\int
[\psi_{,\alpha}\psi^{,\alpha}+V(\psi^2)](-g)^{1/2}\,d^4x 
\lb{eq:KGaction} 
\ee  for a static real scalar field $\psi$.  From it follows the field 
equation
\be
\psi_{,\alpha}{}^{;\alpha} - \psi V'(\psi^2) = 0
\lb{eq:KGeq}
\ee
Assume that the configuration is asymptotically flat and stationary:
$\partial \psi/\partial x^0 = 0$, where $x^0$ is a timelike variable in the
black hole exterior.  Multiply Eq.~(\ref{eq:KGeq})  by $\psi$ and integrate
over the black hole exterior at a given $x^0$ (space ${\cal V}$). 
Integration by parts leads to
\be 
-\int_{\cal V} [g^{ab}\psi_{,a}\psi_{,b} +\psi^2 V'(\psi^2)](-g)^{1/2}\,d^3x
+ \oint_{\partial{\cal V}} \psi\psi^{,\alpha} d\Sigma_\alpha =0
\lb{eq:integral} 
\ee
where $d\Sigma_\alpha $ is the 2-D element of the boundary hypersurface
$\partial {\cal V}$.  The indices $a$ and $b$ run over the space coordinates
only, so that the restricted metric $g^{ab}$ is positive definite in the
black hole exterior.  

Now suppose the boundary ${\partial{\cal V}}$ is taken as a large sphere at
infinity over all time (topology $S^2\times R$) together with a surface close
to the horizon ${\cal H}$, also with topology $S^2\times R$.  Then so long as
$\psi$ decays as $1/r$ or faster at large distances ($r$ is the usual
Euclidean distance), which will be true for static solutions of
Eq.~(\ref{eq:KGeq}), infinity's contribution to the boundary vanishes.  At
the inner boundary we can use Schwarz's inequality to state that at every
point
$|\psi\psi^{,\alpha} d\Sigma_\alpha|\leq (\psi^2\,
\psi^{,\alpha}\psi_{,\alpha}\ d\Sigma^\beta\, d\Sigma_\beta)^{1/2}$.  As the
boundary is pushed to the horizon (a null surface),
$d\Sigma^\beta\, d\Sigma_\beta$ must necessarily tend to zero.  Thus the
inner boundary term will also vanish unless $\psi^2\,
\psi^{,\alpha}\psi_{,\alpha}$ blows up at ${\cal H}$.  But this last
eventuality is usually unacceptable for a black hole. Finiteness of the 
physical scalars $T_{\alpha\beta} T^{\alpha\beta}$ and
$(T_\alpha{}^\alpha)^2$  at ${\cal H}$ tells us that
$\psi^{,\alpha}\psi_{,\alpha}$ and $V$ are both bounded at ${\cal H}$.  Then
if $V$ diverges for large arguments, $\psi$ has to remain bounded, and so
$\psi^2\,
\psi^{,\alpha}\psi_{,\alpha}$ is bounded on ${\cal H}$.  But even if
$V(\infty)\neq \infty$ so that $\psi$ is allowed to diverge, this will almost
certainly cause $\psi^{,\alpha}\psi_{,\alpha}$ to diverge.  In either case
the boundary term vanishes.

Thus for a generic $V$ we conclude that the 4-D integral in
Eq.~(\ref{eq:integral}) must itself vanish. In the case that $V'(\psi^2)$ is
everywhere nonnegative and vanishes only at some discrete values $\psi_j$,
then it is clear that the field
$\psi$ must be constant everywhere outside the black hole, taking on one of
the values $\{0, \psi_j\}$.  The scalar field is thus trivial, either
vanishing or taking a constant value as dictated by spontaneous symmetry
breaking without the black hole !  In particular, the theorem works for the
Klein-Gordon field for which  $V'(\psi^2)=\mu^2$ where $\mu$ is the field's
mass. In that case $\psi=0$ outside the black hole~\cite{Beknohair}. 
Obviously the theorem supports Wheeler's original conjecture by ruling out
black hole parameters having to do with a scalar field. 

One advantage of this type of theorem, in contrast with, say, Chase's, is
that it makes no use of the gravitational field equations.  The inference
that there are no black holes with scalar hair is thus just as true in other
metric theories of gravity.  Another plus is that the theorem is easily
generalizable to exclude massive vector (Proca) field hair~\cite{Beknohair}. 
Because of both features this type of theorem was the state of the art for
many years (see for instance Heusler's monograph~\cite{Heusler}).  But this
should not blind us to its shortcomings.  For example, it does not rule out
hair in the form of a Higgs field with a `Mexican hat' potential, a darling
of particle physicists.  For the Higgs field $V'(\psi^2)$ is negative in some
regime of $\psi$, and the theorem fails.  This gap in the no-hair theorems
remained until fairly late in the subject, as certified by Gibbons'
relatively recent review~\cite{Gibbons}).

\subsection{Hairy black holes ?}\lb{hairy}

When one removes the massive vector field's mass, gauge invariance sets in
and the fundamental field, basically the covariant time component of
$A_\alpha$ (the electromagnetic 4-potential), cannot be required to be
bounded at the horizon because it is {\it not\/} gauge invariant. No no-hair
theorem can be proved:  a black hole can have an electromagnetic field as 
witness the Kerr-Newman family.  By the same logic I
concluded~\cite{Bekthesis} that the gauge invariance of the nonabelian gauge
theories should likewise allow one or more of the gauge field components
generated by sources in a black hole to ``escape'' from it.  Thus gauge
fields around a black hole may be possible in every gauge theory.  Early on
Yasskin ~\cite{Yasskin} exhibited some trivial hairy black holes in
nonabelian gauge theories.   Volkov and Gal'tsov discovery in 1989  of more
interesting black hole solutions with nonabelian gauge
field~\cite{VolkovM,Bizon1} took everybody by surprise.  But it obviously
should not have done so !  

Actually an hairy black hole was known well before.  It is a solution of the
action for a scalar {\it nonminimally\/} coupled to gravity:
\be
S=S_{\rm E} + S_{\rm M} - {\scriptstyle 1\over \scriptstyle 2}\int{\left[
		   \psi_{,\alpha} \psi^{,\alpha} 
	 	  + \xi R\psi^2 + V(\psi^2)
				\right] \sqrt{-g}\, d^4x}.
    \lb{eq:action}
\ee
Here $S_{\rm E}$ stands for the Einstein-Hilbert action, $S_{\rm M}$
for the Maxwell one,  $R$ is the Ricci curvature scalar and $\xi$ measures
the strength of the coupling to the curvature.   One derives the
energy-momentum tensor
\be
T_{\mu \nu}=\psi_{,\mu} \psi_{,\nu} - {\scriptstyle 1\over \scriptstyle 2}
\psi_{,\alpha} \psi^{,\alpha}\,g_{\mu \nu} 
		      - \xi\, \psi \psi_{,\mu;\nu} 
		       + \xi\, \hbox{\rlap{$\sqcap$}$\sqcup$} \psi^2\, g_{\mu \nu} 
		       + \xi\, \psi^2 G_{\mu \nu}
		 - {\scriptstyle 1\over \scriptstyle 2}V g_{\mu \nu} + T_{\mu \nu}^{\rm
(M)}   
\lb{eq:TG}
\ee
Substituting $G_{\mu\nu} = 8\pi GT_{\mu\nu}$ turns this into
\be
T_\mu{}^\nu={\psi_{,\mu} \psi^{,\nu} - {\scriptstyle 1\over \scriptstyle 2}
\psi_{,\alpha} \psi^{,\alpha}\,\delta_\mu{}^\nu 
- \xi\, \psi \psi_{,\mu}{}^{;\nu} 
+ \xi\, \hbox{\rlap{$\sqcap$}$\sqcup$} \psi^2\, \delta_\mu{}^\nu 		    
- {\scriptstyle 1\over \scriptstyle 2}V g_\mu{}^\nu +
T^{(M)}{}_\mu{}^\nu \over 1-8\pi G\xi\psi^2}   
\lb{eq:TGnew}
\ee

For the conformally invariant
coupling $\xi=1/6$ with $V=0$, Bocharova,  Bronnikov and Melnikov (BBM)
~\cite{BBM} and independently I~\cite{Bekscalar} found a black hole solution:  
\bea  
ds^2 &=& -(1-GM/r)^2 dt^2 + (1-GM/r)^{-2} dr^2 +  r^2 (d\theta^2 +
\sin^2\theta\, d\phi^2)\nonumber
\\
 F_{\mu\nu} &=& Q\,r^{-2}\,(\delta_\mu^r \delta_\nu^t - \delta_\mu^t
\delta_\nu^r);\qquad Q < \surd G M
\lb{eq:BBM}
\\
\psi &=& \pm (3\surd G/4\pi)(M^2-G^{-1}Q^2)^{1/2}(r-GM)^{-1}\nonumber
\eea
$M$ and $Q$ are free parameters corresponding to the mass
and charge of the solution.  Note that the metric is an extreme
Reissner-Nordstr\"om metric and that the scalar field blows up at $r= GM$,
the location of the geometry's horizon.

I interpreted this solution as genuine black hole~\cite{Bekbh} because
the apparent singularity of $\psi$ at the horizon  has no deleterious
consequences.   The invariants $T_{\alpha\beta} T^{\alpha\beta}$ and
$(T_\alpha{}^\alpha)^2$ are bounded even at ${\cal H}$ and a particle, even
one coupled to $\psi$, encounters no infinite tidal forces upon approaching
the horizon. Because $M$ and $Q$ are the only independent parameters, with
the scalar field introducing merely a sign choice, I was not alarmed by the
threat posed to no-hair by this solution.  Indeed in one of his last papers
before his tragic death, B. Xanthopoulos in cooperation with
Zannias~\cite{Xanthopoulos}, and then Zannias alone~\cite{Zannias} proved
that there is no nonextremal extension of the BBM black hole which might have
introduced a scalar charge as an extra parameter.  The situation is not
qualitatively changed by the addition of magnetic charge~\cite{Virb}.

Sudarsky and Zannias~\cite{SudZan} have lately claimed that
Eqs.~(\ref{eq:BBM}) are {\it not\/} really a solution of the action
(\ref{eq:action}).  Their point is that (for $Q=0$) although
$T_\alpha{}^\beta$ is finite at ${\cal H}$, if one ``regularizes'' $\psi$ so
that it becomes actually bounded on ${\cal H}$, the resulting finite
$T_\alpha{}^\beta$ does not generate, through Einstein's equations the metric
(\ref{eq:BBM}a).  The argument is rather odd as it creates a problem
(regularizes $\psi$ when $T_\alpha{}^\beta$ is perfectly finite anyway) in
order to solve it.  However, the BBM solution has been shown to be unstable
in linearized theory by Bronnikov (one of its discoverers) and
Kireyev~\cite{BroKir}.  A poor man's way of understanding why can be had by
contemplating Eq.~(\ref{eq:TGnew}) for the energy momentum tensor. Addition
of a bit of matter at the point where the denominator vanishes (which must be
outside the horizon since $\psi^2$ blows up at the horizon) would obviously
lead to a drastic perturbation in the geometry. So the solution
(\ref{eq:BBM}) is unstable.  Only to a purist would an unstable solution
require further discussion, and there hardly seems to be any need to
criticize it on another account.

In fact almost all known hairy black holes in $3+1$ general relativity are 
known to be unstable~\cite{Straumann,Brodbeck,Mavromatos}.  The only one
which is certified to be stable~\cite{Skyrmion_stable}, at least in
linearized theory, is the Skyrmion hair black hole~\cite{Droz}.  It differs
from the Schwarzschild one in that it involves a parameter with properties of
a topological winding number.  This is not an additive quantity among several
black holes, so that the Skyrmion black hole may not represents a true
exception to Wheeler's principle; however, I shall not try to reach a
veredict here.

Because gauge fields seem to produce unstable hairy black holes, one
should look for possible violations of no-hair in the only other direction
left: scalar hair. Thus I proposed~\cite{Sakharov} to shift emphasis to the
``no scalar hair''  conjecture which I will state as: {\it there are no
asymptotically flat, stationary and stable black hole solutions in $3+1$
general relativity which are endowed with scalar fields\/}.   The asymptotic
flatness requirement is introduced to rule out the black holes in de Sitter
background ~\cite{Carter2} as well as the  Achucarro-Gregory-Kuijken black
hole ~\cite{Achu}, a charged black hole transfixed by a Higgs local cosmic
string.  The stability restriction is to exclude the BBM black hole.   This
shift has also been urged by  N\'u\~nez, Quevedo and Sudarsky ~\cite{Nunez}
on the grounds that there are `hairy' solutions (never mind that they are
unstable) and that the hair they sport is never `short' and ignorable. 
 
Consider the case of a theory governed by action (\ref{eq:action}).   How
to prove that scalar hair is excluded even when the assumption $V'>0$ of
Sec.~\ref{updown} is not made ?  By 1995 new techniques to overcome this
problem had been introduced by Heusler~\cite{Heu}, Sudarsky~\cite{Sud} and
myself ~\cite{Bek95}.  These led to various no scalar hair theorems for {\it
spherical\/} black holes and {\it minimally coupled\/} fields.   As
background for these and their natural extension, I will now go into the
generic properties of spherically symmetric stationary black holes with
nonvacuum exteriors.

\subsection{Properties of stationary spherical black holes} \lb{spherical}

For a spherically symmetric and stationary black hole with any kind of matter
and fields in its exterior, the metric may  be taken as
\be
	ds^2 = - e^\nu dt^2 + e^\lambda dr^2 + r^2 (d\theta^2 + \sin^2\theta\,
d\phi^2)  
\lb{eq:metric}
\ee
Here $\nu=\nu(r)$ and $\lambda=\lambda(r)$ with both behaving as $1/r$
for $r\rightarrow \infty$ (asymptotic flatness). The event horizon is at
$r=r_{\cal H}$ with $r_{\cal H}$ being the outermost zero of
$e^{-\lambda}$.  To see why define a family of hypersurfaces with
$S^2\times R$ topology by the conditions $\{\forall t; r=\mbox{\rm
const.}\}$.  Each value of the constant labels a different surface. The
normal to each such hypersurface is $\eta_\alpha = r_{,\alpha} =
\delta_\alpha{}^r$, so that
$\eta_\alpha \eta^\alpha = e^{-\lambda}$  which vanishes at $r=r_{\cal H}$,
but never outside it.  This must thus be location of the horizon which is
defined as a null surface (hence null normal).

Now in order for the black hole solution to be physical, invariants such as
$T_\alpha{}^\alpha$ and  $T_{\alpha\beta}\,T^{\alpha\beta}$ must be bounded
throughout its exterior $r\geq r_{\cal H}$.   In the coordinates of the
metric (\ref{eq:metric})
$T_{\alpha\beta}\,T^{\alpha\beta}=(T_t{}^t)^2+(T_r{}^r)^2+(T_\theta{}^
\theta)^2+ (T_\varphi{}^\varphi)^2$ so that $T_r{}^r$ and $T_t{}^t$ must be
finite for $r\geq r_{\cal H}$.

Let me now introduce two of Einstein's equations:
\bea
e^{-\lambda}(r^{-2}-r^{-1}\lambda')-r^{-2}&=&8\pi G T_t{}^t, 
\lb{eq:Einstein1}  
\\
e^{-\lambda}(r^{-2}+r^{-1}\nu')-r^{-2}&=&8\pi G T_r{}^r.  \lb{eq:Einstein2}
\eea
It follows from the second that $e^\nu$ also has its outermost zero at
$r_{\cal H}$ (Vishveshwara's theorem~\cite{Vishu}).  For assume  that $e^\nu$
vanishes at some point $\bar{r}$. Then $\nu\rightarrow -\infty$ and
$\nu'\rightarrow \infty$ as $r \rightarrow \bar{r}$ from the right. It is
then obvious from Eq.~(\ref{eq:Einstein2}) that $e^{-\lambda}$ must vanish
as $r\rightarrow \bar{r}$ since $T_r{}^r$ must be bounded.   But since as we
move in from infinity, $e^{-\lambda}$ first vanishes at $r=r_{\cal H}$, we
see that $\bar{r}=r_{\cal H}$.  The converse is also true: the horizon
$r=r_{\cal H}$ must always be an infinite redshift surface with $e^\nu=0$.  
For if $e^\nu$  were positive at $r=r_{\cal H}$, then  according to the
metric (\ref{eq:metric}) the $t$ direction would be timelike  there, while the
$\theta$ and $\phi$ directions would be, as always,  spacelike.  But
since the horizon is a null surface, it must have a null tangent direction,
and by time symmetry this must obviously be the $t$ direction.    Thus it  is
inconsistent to assume that $e^\nu\neq 0$ at $r=r_{\cal H}$.

Eq.~(\ref{eq:Einstein1}) may be integrated to get
\be
	e^{-\lambda} =  1 - {r_{\cal H}\over r} + {8 \pi G\over r} \int^r_{r_{\cal
H}} T_t{}^t r^2 dr
\lb{eq:e_lambda}
\ee
The constant of integration has been adjusted so that  $e^{-\lambda}$
vanishes at $r_{\cal H}$.  Obviously $T_t{}^t$ must vanish  asymptotically
faster than
$1/r^3$ in order for
$e^\lambda$ not to diverge at infinity.  Since $T_t{}^t$ must be bounded on
the horizon, we may write the first approximation (in Taylor's sense) near
the horizon  
\be
	e^{-\lambda}=L(r-r_{\cal H})+{\cal O}((r-r_{\cal H})^2);\qquad L\equiv 
r_{\cal H}{}^{-1}+8 \pi G\,r_{\cal H}\,   T_t{}^t(r_{\cal H})  
\lb{eq:e_lambda_first}
\ee
or
\be
	\lambda = {\rm const.} - \ln (r-r_{\cal H})  + {\cal O}(r-r_{\cal H})
\lb{eq:lambda}
\ee
 Since $e^{-\lambda}$ must be nonnegative outside the horizon, we  learn that
 $L\geq 0$, or 
\be
- (8 \pi G r_{\cal H}{}^2)^{-1} \leq  T_t{}^t(r_{\cal H})	
\lb{eq:Ttt_is_bounded}
\ee
so that at every stationary spherically symmetric event horizon, the energy
density, if positive, is limited by the very condition of regularity.  The
inequality is saturated for the extremal black hole.

Eqs.~(\ref{eq:Einstein1}-\ref{eq:Einstein2}) combine to the equation
\be
	e^{- \lambda} (\nu ' + \lambda ') = -8 \pi G (T_t{}^t - T_r{}^r) r.
\lb{eq:eq.rr-eq.tt}
\ee
with integral
\be
	\nu +\lambda = 8 \pi G \int ^\infty _r r'(T_t{}^t-T_r{}^r) e^\lambda dr'
\lb{eq:nu+lambda}
\ee
We have built in the asymptotic requirement $\nu+\lambda\rightarrow 0$
by appropriate choice of the limits of integration.  Obviously in addition to
the asymptotic condition on $T_t{}^t$, $T_r{}^r$  must decrease at least as
fast as
$1/r^3$ so that the integral converges and $\nu+\lambda$ is well defined. By
following the method leading to Eq.~(\ref{eq:e_lambda_first}) and using that
expression for $e^{-\lambda}$ we can get from  Eq.~(\ref{eq:nu+lambda}) 
\be
	\nu+\lambda={\rm const.}-
		8\pi G r_{\cal H}{}^2\ {T_t{}^t(r_{\cal H})-T_r{}^r(r_{\cal H})\over 
	1+8 \pi G T_t{}^t(r_{\cal H}) r_{\cal H}{}^2} \ln (r-r_{\cal H}) +{\cal
O}(r-r_{\cal H})
\lb{eq:nu+lambda_first}
\ee
which in view of Eq.~(\ref{eq:lambda}) gives
\be
	\nu ={\rm const.} + \beta \ln (r-r_{\cal H}) + {\cal O}(r-r_{\cal H});\qquad
\beta \equiv {1+8 \pi G T_r{}^r(r_{\cal H}) r_{\cal H}{}^2\over 1+8 \pi G
T_t{}^t(r_{\cal H}) r_{\cal H}{}^2}
\lb{eq:nu_first} 
\ee

The value of $\beta$ is restricted by the requirement that the scalar 
curvature 
\be
	R=e^{-\lambda} \left(\nu ''+{\scriptstyle 1\over \scriptstyle 2}\nu '^2
			   +{2\over r} (\nu ' -\lambda ')
	                   -{\scriptstyle 1\over \scriptstyle 2}\nu ' \lambda ' 
			   +{2\over r^2} \right)-{2\over r^2}
\ee
be bounded on the horizon (this is the same as boundedness of
$T_\alpha{}^\alpha$).   If we substitute here Eqs.~(\ref{eq:lambda}) and
(\ref{eq:nu_first}) we get 
\be
R = -{2\over r_{\cal H}{}^2} +L\ (r-r_{\cal H} ) \times \left({\scriptstyle 1
\over \scriptstyle 2} {\beta(\beta-1) \over (r-r_{\cal H})^2}+{2\over r_{\cal
H}}{\beta+1 \over (r-r_{\cal H})}+{2\over r_{\cal H}{}^2} \right) 
\lb{eq:u->R} \ee
For a nonextremal black hole $L>0$, so we are left with the condition  
\be
	\beta (\beta -1)=0
\ee
The alternative $\beta=0$ is excluded by the requirement
that $e^\nu=0$ at the horizon.  Thus necessarily $\beta = 1$.  It follows
from Eq.~(\ref{eq:nu_first}) that
\be
T_t{}^t = T_r{}^r \quad {\rm at }\quad  r=r_{\cal H}
\lb{eq:equality}
\ee
\be
	e^\nu = N (r-r_{\cal H}) + {\cal O}((r-r_{\cal H})^2)
\lb{eq:e_nu}
\ee
where $N$ denotes a positive constant.  Equality (\ref{eq:equality}), which
has been derived by several groups~\cite{Achu,Nunez,MB}, can also be proved
for extremal black holes; the factors in $(r-r_{\cal H})$ in
Eqs.~(\ref{eq:e_lambda_first}) and (\ref{eq:e_nu}) are then replaced by
$(r-r_{\cal H})^2$~\cite{MB}. 

\subsection{No minimally coupled scalar hair} \lb{minimal}

Consider a black hole solution of the theory whose action is
\be
S=S_{\rm E}-\int {\cal E}({\cal I}, \psi)\sqrt{-g}\, d^4x;\qquad
{\cal I}\equiv g^{\alpha\beta}\psi_{,\alpha}\psi_{,\beta}
\lb{eq:modified}
\ee
where ${\cal E}$ is some function.  I have dropped the Maxwell action (so
that I only consider electrically neutral black holes), but for later
convenience have generalized the scalar action.  The scalar's energy momentum
tensor turns out to be
\be
T_\mu{}^\nu = 2\partial{\cal E}/\partial{\cal I}\
\psi_{,\mu}  \psi^{,\nu} - {\cal E} \delta_\mu{}^\nu
\lb{eq:Tmunu}
\ee  
Of course not every function ${\cal E}$ leads to a physical theory.  It
is reasonable to restrict attention to fields that bear locally positive
energy density as seen by {\it any\/} physical observer.  Unless ${\cal E}>0$
{\it and\/} $\partial{\cal E}/\partial{\cal I}>0$ for any  $\psi$ and ${\cal
I}>0$, some observer (represented by its 4-velocity $u^\mu$) will see
negative energy density $T_{\mu\nu}u^\mu u^\nu$ somewhere in a stationary
scalar field configuration.   Thus we assume ${\cal E}>0$ and $\partial{\cal
E}/\partial{\cal I}>0$.  The action (\ref{eq:action}) with $\xi=0$ is just
(\ref{eq:modified}) with ${\cal E} = {\scriptstyle 1\over \scriptstyle
2}\left[{\cal I} + V(\psi^2)\right]$ and satisfies both conditions provided
$V$ is positive definite.  In fact, any potential bounded from below will do
because one can add a suitable constant to it to make it nonnegative.

Are there spherically symmetric stationary black hole solutions of the action
(\ref{eq:modified})~\cite{Bek95} ? The $r$ component of the
energy-momentum conservation law
$T_\mu{}^\nu{}_{;\nu}=0$ takes the form~\cite{LLFields} 
\be 
[(-g)^{1/2}\,  T_r{}^r]'-{\scriptstyle 1\over \scriptstyle 2}(-g)^{1/2}\,
(g_{\alpha\beta})'   T^{\alpha\beta}=0,   
\lb{eq:conserve}
\ee
where $'\equiv\partial/\partial r$.  Because of the stationarity and
spherical symmetry, $T_\mu{}^\nu{}$ must be diagonal and
$T_\theta{}^\theta=T_\varphi{}^\varphi$.  These conditions allow us to
rewrite Eq.~(\ref{eq:conserve}) in the form  
\be (e^{\lambda+\nu\over
2}r^2 T_r{}^r)'-{\scriptstyle 1\over \scriptstyle 2}e^{\lambda+\nu\over 2}r^2
\left[\nu' T_t{}^t+\lambda' T_r{}^r + 4 T_\theta{}^\theta/r\right]=0.    
\ee
The  terms containing $\lambda'$ cancel out so that
\be
(e^{\nu/2}r^2
T_r{}^r)'={\scriptstyle 1\over \scriptstyle 2}e^{\nu/2}r^2\left[\nu' T_t{}^t
+ 4 T_\theta{}^\theta/r\right].
\lb{eq:main}
\ee
Eq.~(\ref{eq:Tmunu}) and the symmetries show that
$T_t{}^t=T_\theta{}^\theta=-{\cal E}$. Substituting this in the r.h.s. of
Eq.~(\ref{eq:main}) and  rearranging the derivatives we get our key
expression  
\be (e^{\nu/2}r^2 T_r{}^r)'=-(e^{\nu/2}r^2)'{\cal E}.
\lb{eq:conseq}
\ee
Let us now integrate Eq.~(\ref{eq:conseq}) from $r=r_{\cal H}$ to a generic
$r$.  The boundary term at the horizon vanishes because $e^\nu=0$ and 
$T_r{}^r$ is finite there.    We get
\be 
T_r{}^r(r)=-{e^{-\nu/2}\over r^2}\int_{r_{\cal H}}^r{(r^2 e^{\nu/2})'\,{\cal
E}}dr. \lb{eq:integral2}
\ee

\begin{figure}[ht]
\begin{center}
\psfig{file=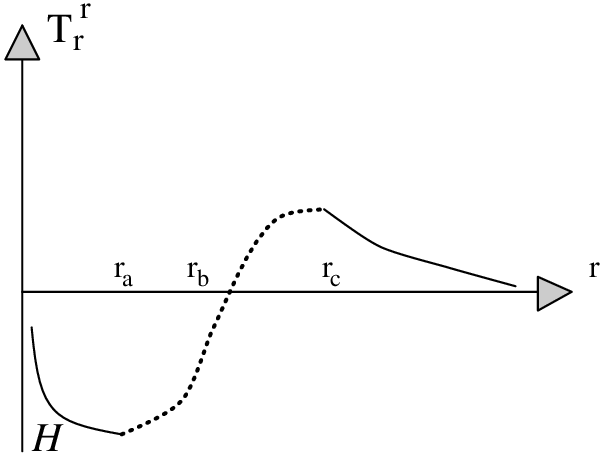,height=5cm}
\label{Fig.1}
\end{center}
\begin{caption}[] 
{Energy momentum conservation reveals the shape of $T_r{}^r$ vs. $r$ near
the horizon $H$ and asymptotically; continuity requires us to complete the
curve so that it rises as well as crosses the $r$ axis.}
\end{caption}
\end{figure}    

Now, since $e^\nu$ vanishes at $r=r_{\cal H}$ and must be positive outside
it, $r^2 e^{\nu/2}$ must grow with $r$ sufficiently near the horizon. It is
then immediately obvious from Eq.~(\ref{eq:integral2}) and the positivity of
${\cal E}$ that sufficiently near the horizon, $T_r{}^r<0$ (see Fig.~1).

Further, carry out the differentiation in Eq.~(\ref{eq:conseq}) and
rearrange terms to get 
\be
(T_r{}^r)'=-e^{-\nu/2}r^{-2}\,(r^2 e^{\nu/2})'\,({\cal E}+T_r{}^r). \lb{eq:D}
\ee
From Eq.~(\ref{eq:Tmunu}) we obtain
\be
{\cal E}+T_r{}^r=2e^{-\lambda}(\partial {\cal E}/\partial{\cal I})\psi_{,r}^2
\lb{eq:sign}
\ee
This is positive by our assumptions.  It then follows from Eq.~(\ref{eq:D})
and our previous conclusion about $r^2 e^{\nu/2}$ that sufficiently near the
horizon $(T_r{}^r)'<0$ as well. 

Since asymptotically $e^{\nu/2}\rightarrow 1$,  Eq.~(\ref{eq:D})  also tells
us that $(T_r{}^r)'<0$ asymptotically.  We mentioned already in connection
with Eq.~(\ref{eq:e_lambda}) that  $T_t{}^t=-{\cal E}$ must decrease
asymptotically faster than $r^{-3}$ to guarantee asymptotic flatness of the
solution.  Thus the integral in Eq.~(\ref{eq:integral2}) converges and
$|T_r{}^r|$ decreases asymptotically as $r^{-2}$.   But since $(T_r{}^r)'<0$
asymptotically, we deduce that $T_r{}^r$ must be positive and decreasing with
increasing $r$ as $r\rightarrow\infty$, as depicted in Fig.~1.   Now we found
that near the horizon $T_r{}^r<0$ and $(T_r{}^r)'<0$.   All these facts
together tell us that in some intermediate interval  $[r_{\rm a}, r_{\rm
b}]$,  $(T_r{}^r)'>0$  and also that $T_r{}^r$ itself changes sign at some
$r_{\rm c}$, with $r_{\rm a}<r_{\rm c}<r_{\rm b}$, being positive in $[r_{\rm
c}, r_{\rm b}]$ (see Fig.~1; there may be several such intervals $[r_{\rm a},
r_{\rm b}]$).   Well, it turns out that this conclusion is  incompatible with
the Einstein equations, to which we now turn.

First we note from Eq.~(\ref{eq:e_lambda}) that $e^\lambda\ge 1$ throughout
the black hole exterior (recall $T_t{}^t=-{\cal E}<0$).  Next we recast
Eq.~(\ref{eq:Einstein2}) in the form
\be
e^{-\nu/2}r^{-2}\,(r^2 e^{\nu/2})'=[4\pi r GT_r{}^r +
(1/2r)]e^\lambda+3/2r> 4\pi rGT_r{}^r e^\lambda+2/r, 
\ee   where the inequality results because $e^\lambda/2 +3/2 > 2$. 
We found that in $[r_{\rm c}, r_{\rm b}]$,  $T_r{}^r>0$. 
Thus $e^{-\nu/2}r^{-2}\,(r^2 e^{\nu/2})'>0$ there.  According to
Eq.~(\ref{eq:D}) this means that $(T_r{}^r)'<0$ throughout $[r_{\rm c},
r_{\rm b}]$. However, we determined that $(T_r{}^r)'>0$ throughout the
encompassing interval $[r_{\rm a}, r_{\rm b}]$.  Thus there is a
contradiction: the solution as we have been imagining it does not exist. 

To escape the contradiction we must have $T_r{}^r=0$ identically in the
black hole exterior.  According to Eq.~(\ref{eq:conseq}) this implies that
${\cal E}=0$ identically.  It then follows from  Eq.~(\ref{eq:sign}) that
$\psi$ must be constant throughout the black hole exterior, taking on a value
which makes $T_{\mu\nu}=0$. Such a values must exist in order that a trivial
solution of the scalar equation be possible in Minkowski spacetime.  It is
precisely this solution which served as an asymptotic boundary condition in
our argument.  By Birkhoff's theorem the spherical stationary black hole
solution of action (\ref{eq:modified}) must be identically Schwarzschild. 
This rules out hair in the form of a neutral minimally coupled scalar field. 
This result can be generalized to many scalar fields
~\cite{Bek95}.

The advantage of this theorem~\cite{Bek95} and those of Heusler~\cite{Heu}
and Sudarsky~\cite{Sud} over the older one of Sec.~\ref{updown} is that
now we can rule neutral Higgs hair provided only $V\geq 0$, without need to
invoke $V'>0$ which is often violated in field theoretic models.  A
disadvantage is that the present theorems work only for spherical symmetry,
and do make use of Einstein's equations.  However, the theorem just described
has been extended to the Brans-Dicke theory~\cite{Bek95} (see also Ayon's
work~\cite{Ayon2}), as well as to electrically charged black
holes~\cite{MB}.  Removal of the static and spherical symmetry assumptions is
a  thing for the future; some headway has been reported by Ayon~\cite{Ayon1}.

\subsection{No curvature coupled scalar hair} \lb{nonminimally}

Consider now a hairy spherically symmetric stationary black hole solution of
the action (\ref{eq:action}) with $\xi\neq 0$ (curvature coupled), and with 
$V\geq 0$ but no electric charge.  The curvature coupled field's energy
density is not necessarily positive definite.  Thus I drop the requirement of
positive energy density, but I shall look only at positive $V$ so that a
suitable limit can be taken to the minimally coupled theory discussed
earlier.  In addition, I shall assume the physical black hole configurations
are such that the dominant energy condition~\cite{Hawk_Ellis} is satisfied
everywhere.  This means the absolute value of the energy density bounds all
the other components of the energy-momentum tensor.  

Both Saa~\cite{Saa1,Saa2} and Mayo and I~\cite{MB} realized that this problem
can be mapped onto the one solved in  Sec.~\ref{minimal} by a conformal
transformation of the geometry.  
\be
	g_{\mu\nu} \rightarrow \bar{g}_{\mu\nu}\equiv g_{\mu\nu}\Omega; \quad \Omega
\equiv 1-8 \pi G \xi \psi^2 
\lb{eq:map1}
\ee
Under this map the action (\ref{eq:action}) is  transformed into  
\bea
	S&=&{1\over 16\pi
G}\int{\bar{R}\sqrt{-\bar{g}}d^4x}-{\scriptstyle 1\over \scriptstyle 2}
\int{\left[(1+f)\bar{g}^{\alpha\beta}\psi_{,\alpha} \psi_{,\beta} + \bar{V}
		\right]\sqrt{-\bar{g}}d^4x}
\nonumber \\	  
	 f&\equiv&48 \pi G \xi^2 \psi^2 (1-8 \pi G \xi \psi^2)^{-2}
\nonumber \\
	\bar{V}&\equiv&V(\psi^2)(1-8 \pi G \xi \psi^2)^{-2}
\eea
The transformed action is of the form (\ref{eq:modified}), and the field
$\psi$ obviously bears positive energy with respect to $\bar{g}_{\mu\nu}$,
not least because of the assumed positivity of $V(\psi^2)$.  Further, the map
leaves the  mixed components $T_\mu^\nu$ unaffected so that the boundedness
of these can be assumed also in the new geometry.   Applying the previous
theorem would seem to allow us to rule out hair coupled to curvature. 
Saa~\cite{Saa1} came to just such a conclusion by a very similar approach.  

But in fact, things are not so straightforward.  Suppose that in the proposed
black hole solution (metric $g_{\mu\nu}$) $\psi$ is such that $\Omega$ can
become negative in some domain outside the horizon, or vanish or blow up at
some exterior point.  Then the new metric $\bar g_{\mu\nu}$ is just not
physical (it has wrong signature, or is degenerate).  One cannot then use the
theorem in Sec.~\ref{minimal} because it refers to physical configurations. 
In his first paper Saa~\cite{Saa1} did not address this issue; in his second
one~\cite{Saa2} he formulated the no-hair theorem to apply only if $|\psi|$
in the proposed solution is bounded everywhere for $\xi<0$ or is bounded by a
number depending on $\xi$ for
$\xi>0$.  But these are not reasonable expectations: nature may decide to
have a solution with very large $|\psi|$ somewhere, and it is not clear
outright that divergence of $|\psi|$ is unphysical.  It is thus best to prove
the no scalar hair theorem by breaking it up into cases and showing for each
that, under natural assumptions, $\Omega$ is well behaved for any physically
reasonable hairy solution, thus allowing use of the theorem in
Sec.~\ref{minimal} to exclude it.  This is done in Mayo and
Bekenstein~\cite{MB}; what follows is a simplified version.

Suppose first $\xi<0$; then $\Omega$ cannot be negative or vanish by
definition.  We prove it cannot blow up in a physical black hole's exterior
as follows.  $\Omega$ can blow up only where $|\psi|$ blows up. In a physical
solution $|\psi|$ should not blow up asymptotically because its value there
has to correspond to the one in a flat spacetime solution.  So suppose
$|\psi|$ blows up at some finite point $r=r_c>r_{\cal H}$.  Then as
$r\rightarrow r_c+\epsilon$,  $\psi_{,r}/\psi\rightarrow -\infty$ and
$\psi_{,rr}/\psi\rightarrow +\infty$.  Now from Eq.~(\ref{eq:TGnew})
calculate
\be
T_t{}^t-T_r{}^r = e^{-\lambda}{  (2\xi-1) \psi_{,r}^2 - \xi (\nu+\lambda)'
\psi \psi_{,r} + 2\xi \psi \psi,_{rr} \over {1-8\pi G \xi \psi^2}}
\lb{eq:Ttt-Trr}
\ee
However, by means of Eq.~(\ref{eq:eq.rr-eq.tt}) we may rewrite 
Eq.~(\ref{eq:Ttt-Trr}) as
\be
T_t{}^t-T_r{}^r={e^{-\lambda}[(2\xi-1)(\psi_{,r}/\psi)^2+2\xi
\psi_{,rr}/\psi]\over 1/\psi^2-8\pi G\xi-8\pi G\xi r  (\psi_{,r}/\psi)}
\lb{eq:newTtt-Trr}
\ee
In light of the mentioned divergences we see that
$T_t{}^t-T_r{}^r\rightarrow +\infty$ as $r\rightarrow r_c+\epsilon$ because
the quantities in the numerator are of like sign for $\xi<0$.  But, as
mentioned in Sec.~\ref{spherical}, divergence of any diagonal component
$T_\mu{}^\nu$ in a spherically symmetric situation is incompatible with a
physical solution.  We conclude that $|\psi|$ cannot blow up at any
$r_c>r_{\cal H}$.

But could $|\psi|$ blow up at the horizon itself in a physical solution ? 
According to Eq.~(\ref{eq:equality}) the r.h.s. of (\ref{eq:newTtt-Trr}) must
vanish at  $r_{\cal H}$.  Were $\psi$ to have a pole or a branch point there,
this vanishing would be impossible in view of the behavior of $e^{-\lambda}$
in Eq.~(\ref{eq:e_lambda_first}).  We conclude that $|\psi|$ cannot blow up
even at $r_{\cal H}$.

Therefore, for $\xi<0$ a physical black hole solution of action
(\ref{eq:action}) defines an everywhere positive and bounded $\Omega$.  The
mapping and use of the theorem in Sec.~\ref{minimal} then excludes this
solution rigorously.  The  discussion assumed the black hole is not extremal. 
In fact it can be generalized to exclude extremal as well as electrically
charged black holes with $\xi<0$ hair~\cite{MB}.

Let us now turn to the case $\xi\geq 1/2$.  The mapping strategy has
here been applied rigorously to rule out electrically charged
holes~\cite{MB}, but does not work well for the neutral ones.  Another line
of argument does the job.  The first point to notice is that we expect $\psi$
to asymptote to a definite finite value
$\psi_0$.  This should be such as to make $1/\psi^2-8\pi G\xi$ positive since
as clear from Eq.~(\ref{eq:TGnew}), $(1-8\pi G\xi\psi^2)^{-1}$ plays the role
of gravitational constant in the asymptotic region, and this should always be
positive regardless of how unconventional the black hole itself may be. 
Unless $\psi_0=0$, $\psi_{,r}/\psi$ must fall off faster than  $r^{-1}$, so
that the denominator in Eq.~(\ref{eq:newTtt-Trr}) is asymptotically positive.
And if $\psi_0=0$, then $\psi_{,r}/\psi$ will behave like $r^{-1}$ so that
the denominator is dominated by
$1/\psi^2$ and is again asymptotically positive. 

Let us now complement Eq.~(\ref{eq:newTtt-Trr}) with
\be
	T_t{}^t-T_\phi{}^\phi = 
	 {\xi e^{-\lambda}(2/r-\nu') (\psi_{,r}/\psi) \over  1/\psi^2-8\pi G\xi}
\lb{eq:Ttt-Tphph}
\ee
As mentioned in Sec.~\ref{spherical}, $\nu' = {\cal O}(r^{-2})$ 
asymptotically. If $|\psi|$ decreases asymptotically towards $|\psi_0|$, so
that that  $\psi_{,r}/\psi<0$ and $\psi_{,rr}/\psi >0$, then it follows
from Eq.~(\ref{eq:Ttt-Tphph}) that in the asymptotic region
$T_t{}^t-T_\phi{}^\phi<0$, while from Eq.~(\ref{eq:newTtt-Trr}) it is clear
that $T_t{}^t-T_r{}^r>0$.  And if
$|\psi|$ increases asymptotically, $\psi_{,r}/\psi>0$ while $(\psi^2)_{,rr}
<0$, so that  asymptotically $T_t{}^t-T_\phi{}^\phi>0$.  In addition,
rewriting Eq.~(\ref{eq:newTtt-Trr}) in the form
\be
T_t{}^t-T_r{}^r={e^{-\lambda}[\xi
\psi^2_{,rr}/\psi^2-(\psi_{,r}/\psi)^2]\over 1/\psi^2-8\pi G\xi-8\pi G\xi r 
(\psi_{,r}/\psi)}
\lb{eq:final}
\ee
shows clearly that $T_t{}^t-T_r{}^r<0$ asymptotically.  In both cases it is
impossible for $|T_t{}^t|$ to dominate in magnitude both $|T_r{}^r|$ and
$|T_\phi{}^\phi|$, as required by the dominant energy condition.  Thus unless
$\psi$ is strictly constant, one cannot even give the black hole a physical
asymptotic region.  We conclude that there are no hairy black holes for
$\xi\geq 1/2$.

The case $0<\xi<1/2$ remains open.  Removal of the spherical
symmetry assumption is yet to be accomplished (but see
Ayon's work~\cite{Ayon1}).

When the scalar $\psi$ becomes complex (Higgs field) and couples to the
black hole's electromagnetic field, things become more complicated.  Theorems
ruling out nonextremal or extremal black holes for {\it any\/} $\xi$ have
been given~\cite{Adler,Lahiri,MB,Ayon-Beato}.

\section{Superradiance} \lb{Superradiance}

\setcounter{equation}{0}
\renewcommand{\theequation}{2.\arabic{equation}}

To the generation that witnessed the emergence of black hole physics in the
1970's, superradiance is a typical black hole phenomenon.  Actually, forms of
superradiance had been identified already in the 1940's in connection
with experimental phenomena like the Cherenkov effect.  And, of course, the
name is also applied to the physics behind the laser and maser, which is not
the sense in which I use it here.  I give here a self-contained review of
various aspects of superradiance, from ordinary objects to black holes. 
Further details can be found in references~\cite{BekSchiff,SaaSchiffer}. 

\subsection{Inertial motion superradiance}  \lb{Inertial}

It follows from Lorentz invariance  and four-momentum conservation that a
free structureless particle moving inertially in vacuum cannot absorb or emit
a photon.  But suppose a particle, possibly with complex structure, moves
inertially through a medium transparent to photons.  Then it can
spontaneously emit photons, even if it started in the ground state !  To see
this let (as in Fig.~2) $E$ and $E'= E -\hbar\omega$ denote the particle's
total energy in the {\it laboratory\/} frame before and after the emission
of a photon with energy $\hbar\omega$ and momentum $\hbar {\bf k}$ (both
measured in the laboratory frame), while ${\bf P}$ and  ${\bf P'}={\bf
P}-\hbar{\bf k}$ denote the corresponding momenta; ${\bf v}=\partial
E/\partial {\bf P}$ is the initial  velocity of the particle. The Lorentz
transformation to the particle's rest frame gives us the rest energy or rest
mass $M=\gamma(E-{\bf v}\cdot {\bf P})$ with $\gamma\equiv (1-{\bf
v}^2)^{-1/2}$.  Immediately after emission $M'=\gamma'(E'-{\bf v'}\cdot {\bf
P'})$.

\begin{figure}[ht]
\begin{center}
\psfig{file=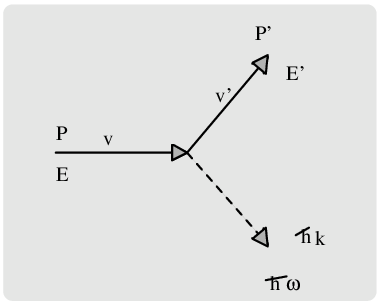,height=5cm}
\label{Fig.2}
\end{center}
\begin{caption}[] 
{Particle with initial energy $E$ and momentum ${\bf P}$ moving through a
transparent medium emits a photon of momentum $\hbar{\bf k}$ and energy
$\hbar \omega$ thereby changing its velocity from ${\bf v}$ to ${\bf v}'$.}
\end{caption}
\end{figure}   

Now substract the formulae for $M'$ and $M$ and neglect terms of order higher
in  ${\cal O}(\omega), {\cal O}({\bf k})$ and ${\cal O}({\bf v'}-{\bf v})$:
\be
M'-M=-\gamma\hbar(\omega -{\bf v}\cdot {\bf k}) +\hbar \omega \cdot 
{\cal O}({\bf
v'}-{\bf v})
\lb{eq:basic}
\ee
The factor ${\cal O}({\bf v'}-{\bf v})$ represents recoil effects; it
is of order $\hbar\omega/M$ and becomes negligible for a sufficiently heavy
particle. In this recoiless limit
\be
M'-M=-\gamma\hbar(\omega -{\bf v}\cdot {\bf k})
\lb{eq:emission}
\ee

Were the particle moving in vacuum, $\,\omega=|{\bf k}|>{\bf v}\cdot {\bf
k}$, so that emission would be possible only with de-excitation ($M'-M<0$),
as plain intuition  would have.  But in the medium intuition receives a
surprise.  Let its index of  refraction be $n(\omega)>1$.  Then $\hbar\omega$
and $\hbar{\bf k}$ are still the energy and momentum of the photon; however
$\omega=|{\bf k}|/n(\omega)$. In the case $v\equiv |{\bf v}|>1/n(\omega)$ the
particle moves faster than the {\it phase\/} velocity of electromagnetic waves
of frequency $\omega$.  If $\vartheta$ denotes the angle between ${\bf k}$
and ${\bf v}$, a photon in a mode with $\cos\vartheta > [v\, n(\omega)]^{-1}$
has $\omega -{\bf v}\cdot {\bf k}<0$, and can thus be {\it emitted\/} only in
consonance with {\it excitation\/} of the object ($M'-M>0$) !  In particular,
a particle in its ground state can emit a photon.  Ginzburg and
Frank~\cite{GinzFr,Ginz}, who pointed out these phenomena, refer to this
eventuality as the {\it anomalous\/} Doppler effect.  The reason for the name
is that in the case  $v<1/n(\omega)$ (subluminal motion for the relevant
frequency) when  $\omega -{\bf v}\cdot {\bf k}>0$ so that by
Eq.~(\ref{eq:basic}) emission can take place only by de-excitation, the
relation between $\omega$ and ${\bf k}$ and the rest frame transition
frequency
$\omega_0\equiv |M-M'|/\hbar$, namely
\be
\omega_0 = \gamma(\omega - {\bf v}\cdot{\bf k}),
\lb{eq:Doppler}
\ee
is the standard Doppler shift formula; indeed Ginzburg and Frank refer to
this case as the normal Doppler effect.  We shall refer to the emission as
spontaneous superradiance.

The energy source for superradiant emission and the associated excitation is 
the bulk motion of the particle.  And this emission is not just allowed by
the conservation laws; it must occur spontaneously, as follows from
thermodynamic reasoning.  The particle in its ground state with no photon
around constitutes a low entropy state; the excitation of the object to one
of a number of possible excited states with emission of a photon with
momentum in a variety of possible directions evidently involves an increase
in entropy.  Thus the emission is favored by the second law of thermodynamics.

The inverse anomalous Doppler effect or superradiant absorption can also
take place: when superluminally moving, the particle can absorb a photon only
by getting de-excited, and cannot absorb while in the ground state !  The
appropriate equation is obtained from Eq.~(\ref{eq:basic}) by reversing the
sign.  Obviously superradiance is not restricted to photons.   All that is
required is that the energy and momentum of a quantum be expressible in terms
of frequency and wavevector in the usual way.  Thus superradiance can take
place for phonons in fluids, plasmons in plasma, etc.

When the particle has no internal degrees of freedom, say a point charge, its
rest mass is fixed.  We may thus set $M'-M=0$ in Eqs.~(\ref{eq:emission}).
The equation cannot then be satisfied for $v < 1/n(\omega)$ since its
r.h.s. would then be strictly positive: again no absorption or emission is
possible from a subluminal particle.  However, for $v > 1/n(\omega)$ the
r.h.s. vanishes for a photon's whose direction makes an angle
$\vartheta$ to the particle's velocity, where $\cos\vartheta=
[v\,n(\omega)]^{-1}$.  Such photons must thus be emitted.    Obviously as the
charge goes by, the front of photons forms a cone with opening angle
$2\Theta_C=2(\pi/2-\vartheta)$, or $\sin\Theta_C(\omega) =
[v\,n(\omega)]^{-1}$. This result makes it clear that one is here dealing
with the famous Cherenkov radiation, which comes out on just such a cone. 
Thus Cherenkov radiation is an example of spontaneous superradiance by a
structureless charge~\cite{Ginz}.  Another example~\cite{BekSchiff} is
furnished by the Mach shock cone trailing a supersonic object, whose opening
angle also corresponds to the condition
$\omega-{\bf v}\cdot {\bf k} = 0$.

\subsection{Superradiant amplification}   \lb{Amplification}  

The above section deals with {\it spontaneous superradiance\/} which occurs
when the Ginzburg-Frank condition
\be
\omega - {\bf v}\cdot {\bf k} <  0
\lb{eq:GinzFr}
\ee
is satisfied.  I mentioned that the radiation must be emitted in order
that the world's entropy may increase.  Einstein's celebrated argument
inextricably connects spontaneous emission with stimulated emission. 
Therefore, when condition (\ref{eq:GinzFr}) is satisfied, there must also
occur amplification of preexisting radiation by an object moving
superluminally (supersonically) in a medium.  Rather than dwell on the simple
particle, I shall show this for an object with complicated structure, so that
it may dissipate energy internally.  The demonstration is thermodynamical
(and basically classical).  For concreteness I suppose the object to move in
a transparent medium filled with electromagnetic radiation.  

Let the radiation be exclusively in modes with frequency near $\omega$ and
propagating within $\Delta {\bf n}$ of the direction ${\bf n}$.  Also let
$I(\omega, {\bf n})$ denote the corresponding intensity (per unit area, unit
solid angle and unit bandwidth). Experience tells us that the body will
absorb power $a(\omega, {\bf n})\,\Sigma({\bf  n})\,I(\omega, {\bf
n})\,\Delta\omega\Delta {\bf n}$, where $\,\Sigma({\bf  n})\,$ is the
object's geometric crossection orthogonal to direction ${\bf n}$, and
$a(\omega, {\bf n})<1$ is its absorptivity for the mentioned photons.
The remainder power, $[1-a(\omega, {\bf n})]\,\Sigma({\bf  n})\,I(\omega, 
{\bf n})\,\Delta\omega\Delta {\bf n}$, will be scattered.  In addition the
object may emit spontaneously some power $W$, say by thermal emission.  By
conservation of energy, absorption and emission cause the object's total
energy (in the laboratory frame) $E$ to change at a rate  
\be    
dE/dt= a\,\Sigma\,I\,\Delta\omega\Delta {\bf n} - W 
\lb{eq:Erate}
\ee 

Now the linear momentum conveyed by the radiation is ${\bf k}/\omega$ times
the energy conveyed, where ${\bf k}={\bf n}\,\omega n(\omega)$.  This is
clear if we think of the radiation as composed of quanta, each with energy
$\hbar\omega$ and  momentum $\hbar{\bf k}$ with $\omega\,  n(\omega)  =|{\bf
k}|$.  The result can also be derived from the temporal-spatial and
spatial-spatial components of the energy-momentum tensor for the
electromagnetic field in a medium.  Thus absorption and emission
cause the linear momentum ${\bf P}$ of the body to change at a rate
\be   
d{\bf P}/dt= ({\bf k}/\omega)\,a\,\Sigma\,I\,\Delta\omega\Delta {\bf
n} - {\bf U}  
\lb{eq:Prate} 
\ee
where ${\bf U}$ signifies the rate of spontaneous momentum emission.

In calculating the rate of change of rest mass $M$ of the body,
I ignore the effects of elastic scattering because in the frame of
the body waves are scattered with no Doppler shift (since there is no
motion),so they contain the same energy before and after the
scattering.  Thus the scattering cannot contribute to $dM/dt$.  Obviously
the change in $M$ is obtained by a Lorentz transformation:
\be
dM/dt = \gamma(dE/dt - {\bf v}\cdot d{\bf P}/dt)
\lb{eq:Mrate}
\ee
Of course, a change in the proper mass means that the number of microstates
accessible to the object has changed, i.e., that its entropy $S$ has
changed. Recalling the definition of temperature $T=\partial M/\partial S$
and Eqs.~(\ref{eq:Erate})--(\ref{eq:Prate}), we see that
\be 
{dS\over dt}= \gamma T^{-1}\,[\omega^{-1}\,(\omega - {\bf v}\cdot {\bf k})\,
a\,\Sigma\,I \,\Delta\omega\Delta
{\bf n} - W +{\bf v}\cdot{\bf U}\,]
\lb{eq:Srate}
\ee

The second law does not allow the claim that this last expression is positive
because there is also a change in the entropy in the radiation.  But one can
put an upper bound on the rate of change of the radiation entropy, $d{\cal
S}/dt$ by ignoring any entropy carried {\it into\/} the object by the
radiation.   Now the entropy in a single mode of a field containing on the
mean $N$ quanta is at most
~\cite{LLSP1} 
\be
{\cal S}_{\rm max}= (N+1)\ln (N+1) - N\ln N \approx \ln N
\lb{eq:Smax}
\ee
where the approximation applies for $N\gg 1$.   The scattered waves carry a
mean number of quanta proportional to $I(\omega, {\bf n})$. Hence for large
$N$ the outgoing radiation's contribution to $d{\cal S}/dt$ is bounded from
above by a quantity of $O[\ln I(\omega, {\bf n})]$.  There is an additional
contribution of ${\cal O}(W)$ to $d{\cal S}/dt$  coming from the spontaneous
emission.  Hence
\be
d{\cal S}/dt < {\cal O}[\ln I(\omega, {\bf n})] + {\cal O}(W)
\lb{eq:Srad}
\ee  

Because the object dissipates energy, the second law of thermodynamics
demands $dS/dt+d{\cal S}/dt > 0$.  As $I(\omega, {\bf n})$ is made  larger
and  larger, the total entropy rate of change becomes dominated by the term
proportional to $I(\omega, {\bf n})$ in Eq.~(\ref{eq:Srate}) because $W$ and
${\bf U}$ are kept fixed.  Positivity of $dS/dt+d{\cal S}/dt$ then requires
\be
(\omega - {\bf v}\cdot {\bf k})\, a(\omega, {\bf n}) > 0
\lb{eq:acondition}
\ee
Thus when the Ginzburg-Frank condition is fulfilled,  $a(\omega, {\bf n}) <
0$. This result was obtained by assuming $a\,\Sigma\,I\,\Delta\omega\Delta
{\bf n} \gg W$.  But since---barring nonlinear effects---$a$ must be
independent of the incident intensity, the result must be true for any
intensity which can still be regarded as classical.  Now $a < 0$ means that
the scattered wave, with power proportional to $1- a$, is stronger than the
incident one (which is represented by the  ``1'' in the previous expression). 
Thus the moving object must amplify preexisting radiation in modes satisfying
the Ginzburg-Frank condition. Superradiant amplification is mandatory.  For
modes with $\omega - {\bf v}\cdot {\bf k} > 0$, $a > 0$ and so the object
absorbs on the whole.

Obviously $a$ switches sign at $\omega={\bf v}\cdot{\bf k}$.  This switch
cannot take place by $a$ having a pole since $a <1$. {\it If\/} $a$ is
analytic in  $\omega-{\bf v}\cdot{\bf k}$, it must thus have the expansion  
\be
a\,=\,\alpha({\bf v}, {\bf n})\,(\omega-{\bf v}\cdot{\bf
k}) + \cdots
\lb{eq:expansion}
\ee
in the vicinity of the superradiant treshold $\omega\,=\,{\bf
v}\cdot{\bf k}$.  However, we must emphasize that thermodynamics does not
require the function $a$ to be continuous at $\omega\,=\,{\bf v}\cdot{\bf
k}$.

As an example of both spontaneous superradiance and
superradiant amplification we rederive Landau's critical velocity for
superfluidity ~\cite{LLSP2}.  A superfluid can flow through thin channels
with no friction.  However, when the speed of flow is too large, the
superfluidity is destroyed.  As Landau did, I phrase the argument in the rest
frame of the fluid with respect to which the walls of the channel are in
motion.  The walls play the role of the object in our superradiance argument,
and the waves of frequency $\omega=\varepsilon/\hbar$ and wavenumber ${\bf
k}={\bf p}/\hbar$ associated with the quasiparticles in the fluid are
surrogates of the electromagnetic waves in both our above arguments.  In
superfluid He$^4$ the dispersion relation $\varepsilon({\bf p})$ has a
nonvanishing minimum:  
$v_c \equiv {\rm min}\ \varepsilon({\bf p})/|{\bf p}|>0$.

When the walls move with speed $v>v_c$, the quantity $\omega - {\bf
v}\cdot{\bf k}=  (\varepsilon - {\bf v}\cdot{\bf p})/\hbar\ $ becomes
negative for at least one quasiparticle mode. According to
Sec.~\ref{Inertial} the wall material will then become excited {\it and\/}
simultaneously create quasiparticles in those modes.  Furthermore
(Sec.~\ref{Amplification}), the quasiparticles thus created can undergo
superradiant multiplication while impinging on other parts of the walls.  As
a consequence, an avalanche of quasiparticle formation ensues, which acts to
convert the superfluid into a  normal fluid.  Thus the transition away from
superfluidity is a literal  example of the superradiance phenomenon.  In this
phenomenon the speed $v_c$, of order the speed of sound, plays the role of
the speed of light in our original arguments.

\subsection{Gravitational generation of electromagnetic waves}  
\lb{GtoEM}

Now for our first black hole example.  Consider an  electrically neutral
black hole of mass $M$ moving with uniform velocity $\bf v$  through a
uniform and isotropic transparent dielectric with index of refraction
$n(\omega)$ made of material with atomic mass number $\tilde A$ and  pervaded
by a spectrum of electromagnetic waves.  We could be thinking about an
astronomical sized black hole moving through a cloud of gas, or about a
microscopic black hole whizzing through a solid state detector.  Anyway, I
assume the hole does not accrete material; however, its gravitational field
certainly influences the dielectric. 

In applying the argument of Sec.~\ref{Amplification}, the entropy of
the object is replaced by the black hole entropy together with entropy of the
surrounding dielectric.  Now black hole entropy is proportional to the
horizon area, and  Hawking's area theorem~\cite{Hawking_area} tells us that
black hole area will increase in any classical process, such as absorption of
electromagnetic waves by the hole.   If the dielectric is ordinary
dissipative material, it will also contribute to the increase in entropy
through changes it undergoes in the vicinity of the passing hole.   Thus an
argument like that in in Sec.~\ref{Amplification} tells us that the black
hole plus surrounding dielectric will amplify radiation modes obeying the
Ginzburg-Frank condition at the expense of the hole's kinetic energy. 
Likewise, even if there are no waves to start with, an argument like that in
Sec.~\ref{Inertial} tells us that the black hole plus dielectric will
spontaneusly emit electromagnetic waves in modes that obey the condition.  

The process in question is distinct from the standard Cherenkov effect
because the hole is neutral. Now waves cannot classically emerge from within
the hole, so what is their source ?  The hole's gravity pulls on the
positively charged nuclei in the dielectric stronger than on the enveloping
electrons.  As a result the array of nuclei sags with respect to the
electrons, and produces an electrical polarization of the dielectric
accompanied by an electric field which ultimately balances the tendency of
gravity to rip out nuclei from electrons.  It is this electric structure
which is to be viewed as the true source of the waves.  If one is interested
in the intensity of this gravitationally induced electromagnetic radiation,
one may map the present problem onto the Cherenkov one by noting that the
induced electric field ${\bf E}$ is related to the gravitational one, ${\bf
g}$ by $e{\bf E}=-\delta \mu\, {\bf g}$ where $\delta \mu\approx \tilde A
m_p$ is the nuclei-electron mass difference, and $e>0$ the unit of charge. 
From the gravitational Poisson equation it follows that $\nabla\cdot{\bf
E}=4\pi G M(\delta \mu/e)\delta({\bf r}-{\bf r}_0)$ where ${\bf r}_0$ denotes
the momentary black hole position.  The electric field accompanying the black
hole is thus that of a pointlike charge $Q\equiv G\tilde AMm_p/e$.  This
assumes, and this is no trivial assumption~\cite{BekSchiff,SaaSchiffer}, that
the dielectric has time to relax to allow for the generation of the
compensating field.  If so, the electromagnetic radiation will be Cherenkov
radiation of a charge $Q$ moving with velocity ${\bf v}$.  In units of $e$,
$Q$ amounts to about $10^3\tilde A$ times the gravitational radius of the
hole measured in units of the classical radius of the electron.  Hence a
relativistically moving $10^{15}$ g primordial black hole would radiate just
like particle with  $\sim 10^3 \tilde A$ elementary charges.

\subsection{Rotational superradiance}\lb{rotational}

Zel'dovich came upon the notion of black hole superradiance by examining what
happens when scalar waves impinge upon a rotating absorbing object
~\cite{Zeld1}.  His later thermodynamic proof~\cite{Zeld2} that this
superradiance is a general feature of rotating objects and any waves provides
the inspiration for the argument given in Sec.~\ref{Amplification}.  Here I
just elaborate on Zel'dovich's original proof by taking into account the
radiation entropy, which he neglected.

I focus on an axisymmetric macroscopic object rotating rigidly in vacuum
with constant angular velocity $\Omega$ about a constant axis. Axisymmetry is
critical; otherwise precession of the axis would arise.   I consider the
object to have many internal degrees of freedom, so that it can internally
dissipate absorbed energy, and that it rapidly reaches equilibrium with well
defined entropy $S$, rest mass $M$ and temperature $T$. 

Let the object be exposed to external radiation.  By the symmetries we may
classify the radiation modes by frequency $\omega$ and azimuthal number $m$. 
This last refers to the axis of rotation.  Suppose that in the modes with
azimuthal number $m$ and frequencies in the range in
$\{\omega, \omega + \Delta\omega\}$,  power
$I_m(\omega)\,\Delta\omega$ is incident on the body.   Then, as is easy to
verify from the energy-momentum tensor, or from the quantum picture of
radiation, the radiative angular momentum is incident at rate
$(m/\omega)I_m(\omega)\,\Delta\omega$.  If  $I_m(\omega)$ is large enough, we
can think of the radiation as classical.  Experience tells us that the body
will absorb a fraction $a_m(\omega)$ of the incident power and angular
momentum flow in the modes in question, where
$a_m(\omega)<1$ is a characteristic coefficient of the body. A fraction
$[1-a_m(\omega)\,]$  will be scattered back into modes with the same $\omega$
and
$m$.   We may thus replace Eqs.~(\ref{eq:Erate})-(\ref{eq:Prate}) by
\bea    
{dE\over dt}&=& a_m\,I_m\,\Delta\omega - W  
\lb{eq:newErate}
\\
{dJ\over dt}&=& (m/\omega)\,a_m\,I_m\,\Delta\omega - U_J 
\lb{eq:newJrate} 
\eea
where $J$ is the body's angular momentum and $U_J$ is the overall rate of
spontaneous angular momentum emission in waves.

Now the energy $\Delta E_0$ of a small system measured in a frame rotating
with angular frequency ${\bf \Omega}$ is related to its energy $\Delta E$ and
angular momentum $\Delta{\bf J}$ in the inertial frame by~\cite{LLMech}
\be
\Delta E_0 = \Delta E - {\bf \Omega}\cdot\Delta{\bf J}
\lb{eq:restenergy}
\ee
Thus, when as a result of interaction with the radiation, the energy of
our rotating body changes by $dE/dt\times \Delta t$ and its angular momentum
in the direction of the rotation axis by $dJ/dt\times \Delta t$,  its
mass-energy in its rest frame changes by $(dE/dt - \Omega\, dJ/dt)\times
\Delta t$.  From this we infer, in parallel with the derivation of
Eq.~(\ref{eq:Srate}), that the body's entropy changes at a rate
\be 
{dS\over dt}= T^{-1}\left[\omega^{-1}\,(\omega - m
\Omega)\,a_m\,I_m\,\Delta\omega - W + \Omega\,U_J\, \right]
\lb{eq:newSrate}
\ee

As in the discussion involving Eqs.~(\ref{eq:Smax})-(\ref{eq:Srad}) we would 
now argue that when $I_m(\omega)$ is large, the term proportional to $(\omega -
m \Omega)\,a_m(\omega)$ in Eq.~(\ref{eq:newSrate}) dominates the overall
entropy balance.   The second law thus demands that
\be
(\omega - m \Omega)\,a_m(\omega) > 0
\lb{eq:newcondition}
\ee
Thus whenever the condition
\be
\omega-m\Omega < 0
\lb{eq:supercriterion}
\ee
is met, $a_m(\omega) <0$ necessarily.  As in Sec.~\ref{Amplification}, we can
argue that the sign of $a_m(\omega)$ should not depend on the strength of the
incident radiation if nonlinear radiative effects do not intervene.  Hence,
independent of the strength of $I_m(\omega)$, condition
(\ref{eq:supercriterion}) is the generic condition for rotational
superradiance.  It was first found in the context of ordinary objects by
Zel'dovich~\cite{Zeld1}.

Evidently $a_m(\omega)$ switches sign at $\omega=\Omega m$.  This
switch cannot take place by $a_m(\omega)$ having a pole there since 
$a_m(\omega)<1$. {\it If\/} $a_m(\omega)$ is analytic in  $\omega-\Omega
m$, it must thus have the expansion  
\be
a_m(\omega)=\alpha_m(\Omega)\,(\omega-\Omega m)+\cdots
\lb{eq:expansion2}
\ee
in the vicinity of the superradiance treshold $\omega=\Omega m$.  However, we
must again stress that thermodynamics does not demand continuity of
$a_m(\omega)$ at
$\omega-\Omega m=0$.  Specific examples like that of the rotating cylinder
~\cite{Zeld2,BekSchiff} do show continuity.

\subsection{Black hole superradiance}  \lb{BHrotational}

By analogy with the results described in Sec.~\ref{rotational},
Zel'dovich~\cite{Zeld2} conjectured that a Kerr black hole should also
superradiate with respect to modes obeying condition
(\ref{eq:supercriterion}).  This was established directly by
Misner~\cite{Misner} for the scalar field case (so that I refer to
(\ref{eq:supercriterion}) as the Zel'dovich-Misner condition), and some
approximate formulae for the gain were worked out by Starobinskii and
Churilov~\cite{Starobinskii} (they confirm the rule (\ref{eq:expansion2})). 
One can give an illuminating and quick derivation of the necessity for
black hole superradiance ~\cite{Bek_super} starting from Hawking's area
theorem~\cite{Hawking_area}.  In the present subsection I take units for
which $G=c=1$.

Consider a Kerr black hole of mass $M$ and angular momentum $J$.  Its horizon
area is
\be
A = 4\pi\left[\left(M+\sqrt{M^2-(J/M)^2}\right)^2+(J/M)^2\right]
\lb{eq:AreaK}
\ee
and small changes of it are given by
\bea
d A &=& \Theta_K{}^{-1}\cdot\left(d M - \Omega\, d J\right)
\lb{eq:dAreaK}
\\
\Theta_K &\equiv& {\scriptstyle 1\over \scriptstyle 2} A^{-1}
\sqrt{M^2-(J/M)^2}
\\
\Omega &\equiv& {J/M\over {r_{\cal H}}^2  + (J/M)^2}
\lb{eq:coeff}
\eea
Let these changes be caused by absorption from a wavemode whose angular and
temporal behavior is ${\cal Y}_{\ell m}(\theta, \phi) e^{-\imath\omega t}\sim
{\cal P}(\theta) e^{\imath m\phi -\imath\omega t}$, with ${\cal Y}_{\ell m}$
the spheroidal harmonics (close cousins to the spherical harmonics) relevant
to the parameter $J/M$~\cite{Starobinskii,PressTeuk1}.  As in
Sec.~\ref{rotational}, the overall changes $d M$ and $d J$ must stand in the
ratio $\omega/m$.   Thus 
\be
d M - \Omega\, d J\propto a_m(\omega)(\omega-m\Omega)
\lb{eq:Achange}
\ee 
where $a_m(\omega)$ is the absorption coefficient of the black hole and the
coefficient of proportionality is positive.  Substituting this in
Eq.~(\ref{eq:dAreaK}) and demanding that $d A>0$ tells us that here, as with
ordinary rotators, superradiance ensues [$a_m(\omega) <0$] when the
Zel'dovich-Misner condition holds.

The argument just reviewed differs from that
spanning Eqs.~(\ref{eq:newErate})--(\ref{eq:supercriterion}) in that no
cognizance need be taken of the radiation entropy.  This is because Hawking's
theorem is purely a dynamical one, not a thermodynamic one: {\it
classically\/} horizon area increases regardless of what happens to the
radiation outside the hole.  In particular, one does not have to assume
high incident intensity to get the proof to work as was the case for the
ordinary rotator.  However, suppose the intensity of a superradiant mode
illuminating the hole is so low that photons hit it one at a time.
Occasionally a photon will tunnel through the potential barrier guarding the
black hole and be absorbed.  A look at Eqs.~(\ref{eq:dAreaK}) and
(\ref{eq:supercriterion})  shows that horizon area will necessarily decrease
this time !  Thus this purely quantum process violates Hawking's area
theorem.  Now in the framework of semiclassical gravity the only thing that
can be going wrong is the theorem's assumption that the weak energy condition
is valid.  It apparently is not for a one-photon quantum state.  

This immediately opens the door to the Hawking evaporation.  For
Hawking's area theorem forbids spontaneous emission from a Kerr black hole
only in modes {\it not\/} satisfying the Zel'dovich-Misner condition since
such emission would be  tantamount to a decrease in horizon area [look at
Eq.~(\ref{eq:dAreaK})].  The moment the theorem can be sidestepped by quantum
processes, spontaneous emission in such modes becomes a possibility.  As we
know it really happens (Hawking radiance) when the fields are in a particular
quantum state (Unruh vacuum).  The failure of the area theorem does not
destroy the argument for superradiance.  One has only to use the argument of
Sec.~\ref{rotational} with the role of the object's entropy played by black
hole entropy and that of the second law by the generalized second
law~\cite{Bekthesis,BekEntropy}.  One then recovers the proof for
superradiance in the Zel'dovich-Misner modes even in the limit of low
incident power where one expects that quantum effects foul up the area
theorem.  We already mentioned that superradiance is a manifestation of
stimulated emission.  Thus we also expect a corresponding spontaneous
emission purely in the superradiant modes.  This is Unruh's nonthermal
radiance~\cite{Unruh} which emerges from a Kerr black hole, and is distinct
from Hawking's.  Unruh's radiance does not appear in the nonsuperradiant
modes.

One other black hole superradiance should be mentioned, namely charge
superradiance. Whenever a black hole bears some electric charge and horizon
electric potential $\Phi$ (see Eq.~(\ref{eq:Phi}) below), it can superradiate
in any mode of a charged bosonic field,  e.g. a pion field, which obeys the
condition $\omega-(e/\hbar)\Phi<0$, where $e$ denotes the field's elementary
charge.  The proof~\cite{Bek_super} is similar to that for rotational black
hole superradiance.  Of course, hybrid superradiance involving charged bosons
and a Kerr-Newman black hole can also happen.  The appropriate
Zel'dovich-Misner criterion is left as an exercise to the reader !

\subsection{Zel'dovich's superradiating cylinder}  \lb{cylinder}

In Sec.~\ref{rotational} we saw that the second law of
thermodynamics requires that a rotating object superradiate. Now if
electromagnetic waves are the issue, how do Maxwell's equations know that
they have to engender superradiance ? This question is analogous to the
question how do Einstein's classical equations know to enforce superradiance
as required by the generalized second law of thermodynamics (answer: because
they imply the area theorem).  In his pioneering paper
Zel'dovich~\cite{Zeld2} remarked that if one is concerned with a steadily 
rotating weakly conducting cylinder, the electric current induced in it by an
incident wave obeying the Zel'dovich-Misner condition has opposite sign to
the electric field, so Ohmic dissipation is negative: rather than the wave
dissipating, it is enhanced.  Zel'dovich's calculation is skimpy and leaves
unanswered the question of how things would work out for large conductivity,
or for a {\it dielectric\/} cylinder which dissipates.  I concentrate on the
dielectric cylinder here; the more general question is dealt with in my paper
with Schiffer~\cite{BekSchiff}.

I consider a very long dielectric cylinder of radius $R$ made of material
with permittivity $\epsilon$ (complex so that the material can dissipate
energy) and which rotates steadily with angular frequency $\Omega$.  In a
dielectric in flat spacetime, Maxwell's equation take the form
\bea
F_{[\alpha\beta,\gamma]} &=& 0
\lb{eq:Maxwellh}
\\
H^{\alpha\beta}{}_{,\beta} &=& 0
\lb{eq:Maxwelli}
\eea
where $H^{\alpha\beta}$ is an antisymmetric tensor built in the style of
$F^{\alpha\beta}$, but with the electric displacement ${\bf D}$
replacing ${\bf E}$.  Although we shall assume the material is nonmagnetic,
the space-space components of $H^{\alpha\beta}$ differ from those of
$F^{\alpha\beta}$ unless the medium is stationary. If $u^\alpha$ is the
medium's four velocity, the constitutive relations are $ H^{\alpha\beta}
u_\beta = \epsilon F^{\alpha\beta} u_\beta$, where  $\epsilon$ must be
evaluated in the rest frame of the material.  A complex relation between
field and displacement components is meaningful if we are talking about
Fourier components which are complex anyway.  I shall assume $\epsilon$ is
constant throughout the cylinder.    

In ordinary cylindrical coordinates $\{x^0, x^1, x^2, x^3\}=\{t, r, \phi,
z\}$ we have  $u_\beta = (-1, 0, \Omega r^2, 0)\gamma$ with $\gamma\equiv
(1-\Omega^2 r^2)^{-1/2}$.  The  important constitutive relations are
\bea
H^{31}&=&F^{31}\equiv B_\phi
\nonumber
\\
H^{23}-\Omega H^{03} &=& F^{23}-\Omega  F^{03} \equiv (r\gamma)^{-1} B_r
\lb{eq:const}
\\
(H^{03}-\Omega r^2 H^{23})\epsilon^{-1}&=& F^{03}-\Omega r^2 F^{23}
\equiv  \gamma^{-1} E_z
\nonumber
\eea
where $E_z$, $B_\phi$ and  $B_r$  denote the corresponding physical
components of the electric field and magnetic induction {\it in the rotating
frame\/}.  Relations~(\ref{eq:const}) just say that $\epsilon$ is the ratio
of electric displacement to electric field in the frame of the dielectric. 

Because the rotation is assumed to be a steady one, and there is axisymmetry,
one is entitled to write $F^{03}=f(r)e^{\imath(m\phi-\omega t)}$ where $m$
is the azimuthal (integer) quantum number and $\omega$ is the frequency as
seen in  the stationary frame (I exclude by fiat the possibility of a $z$
variation of the phase).   Assuming that all field components behave as
$e^{\imath(m\phi-\omega t)}$, I get from
Eqs.~(\ref{eq:Maxwellh}--\ref{eq:Maxwelli}) the components (the rest are not
useful for the present discussion)
\bea
\partial  F^{03}/\partial r +\imath\omega F^{31}  = 0
\nonumber
\\
\imath\omega F^{23} -\imath m r^{-2} F^{03} = 0
\lb{eq:system}
\\
\partial(H^{31} r)/\partial r -\imath m r H^{23} +\imath\omega r H^{03} = 0
\nonumber
\eea
where I have used the flat metric in cylindrical coordinates.  The first two
equations determine  algebraically $F^{31}$ and $F^{23}$ in terms of the
complex amplitude $f(r)$.  With help of the constitutive relations
(\ref{eq:const}) one can eliminate $H^{23}$ and $H^{03}$ from the last
equation, being left with
\be
r^2 f'' + rf' -m^2 f-\left[\omega^2 +(1-\epsilon)(\omega-m\Omega)^2
\gamma^2\right] r^2 f=0
\lb{eq:radial2}
\ee
which is evidently the radial equation for the problem.  The fact that the
components $F^{12}, F^{01}$ and $F^{02}$ do not occur in the system
(\ref{eq:system}) means that they can only put in an appearance in a
different mode (polarization) with the same $\omega$ and $m$.  We can thus
set them to zero if we are interested only in the mode governed by $f$. 

To determine when superradiance occurs we must have an expression for the
radial energy flux.  Whether in vacuum or in matter this is given
by~\cite{LLECM}  $S_r = ({\bf E}\times {\bf H})_r/4\pi$ so that here 
\be
 S_r = (F^{02} H^{12} -F^{03} H^{31})/4\pi = -F^{03} H^{31}/4\pi
\lb{eq:radialflux}
\ee
This is the instantaneous flux; of more interest is the time averaged
flux which can be obtained by first replacing the complex fields by
corresponding real expressions~\cite{LLECM}
\bea
F^{03} &\rightarrow& \left[f e^{\imath(m\phi-\omega t)} +
f^*e^{-\imath(m\phi-\omega t)}\right]/2
\lb{eq:equationa}
\\
F^{31} &\rightarrow&\left[\imath f'
e^{\imath(m\phi-\omega t)} -\imath f^{* '} e^{-\imath(m\phi-\omega
t)}\right]/2\omega
\eea
In the course of time averaging two terms involving exponents $e^{\pm
2\imath(m\phi-\omega t)}$ average out. Using Eqs.~(\ref{eq:const}) one gets
\be
\overline S_r =\imath (f f^{*'} -f^* f' )/16\pi\omega
\lb{eq:energyflux}
\ee

This expression is clearly real, but its sign is none too clear.  To find it
out, I calculate with help of the radial equation that
\be
{d\over dr}[r(f f^{*'} -f^* f' )] = 2\imath r (\omega-m\Omega)^2 |f|^2
\gamma^2\Im\epsilon
\lb{eq:Wronskian}
\ee
where $\Im$ means ``take the imaginary part''. By integrating this equation
over $r$ from $r=0$ to $r=R$, and relying on the fact that $\overline S_r$
must surely be bounded at $r=0$, I get
\be
\overline S_r(r=R) ={-1\over 32\pi\omega R}\int_0^R r (\omega-m\Omega)^2
|f|^2
\gamma^2\Im\epsilon\, dr
\lb{eq:flux2}
\ee
By conservation of energy the flux at large distances from the cylinder
scales from $\overline S_r(r=R)$ according to $R/r$  (no sources at $r>R)$.

Now there is a theorem~\cite{LLECM} that $\Im\epsilon$ must be an odd
function of frequency and positive for positive frequency.  This is a
requirement of thermodynamic origin.  In our case frequency means frequency
in the  rotating frame.  Now the correct azimuthal coordinate in the rotating
frame is
$\tilde\phi=\phi-\Omega t$, so if the phase is to have the form
$m\tilde\phi-\tilde\omega t$, then $\tilde\omega$, the frequency as seen {\it
in the rotating frame\/}, must be $\omega-m\Omega$. Therefore, the integral
above must be negative for $\omega-m\Omega>0$ and positive for
$\omega-m\Omega<0$.  This means that superradiance (net energy outflux) sets
in if and only if the Zel'dovich-Misner condition is satisfied.  This is in
agreement with the thermodynamic argument of Sec.~\ref{rotational}, but shows
what feature is ``microscopically'' responsible for the superradiance.

\section{Adiabatic invariance} \lb{adiabatic}

\setcounter{equation}{0}
\renewcommand{\theequation}{3.\arabic{equation}}

An important turning point in black hole physics occurred with
the realization  of Christodoulou~\cite{Christodoulou}, of Penrose and
Floyd~\cite{Penrose} and of Hawking~\cite{Hawking_area} that transformations
of a black hole generically have an irreversible character.  That is, the
black hole cannot afterward be brought to its original state. Nowdays we
summarize this lore with the rule that horizon area tends to grow, a rule
which has gotten identified with the second law of thermodynamics through the
correspondence {\bf horizon\ area} $\leftrightarrow$ {\bf entropy}.  But
equally important is the feature, stressed originally by Christodoulou
~\cite{Christodoulou,CR}, that some special processes involving a black hole
are truly reversible.  These reversible processes give to black hole
dynamics a more mechanical flavor than would be the case if horizon area
grew under any change of the black hole; they are the analogs of adiabatic
changes of a mechanical system.  Further details may be found in my
contribution to the Festschrifft for Vishveshwara~\cite{BHTrail} and in the
paper by Mayo~\cite{Mayo}.

In this section I use units with $G=c=1$.

\subsection{Adiabatic invariants in general}\lb{invariants}

In mechanics the evolution of a system is dictated by its Hamiltonian
$H(q,p)$ (I write only one degree of freedom; there might be many).  It may
be the case that this Hamiltonian depends on an external parameter $\lambda$:
$H(q,p,\lambda)$.  For example a charged particle can find itself in an
external magnetic field  ${\bf B}$ which then plays the role of
$\lambda$.  Things get interesting when $\lambda=\lambda(t)$ whereupon the
Hamiltonian ceases to be conserved.  Now suppose the system has a timescale
$T$ for a motion which crudely brings it back to the original state or
dimensions (quasiperiodic motion).  If $\lambda$ changes on a timescale much
longer than $T$, the process is called an adiabatic one.  Any mechanical
quantity, ${\cal A}[q,p]$, which is found to change on a timescale much longer
than that of $\lambda$ is called an {\it adiabatic invariant\/}.

Ehrenfest~\cite{Ehrenfest} proved that for a system where a particular
degree of freedom is separable, the corresponding integral $\oint p dq$ taken
around one orbit, usually called an action variable or Jacobi action, is
necessarily an adiabatic invariant.  Some examples will clarify this.  If a
particle bounces between two parallel walls whose separation $L(t)$ grows
linearly with time, and no forces act on it between bounces, then $\oint p dq
=(|p_\rightarrow|L_1+|p_\leftarrow|L_2)$, where $L_1$ is the separation at
the end of the rightward motion,  etc.   This quantity is exactly the same
from cycle to cycle (a very good adiabatic invariant).  We can say
approximately that $\overline{|p(t)|} \propto L(t)^{-1}$, a result of great
importance in understanding adiabatic cooling of a gas.  If the string of a
swinging pendulum of small angular amplitude
$\theta$ and frequency $\omega$ is paid out slowly on the timescale $
2\pi/\omega$, then $\oint p_\theta d\theta\approx 2\pi E/\omega$ with $E$
being a typical value of the total energy in the oscillation.  This quantity
varies little from one swing to the next, a result which is useful in
understanding why photon occupation number is conserved under slow expansion
of a radiation filled box.  Finally for the charge $e$ moving in a spatially
uniform field ${\bf B}(t)$ which varies little in the course of a Larmor
orbit, $\oint p_\varphi d\varphi\approx   2e\pi B R^2$ where $B$ is a
typical magnitude of ${\bf B}$ while $R$ is the Larmor radius of the orbit. 
Again this quantity varies much slower than $B$ from orbit to orbit, allowing
us to conclude that the magnetic flux through the orbit is
approximately conserved.  This result has many implications from plasma
physics to astrophysics to condensed matter physics.

The rate of change of a Jacobi action of a system with a smooth Hamiltonian
falls off exponentially rapidly as $\dot \lambda\rightarrow 0$~\cite{LLMech}. 
Without smoothness this is not true; for instance, in the example of the
particle bouncing between separating walls, if the motion is not linear in
time,  $\oint p dq$ varies as a low power of $\dot \lambda$ as $\dot
\lambda\rightarrow 0$.   And there is nothing in mechanics which forbids
adiabatic invariants that are {\it not\/} Jacobi actions.  We learn that
there may be adiabatic invariants which approach constancy only as power laws
in $\dot\lambda\rightarrow 0$.  In light of this, a useful definition of an
adiabatic invariant is that ${\dot{\cal A}/\dot\lambda}\rightarrow 0$
as $\dot\lambda\rightarrow 0$.  This is at variance with the much tighter
definition given in mathematically rigorous treatises~\cite{Arnold}.

Now, does a black hole in near equilibrium have adiabatic invariants, namely
quantities which vary very slowly compared to variations of the external
perturbations on the black hole ?  I will not look for quantities analogous
to the Jacobi actions.   The Christodoulou reversible processes suggest that
horizon area might be an adiabatic invariant.  Let us see how with the
simplest example. 

\subsection{Particle absorption by charged black hole}\lb{particleabs}

Consider a Reissner-Nordstr\"om black hole of mass $M$ and positive charge
$Q$. The exterior metric is
\be
ds^2 = - \chi\, dt^2 +  \chi^{-1}\, dr^2 + r^2 (d\theta^2 + \sin^2\theta 
d\varphi^2),
\lb{eq:RNmetric}
\ee
with
\be
\chi \equiv 1 - 2M/r + Q^2/r^2.
\lb{eq:chi}
\ee
One shoots in  radially from far away a classical point particle of mass $m$
and positive charge $\varepsilon$ with total relativistic energy adjusted to
the value 
\be
E=\varepsilon Q/r_{\cal H}.
\lb{eq:poised}
\ee
where $r_{\cal H}$ is the $r$ coordinate of the event horizon,
\be
r_{\cal H} = M + \sqrt{M^2 - Q^2}
\lb{eq:r_H}
\ee
In Newtonian terms this particle should marginally reach the horizon where
its potential energy just exhausts the total energy.   The relativistic
equation of motion leads to the same conclusion.

The relativistic action for radial motion is
\be
S = \int L\,d\tau= \int{ \left[-m\,\sqrt{\chi\, (dt/d\tau)^2 -
(dr/d\tau)^2/\chi}  -
\varepsilon A_t\  dt/d\tau\,\right]\,d\tau},
\lb{eq:raction}
\ee
where $\tau$, the proper time, acts as a path parameter, and $A_t = Q/r\ $ is
the only nontrivial component of the electromagnetic 4-potential.  The
stationary character of the background metric and field means that there
exists a conserved quantity, namely
\be
E = - {\partial L\over \partial(dt/d\tau)} =  {m\,\chi\over \sqrt{\chi\,
(dt/d\tau)^2 - (dr/d\tau)^2/\chi}}\ {dt\over d\tau} + {\varepsilon\,Q\over r}.
\lb{eq:conserved}
\ee
Since the norm of the 4-velocity is conserved, the square root in this above
equation has to be unity.  Substituting $dt/d\tau$ from this condition back
in Eq.~(\ref{eq:conserved}) gives
\be
E =  m\,\sqrt{\chi + (dr/d\tau)^2}+ {\varepsilon\,Q\over r}.
\lb{eq:newenergy}
\ee
It is easy to see that this is precisely the total energy of the
particle, for at large distances from the hole, $E \approx m+m\upsilon^2/2 -
m\,M/r + \varepsilon\,Q/r$ (sum of rest, kinetic, gravitational and
electrostatic potential energies).  Setting
$E =
\varepsilon\,Q/r_{\cal H}$ shows that the radial motion has a turning point
($dr/d\tau = 0$) precisely at the horizon
$[\chi(r_{\cal H}) = 0]$. 
  
Because the particle's motion has a turning point at the horizon, it
gets accreted by it.  The area of the horizon is originally 
\be
A = 4\pi {r_{\cal H}}^2  = 4\pi\left(M+\sqrt{M^2-Q^2}\right)^2,
\lb{eq:AreaRN}
\ee
and the (small) change it incurs upon absorbing the particle is
\be
d A =\Theta_{RN}{}^{-1} (d M - Q\,d Q/r_{\cal H}) 
\lb{eq:dAreaRN}
\ee
with
\be
\Theta_{RN} \equiv {\scriptstyle 1\over \scriptstyle 2} A^{-1} \sqrt{M^2-Q^2} 
\lb{eq:thetaRN}
\ee
Thus if the black hole is not extremal so that $\Theta_{RN}\neq 0$, 
$d A=0$ because
$d M=E=\varepsilon Q/r_{\cal H}$ while $d Q=\varepsilon$. 
Therefore, the horizon area is invariant under the accretion of the particle
from a turning point (more precisely, $d A$ is of higher order of
smallness than $d M$).

To a momentarily radially stationary local inertial observer, the particle
in question hardly moves radially as it is  accreted.  Thus its assimilation
is adiabatic.  By contrast, if $E$ were larger than in Eq.~(\ref{eq:poised}),
the particle would not try to turn around at the horizon, and the local
observer would see it moving radially at finite speed and being assimilated
quickly.  And the horizon's area would increase upon its accretion, as is
easy to check from the previous argument. Thus invariance of the horizon area
goes hand in hand with adiabatic changes at the black hole, as judged by
local observers at the horizon.

The above conclusions fail for the extremal Reissner-Nordstr\"om
black hole.  When $Q=M$, $\sqrt{M^2-Q^2}$ in Eq.~(\ref{eq:AreaRN}) is
unchanged to ${\cal O}(\varepsilon^2)$ during the absorption, so that $d A
=8\pi ME$.  This is not a small change, so the horizon's area is not an
adiabatic invariant.  Thus extremal black holes behave differently from
generic black  holes in this as in other phenomena. 
 
Christodoulou actually first worked out the ``reversible process'' for a Kerr
black hole~\cite{Christodoulou}; that calculation is more complicated than
the above.  The generalization to the Kerr-Newman black hole was made by
Christodoulou and Ruffini~\cite{CR}.  We shall return to it in Sec.~5.  

In all the above the particle model of matter is used.  What would happen if
we let the black hole interact with waves ?  One can consider the addition to
the black hole of charge by means of a charged wave, and demonstrate the
adiabatic invariance of the horizon area under suitable circumstances.  The
idea will be clear, especially against the background provided by the last
paragraph of Sec.~\ref{BHrotational}, when we consider the addition of
angular momentum to a Kerr black hole via waves.   

\subsection{Wave absorption by rotating black hole}\lb{waveabs}

Consider a Kerr black hole of mass $M$ and angular momentum $J$.  Its
rotational angular frequency $\Omega$ is given by Eq.~(\ref{eq:coeff}); it is
the angular velocity with which every observer near the horizon gets dragged
azimuthally.  Let distant sources irradiate the black hole with a {\it
weak\/} scalar wavemode of frequency $\omega$, ``orbital'' angular momentum
${\ell}$ and azimuthal ``quantum'' number $m$.  In the spirit of perturbation
theory I neglect the gravitational waves so produced.   The black hole
geometry will eventually be changed by interaction with this wave, but since
the latter is taken to be weak, I shall assume that the change amounts to a
transition from one Kerr geometry to another with slightly different $M$ and
$J$.  In the final analysis such assumption is justified by the stability of
the Kerr geometry and the no-hair theorems. Since the geometry thus remains
axisymmetric and stationary after the change, the wave preserves its
angular-temporal form ${\cal Y}_{\ell m}(\theta, \varphi) e^{-\imath\omega
t}$ over all time (here ${\cal Y}_{\ell m}(\theta, \varphi)$ denotes  a
spheroidal harmonic function~\cite{Starobinskii}, a cousin of the spherical
harmonic $Y_{\ell m}(\theta, \varphi)$).

According to Sec.~\ref{BHrotational} the hole's absorptivity to scalar waves,
$a_m(\omega)$, must have the sign of
$\omega-m\Omega$: the hole absorbs energy for $\omega-m\Omega>0$ and gives up
energy for $\omega-m\Omega<0$.  As
$\omega\rightarrow m\Omega$, $a_m$ must pass through zero because passage
through a pole is unthinkable ($a_m<1$ always). In fact the general argument
leading to Eq.~(\ref{eq:expansion2}) is applicable here and tells us that
$a_m \sim \omega-m\Omega$ near the neutral point.  Indeed, Starobinskii and
Churilov~\cite{Starobinskii} calculated
\be
a_m \approx K_{\omega\ell}\cdot \left(\omega -\Omega\, m\right),
\lb{eq:Gamma}
\ee
where $K_{\omega\ell} (M, J)$ is a positive coefficient.  It follows from
this and by analogy with Eqs.~(\ref{eq:newErate})-(\ref{eq:newJrate})
that the changes in $M$ and $J$ are 
\bea
d M\propto \omega \left(\omega -\Omega\, m\right)
\\
d J\propto m \left(\omega -\Omega\, m\right)
\lb{eq:changeM}
\eea
with a common positive proportionality  constant. By substituting these in
Eq.~(\ref{eq:dAreaK}) we obtain
\be
d A\propto  \left(\omega -\Omega\, m\right)^2,
\lb{eq:DeltaA}
\ee
again with positive coefficient.  The fact that $d A>0$ is in harmony
with Hawking's area theorem~\cite{Hawking_area}.

For small $\omega-m\Omega$, say on the scale $M^{-1}$, the long term changes
of the system (black hole) are governed by changes in $M$ and $J$ which are
seen to be of ${\cal O}(\omega-m\Omega)$.  By contrast the horizon area
change is of ${\cal O}\big((\omega-m\Omega)^2\big)$ so that the horizon area
behaves like an adiabatic invariant.  

In Sec.~\ref{particleabs} we saw that for the Reissner-Nordstr\"om case
the process may be termed adiabatic because the particle gets
assimilated very slowly by the black hole.  For waves in the Kerr case the
meaning of ``adiabatic'' needs to be refined.  It is known that a static
($\omega=0$) but nonaxisymmetric perturbation of a Kerr black hole, such as
would be caused by field sources held in its vicinity at rest with respect to
infinity, necessarily causes an increase in horizon
area~\cite{HawkingHartle}.   However, static perturbations in this sense are
not adiabatic from the local point of view.  Because of the dragging of
inertial frames~\cite{MTW}, any nonaxisymmetric static field is perceived by
momentarily radially stationary local inertial observers as endowed
with temporal variation as these observers are necessarily dragged through
the field's spatial inhomgeneity.  At the horizon the dragging frequency is
the hole's rotational frequency $\Omega$, and a field component with
azimuthal ``quantum'' number $m$ is seen to vary with temporal frequency
$m\Omega$ which need not be small.  Evidently, ``adiabatic'' must here mean
that {\it according to momentarily radially stationary local inertial
observers\/}, the perturbation has only low frequency Fourier components.  As
we saw in Sec.~\ref{cylinder}, the frequency of a wave like ${\cal Y}_{\ell
m}(\theta,\varphi)\,e^{-\imath\omega t}\,\propto e^{\imath m\phi-\imath\omega
t}$ as sensed by observers rotating with the hole at the horizon is precisely
$\omega-m\Omega$, and so it is this frequency which must be small in order
for the process to be considered adiabatic.  As we just saw, only
perturbations with small $\omega-m\Omega$ leave the horizon area invariant to
a higher order than other corresponding changes in the black hole.

\begin{figure}[ht]
\begin{center}
\psfig{file=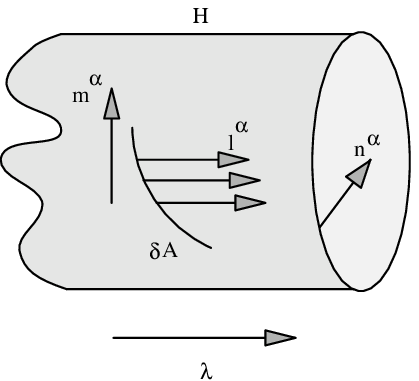,height=5cm}
\label{Fig.3}
\end{center}
\begin{caption}[] 
{The patch of area $\delta A$ on ${\cal H}$ is formed by
null generators whose tangents are $l^\alpha=dx^\alpha/d\lambda$,
where $\lambda$ is an affine parameter along the generators; shown also are
the spacelike vector $m^\alpha$ and the ingoing null vector $n^\alpha$ of the
same Newman-Penrose tetrad.}
\end{caption}
\end{figure}    

These conclusions are inapplicable to the extremal Kerr black
hole ($J=M^2$).  In this case $\Theta_K = 0$ (see Sec.~\ref{BHrotational}),
so one cannot use Eq.~(\ref{eq:dAreaK}) to calculate the change in area, but
must work directly with Eq.~(\ref{eq:AreaK}).  From Eq.~(\ref{eq:coeff}) one
learns that $\Omega=1/2M$ so that $\Delta J =\Omega^{-1} \Delta M = 2M\Delta
M$.  Replacing $M\rightarrow M+\Delta M$ and $J\rightarrow J+2M\Delta M$ in
Eq.~(\ref{eq:AreaK}), and substracting the original expression gives 
\be
\Delta A = 8 \pi(2+\surd 2)  M\Delta M + {\cal O}((\Delta M)^2).
\ee
Since a generic addition of mass $\Delta M$ will give a $\Delta A$ of the
same order, the horizon area of an extremal Kerr hole is {\it not\/} an
adiabatic invariant.

\subsection{Dynamics of horizon area}\lb{areatheorem}

Before delving further into the subject let us review the central
result in the field, Hawking's area theorem~\cite{Hawking_area}, and the
horizon dynamics upon which it is based.

As usual, we denote the event horizon by ${\cal H}$.  Consider a small patch
of ${\cal H}$'s area $\delta A$; it is formed by null generators whose
tangents are $l^\alpha=dx^\alpha/d\lambda$, where $\lambda$ is an affine
parameter along the generators (see Fig.~3).  By definition of the 
convergence $\rho$ of the generators ~\cite{MTW}, $\delta A$ changes at a 
rate
\be
d\delta A/d\lambda = -2\rho\delta A. 
\lb{eq:changeA}
\ee
Now $\rho$ itself changes at a rate given by the optical analogue of the
Raychaudhuri equation (with Einstein's equations already incorporated)
~\cite{NP,Pirani}
\be
d \rho/d\lambda = \rho^2 + |\sigma|^2 + 4\pi T_{\alpha\beta}\,l^\alpha
l^\beta, 
\lb{eq:changerho}
\ee
where $\sigma$ is the shear of the generators, and $T_{\alpha\beta}$ the 
energy momentum tensor.   The shear evolves according to
\be
d  \sigma/d\lambda = 2\rho\sigma  + C_{\alpha\beta\gamma\delta}\, l^\alpha
m^\beta l^\gamma m^\delta, 
\lb{eq:changesigma}
\ee
where $C_{\alpha\beta\gamma\delta}$ is the Weyl conformal tensor, and
$m^\alpha$  one of the spacelike Newman-Penrose tetrad legs which lies in
${\cal H}$.

Many types of classical matter obey the weak energy condition
\be
T_{\alpha\beta}\,l^\alpha l^\beta \geq 0. 
\lb{eq:energycondition}
\ee
We have seen in Sec.~\ref{BHrotational} that matter in
certain quantum states can violate this condition.  In the discussion
of adiabatic invariance I take a completely classical view, and will assume
that Eq.~(\ref{eq:energycondition}) is always true.  Then $\rho$
can---according to Eq.~(\ref{eq:changerho})---only {\it grow\/} or remain
unchanged along the generators.  Now were $\rho$ to become positive at any
event along a generator of our patch, then by Eq.~(\ref{eq:changerho}) it
would remain positive henceforth, and indeed grow bigger.
 Eq.~(\ref{eq:changeA}) then shows that $\delta A$ would shrink to nought in
a {\it finite\/} span of $\lambda$~\cite{Hawking_area,MTW} thus implying
extinction of generators. But it is an axiom of the
subject~\cite{Hawking_area,MTW} that ${\cal H}$'s generators cannot end in
the future.  The only way out is to accept that $\rho\leq 0$ everywhere along
the generators, which by Eq.~(\ref{eq:changeA}) signifies that the patch's
area can never decrease.  This is the essence of Hawking's area theorem.

Under what conditions is ${\cal H}$'s area constant ? 
Hawking~\cite{Hawking_area} and Hawking and
Hartle~\cite{HawkingHartle} consider this to be possible only if the black
hole it exactly stationary.  The examples in
Secs.~\ref{particleabs}-\ref{waveabs} show that there are slightly
nonstationary situations where the increase in horizon area is
imperceptible.  Let us characterize the situations where no change in area
occurs.

By Eq.~(\ref{eq:changeA})  this requires that $\rho=0$.  But then
Eq.~(\ref{eq:changerho})  implies that also $\sigma=0$  while
$T_{\alpha\beta}\,l^\alpha l^\beta=0 $  on ${\cal H}$.  Then 
Eq.~(\ref{eq:changesigma}) implies that  $C_{\alpha\beta\gamma\delta}\,
l^\alpha m^\beta l^\gamma m^\delta $   vanishes on ${\cal H}$.  The
particular Weyl tensor component in question describes gravitational waves
crossing the horizon inward bound.  These will not occur if the situation is
quasistationary, since gravitational waves are generated by matter only to
${\cal O}(\upsilon^5)$ where $\upsilon$ is the velocity of the matter sources.
Thus as a minimum we must have an approximate time Killing vector and slow
motion of matter. This granted, preservation of ${\cal H}$'s area requires in
addition
\be
T_{\alpha\beta}\,l^\alpha l^\beta  = 0 \quad {\rm on\ \cal H}.
\lb{eq:zerocondition}
\ee
Is this condition always satisfied in a quasistationary situation even when
sources of nongravitational fields reside in the vicinity of the black hole ? 
If not, then there is no hope for an adiabatic theorem because the area will
be found to increase even in situations which look like requiring no changes
of the black hole.

Computations, some of them arduous, show that the condition indeed holds. 
It is true for the energy-momentum tensor of either minimally or conformally
coupled scalar fields from static sources in a Schwarzschild~\cite{BHTrail}
or Reissner-Nordstr\"om~\cite{Mayo} black hole's vicinity, for that from
minimally coupled scalar field's sources axisymmetrically arranged around a
Kerr black hole~\cite{Mayo}, and for the electromagnetic field's $T_{\mu\nu}$
from charges arranged statically about a Schwarzschild black hole. 
Energy-momentum conservation is the common reason for the enforcement of
Eq.~(\ref{eq:zerocondition}) in a (nearly) stationary situation.  The
argument is very simple.

Assume that the exterior geometry has a time translation Killing vector
$\xi^\alpha$.  This might be the only Killing vector as when a Schwarzschild
black hole is perturbed by static field sources placed with no particular
symmetry around it.  Or the situation might also be axisymmetric  (additional
Killing vector $\eta^\alpha$) while still static if the array is made
axisymmetric.  A third case is that of a nonstatic but still stationary and
axisymmetric situation where the black hole rotates with angular
frequency $\Omega$.   Because ${\cal H}$ is a Killing horizon, the tangent to
any of its generators, $l^\alpha$, must be along a Killing vector, itself a
linear combination of the above Killing vectors.  In a truly static situation
$l^\alpha\propto \xi^\alpha$, but if the black hole rotates,
$l^\alpha\propto (\xi^\alpha +\Omega\eta^\alpha)$.  The Killing 
vector $\zeta^\alpha\equiv \xi^\alpha +\Omega\eta^\alpha$ (with
$\Omega\neq 0$ or, if appropriate, $\Omega=0$) defined over all the black
hole exterior is an extension of
$l^\alpha$ off ${\cal H}$.  Now because $T^{\alpha\beta}{}_{;\beta}=0$ and
$T^{\alpha\beta}=T^{\beta\alpha}$ as well as the Killing equation
$\zeta_{\alpha;\beta}+\zeta_{\beta;\alpha}=0$, $(T^{\alpha\beta}
\zeta_\alpha)_{;\beta}=0$.  Gauss's law then gives
\be
\int (T^{\alpha\beta}\zeta_{\alpha})_{;\beta}\,(-g)^{1/2} d^4x = \oint
T^{\alpha\beta} \zeta_\alpha\, d\Sigma_\beta = 0
\lb{eq:condition}
\ee
where the second integral is taken over any closed orientable 3-surface, and
$d\Sigma_\beta$ is the outward pointing element of 3-volume on it. 

\begin{figure}[ht]
\begin{center}
\psfig{file=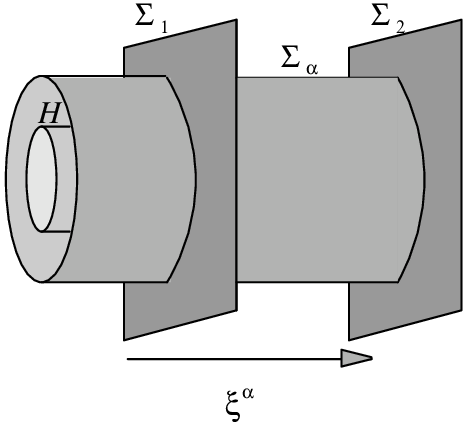,height=5cm}
\label{Fig.4}
\end{center}
\begin{caption}[] 
{The bounding 3-hypersurface is composed of the section of the horizon $H$
between two constant-time hypersurfaces, $\Sigma_1$ and $\Sigma_2$, the two
hypersurfaces themselves, and the part $\Sigma_{\rm a}$ between $\Sigma_1$ and
$\Sigma_2$ of a spacelike hypersurface in the asymptotically flat region far
from the black hole.}
\end{caption}
\end{figure}     

As shown in Fig.~4, let us take this 3-surface to be composed of the
section of ${\cal H}$ between two constant-time hypersurfaces, $\Sigma_1$ and
$\Sigma_2$, the two hypersurfaces themselves, and the part $\Sigma_{\rm a}$
between $\Sigma_1$ and $\Sigma_2$ of a spacelike hypersurface with $S^2\times
R$ topology in the asymptotically flat region far from the black hole.  The
contribution to the integral from $\Sigma_2$ cancels that from $\Sigma_1$
because the time translation maps one into the other while leaving
$T^{\alpha\beta}$ unchanged, and because the sign of $d\Sigma_\alpha$ is
opposite on $\Sigma_1$ and $\Sigma_2$.  The $d\Sigma_\beta$ on $\Sigma_{\rm
a}$ points in the radial direction in suitable coordinates.  In the static
situation with
$\zeta^\alpha=\xi^\alpha$, $T^{\alpha\beta} \zeta_\alpha
d\Sigma_\beta$ at $\Sigma_{\rm a}$ represent energy flow inward at
$\Sigma_{\rm a}$.  If no energy influx exists, for example because the ropes
supporting various objects that perturb the black hole are not moving, then
the contribution of $\Sigma_{\rm a}$ vanishes and we get
\be
\int_{\cal H} T^{\alpha\beta} l_\alpha d\Sigma_\beta = 0
\lb{eq:horizoncondition}
\ee
In the rotating case when $\zeta^\alpha\equiv \xi^\alpha +\Omega\eta^\alpha$, 
$T^{\alpha\beta} \zeta_\alpha\, d\Sigma_\beta$ at
$\Sigma_{\rm a}$ contains an additional term representing inflow of angular
momentum ($T_{r\varphi}\,dt\,r^2 d\theta\, d\varphi$ in the usual
coordinates).  In other words, the new term represents a torque on the black
hole.  If the sources disturbing it are arranged axisymmetrically and
coaxially with its rotation, there will be no such torque.  In this case we
recover Eq.~(\ref{eq:horizoncondition}).

If the weak energy condition (\ref{eq:energycondition}) is satisfied at
$\Sigma_1$, it is preserved between $\Sigma_1$ and $\Sigma_2$ by the
time translation symmetry.  Thus the integral in
Eq.~(\ref{eq:horizoncondition}) is a sum of positive semidefinite
contributions, one for each horizon patch.   Hence
Eq.~(\ref{eq:horizoncondition}) implies that Eq.~(\ref{eq:zerocondition}) is
true everywhere on the horizon. There is thus no reason for the area to
increase, even secularly.  Thus if external sources disturb a Schwarzschild
black hole in a static way or a Kerr black hole in a stationary and
axisymmetric way, they do not cause the horizon area to grow.  This is as we
would have liked to believe, but it is reassuring to have a proof that mere
presence of matter fields at the horizon does not cause its area to
increase.  There is thus no impediment of principle to an adiabatic theorem
for black holes.  

\subsection{Black hole disturbed by scalar charges}
\lb{scalarcharges}

In Sec.~\ref{waveabs} I demonstrated the adiabatic invariance of horizon area
for a Kerr black hole under the influence of scalar waves.  Here I
demonstrate the invariance for a Schwarzschild black hole subject to low
frequency scalar perturbations originating from sources ``rattling'' in the
hole's vicinity.  

Consider a Schwarzschild black hole with exterior metric
\be
ds^2 = - (1-2M/r) dt^2 + (1-2M/r)^{-1} dr^2 + r^2 (d\theta^2 +
\sin^2\theta\, d\varphi^2).
\lb{eq:Schwarzschild}
\ee
Suppose sources of a {\it minimally coupled\/} scalar field $\Phi$ have been
brought to a finite distance from the hole and are there caused to perform
some motion at low frequencies. Does this influence cause an increase in
${\cal H}$'s area ?   

If the scalar's sources are weak, one may regard $\Phi$ as a quantity of
first order, and proceed by perturbation theory.  The scalar's
energy-momentum tensor,
\be
	T_\alpha{}^\beta  = \nabla_\alpha \Phi \nabla^\beta \Phi - {1\over
2}\,\delta^\beta_\alpha\, \nabla_\gamma\Phi\,\nabla^\gamma \Phi,
\lb{eq:energytensor}
\ee
will be of second order of smallness.  I shall suppose the same is true of
the energy-momentum tensor of the sources themselves.  Thus to first order
the metric (\ref{eq:Schwarzschild}) is unchanged.   The
scalar equation outside the scalar's sources can be written
\be
-{r^4\over (r^2-2Mr)}{\partial^2\Phi \over \partial t^2} +
{\partial\over \partial r} \left[ (r^2-2M) {\partial\Phi\over\partial r}
\right] - \hat L^2\,\Phi = 0.
\lb{eq:tdependence}
\ee
where $\hat L^2$ is the usual squared angular momentum operator (but without
the $\hbar^2$ factor).  This equation suggests looking for a solution of the
form~\cite{MTW}
\be
\Phi = \Re \int_{0}^{\infty} d\omega\sum_{\ell=0}^\infty\,
\sum_{m=-\ell}^\ell C_{\ell m}(\omega)\, f_{\ell m}(\omega,r)\, Y_{\ell
m}(\theta,\varphi) e^{-\imath\omega t}.
\lb{eq:Phitd}
\ee
where the $Y_{\ell m}$ are the familiar spherical harmonic (complex)
functions. Since the  $Y_{\ell m}$ form a complete set in angular space, any
function
$\Phi(r,\theta,\varphi,t)$ can be so expressed with the help of a Fourier
decomposition in the time variable.  The constant coefficients $C_{\ell
m}(\omega)$ are to be used to match $\Phi$ to the prescribed
sources; their presence allows for arbitrary normalization of the $f_{\ell
m}$. Since $\hat L^2 Y_{\ell m} = \ell(\ell+1) Y_{\ell m}$, the radial and
angular variables separate. In terms of Wheeler's ``tortoise'' coordinate
$r^* \equiv r + 2M\ln(r/2M-1)$, for which the horizon resides at
$r^*=-\infty$, and the new radial function $H_{\ell m}(\omega, r^*) \equiv r
f_{\ell m}(\omega, r)$, one finds for $H_{\ell m}$ the equation
\be
-{d^2H_{\ell m}\over dr^{*2}} + \left(1-{2M\over r}\right)\left({2M\over
r^3} + {\ell(\ell+1)\over r^2}\right) H_{\ell m}= \omega^2
H_{\ell m}. 
\lb{eq:Feq}
\ee
Since the index $m$ does not figure here, I write just
plain $H_\ell(\omega, r)$; one may obviously pick  $H_\ell$ to be real.

The resemblance between Eq.~(\ref{eq:Feq}) and the Schr\"odinger eigenvalue
equation permits the following analysis~\cite{MTW} of the effects of distant
scalar sources on the black hole horizon.  Waves with ``energy'' $\omega^2$
on their way in from a distant source run into a positive potential, the
product of the two parentheses in Eq.~(\ref{eq:Feq}).  The potential's peak is
situated at $r\approx 3M$ for all $\ell$.  Its  height is $0.0264 M^{-2}$
for $\ell=0$,
$0.0993 M^{-2}$ for $\ell=1$ and $\approx 0.038\,\ell(\ell+1) M^{-2}$ for
$\ell\ge 2$. Therefore, waves with any $\ell$ and $\omega < 0.163 M^{-1}$
coming from sources at $r\gg 3M$ have to tunnel through the potential barrier
to get near the horizon.  As a consequence, the wave amplitudes that
penetrate to the horizon are small fractions of the initial amplitudes, most
of the waves being reflected back.  In fact, the tunnelling coefficient
vanishes in the limit $\omega\rightarrow 0$~\cite{MTW}. This means that
adiabatic perturbations by distant sources (which surely means they only
contain Fourier components with $\omega\ll M^{-1}$) perturb the horizon very
weakly (this is just the inverse of Price's theorem~\cite{MTW} that a totally
collapsed star's asymptotic geometry preserves no memories of the star's
shape).  Thus one would not expect significant growth of horizon area from
adiabatic scalar perturbations originating in distant sources.

What if the scalar's sources are moved into the region $2M < r < 3M$ inside
the barrier ?  They will now be able to perturb the horizon; do they change
it's area ?  To check let us look for the solutions of Eq.~(\ref{eq:Feq}) in
the region near the horizon where the potential is small compared to
$\omega^2$; according to the theory of linear second order differential
equations they are of the form
\be
H_{\ell\omega}(r^*) = \exp(\pm \imath\omega r^*)\times[1 +  {\cal O}(1-2m/r)].
\lb{eq:solution}
\ee
The Matzner boundary condition~\cite{Matzner} that the physical solution be
an ingoing wave, as appropriate to the absorbing character of the horizon,
selects the sign in the exponent as negative.  Hence the typical term in
$\Phi$ is
\be
{1 +  {\cal O}(1-2m/r)\over r} P_\ell(\cos\theta)\,\cos\psi;\qquad \psi\equiv
\omega(r^*+t)-m\varphi.
\lb{eq:term}
\ee  

We obviously require that the event horizon remain regular under the scalar's
perturbation; otherwise the black hole would be destroyed.  A minimal
requirement for regularity is that physical invariants like $\Upsilon_1\equiv
T_\alpha{}^\alpha$,  $\Upsilon_2\equiv T_\alpha{}^\beta T_\beta{}^\alpha$, 
$\Upsilon_3\equiv T_\alpha{}^\beta T_\beta{}^\gamma T_\gamma{}^\alpha$, 
etc., be bounded, for divergence of any of them would surely induce
curvature singularities via the Einstein equations.  By
Eq.~(\ref{eq:energytensor}) the invariant $\Upsilon_k$ is always proportional
to $(\Phi_{,\alpha}\Phi^{,\alpha})^k$.  For a single mode like that in
Eq.~(\ref{eq:term}), an explicit calculation on the Schwarzschild background
using $dr^*/dr = (1-2M/r)^{-1}$ gives, after a miraculous cancellation of
terms divergent at the horizon (pointed out by A. Mayo),
\be
\Phi_{,\alpha}\Phi^{,\alpha} \propto
{m^2 P_\ell{}^2\sin^2\psi\over r^4\sin^2 \theta}
 +\left({dP_\ell\over d\theta}\right)^2{\cos^2\psi\over r^4}
+{\omega \sin(2\psi)\over r^3}P_\ell^2 + \cdots\, ,
\lb{eq:miracle}
\ee
where ``$\ \cdots\ $'' here and henceforth denote terms that vanish as
$r\rightarrow 2M$.  This expression is bounded at the horizon.    Now suppose
$\Phi$ is the sum of two modes like (\ref{eq:term}), which we label with
subscripts ``1'' and ``2''.  Then a calculation
gives $\Phi_{,\alpha}\Phi^{,\alpha}$ as consisting of three groups of terms,
two of them of form (\ref{eq:miracle}) with subscripts 1 and 2, respectively,
and a third of the form
\bea
{m_1 m_2 P_{\ell_1}P_{\ell_2}\sin\psi_1\sin\psi_2\over
r^4\sin^2 \theta}
 +\left({dP_{\ell_1}\over d\theta}\right)\left({dP_{\ell_2}\over
d\theta}\right){\cos\psi_1\cos\psi_2 \over r^4} +
\nonumber \\
+{\omega_1 \sin\psi_1\cos\psi_2 +\omega_2 \sin\psi_2\cos\psi_1\over
r^3}P_{\ell_1}P_{\ell_2} + \cdots
\eea
This is also bounded. By induction any $\Phi$ of form (\ref{eq:Phitd}) will
give a bounded $\Phi_{,\alpha}\Phi^{,\alpha}$. Thus all the $\Upsilon_k$ are
bounded at $r=2M$, and a generic scalar perturbation does not disturb the
horizon unduly.

The extent to which the black hole is perturbed must be linear in the
magnitude of the invariant $\Upsilon_1$ (Einstein's equations have
$T_{\alpha\beta}$ as source, not $T_\alpha{}^\gamma T_\gamma{}^\beta$). It is
then clear from both our results that this perturbation is of order
${\cal O}(\omega^0)$ generically, and of ${\cal O}(\omega)$ in the monopole
case. As we shall now see, the change in the horizon area is of ${\cal
O}(\omega^2)$, so that for small $\omega$ the area is (relatively) invariant.

A 3-D hypersurface of the form $\{\forall t, r={\rm const.}\}$ has as
tangent the Killing vector $\xi^\alpha = \delta_t{}^\alpha$ with norm
$-(1-2M/r)$, and as normal $\eta_\alpha = \partial_\alpha (r-{\rm const.})
=\delta_\alpha{}^r$ with norm $(1-2M/r)$.  The vector $N^\alpha\equiv
\xi^\alpha+(1-2M/r) \eta^\alpha$ is obviously null, and as $r\rightarrow 2M$
both its covariant and contravariant forms remain well defined, so that it
must there be proportional (with finite nonvanishing proportionality
constant) to $l^\alpha$, the tangent to the horizon generator.   This can be
verified by remarking that $N^\alpha$, just as $l^\alpha$, is null, future
pointing ($N^t>0$) as well as outgoing ($N^r>0$). 

Now
\be
T_{\alpha\beta}  N^\alpha N^\beta = (T_r{}^r - T_t{}^r)\, N_r\,N^r + 2
T_t{}^r N_r N^t.
\lb{eq:TNNnew}
\ee
From $N^\alpha$'s definition we have $N_rN^r = 1-2M/r$ and  $
N_r N^t=1$. And from Eq.~(\ref{eq:energytensor}) it is clear that $T_r{}^r -
T_t{}^t=
\Phi_{,r}\Phi^{,r}-\Phi_{,t}\Phi^{,t}$ while $T_t{}^r = \Phi_{,t}\Phi^{,r}$. 
Thus
\be
T_{\alpha\beta} N^\alpha N^\beta = [\Phi,_t + (1-2M/r)\Phi,_r]^2.
\ee
If one now substitutes a $\Phi$ made up of a single mode like in
Eq.~(\ref{eq:term}),  one concludes that
\be
T_{\alpha\beta} l^\alpha l^\beta \propto {\omega^2 P_\ell^2\sin^2\psi\over
r^2} + \cdots\, .
\lb{eq:term2} 
\ee
A quick way to this result is to recognize that
$l^\alpha\propto \xi^\alpha\equiv (\partial/\partial t)^\alpha$ because the
horizon generators must lie along the only Killing vector field of the
problem.  In view of Eq.~(\ref{eq:energytensor}) and the null character of
$l^\alpha$,
\be
T_{\alpha\beta} l^\alpha l^\beta \propto (\Phi,_\alpha \xi^\alpha)^2 =
(\partial\Phi/\partial t)^2, 
\ee
which reproduces Eq.~(\ref{eq:term2}). And if one substitutes the generic
$\Phi$, the proportionality to the square of frequency will obviously remain.

Thus, when scalar field sources are moved inside the barrier, they perturb
the geometry by an amount which does not, in general, vanish as the
perturbations is made to change slower and slower.  By contrast, the rate of
change of the horizon area vanishes as the square of the typical Fourier
frequency of the perturbation.  In this sense the horizon area is an adiabatic
invariant.  The result has been generalized by Mayo~\cite{Mayo} for
electromagnetic fields from charges near a Schwarzschild black hole.  That
calculation was harder than the one above, and succeeded only by judicious
use of the analogy between electrodynamics in a curved spacetime and in a
flat spacetime filled with a medium with appropriately varying permittivity
and permeability~\cite{Volkov}.

\subsection{Sketch of a proof of the adiabatic theorem}
\lb{theorem}
 
The above permits us to discuss a process quite different from those treated
in Secs.~\ref{particleabs}--\ref{waveabs}.  In those cases the black hole is
transformed by the process, its charge or angular momentum changing, so that
the adiabatic process converts one Kerr-Newman black hole into another.  But
imagine instead slowly bringing a scalar charge from a distance to near a
Schwarzschild's black hole horizon, and then withdrawing it slowly as well. 
The horizon undergoes a perturbation which is then relaxed.  According to
Sec.~\ref{scalarcharges}, the horizon's area does not change appreciably, so
the black hole must return to its original Schwarzschild state.  We now
sketch a simple proof of the adiabatic theorem for the same kind of
situation, but for any sort of matter perturbations, which need not be small.

We assume a static black hole is surrounded by charges of some sort which
perturb it via their fields; these charges are assumed supported in some
way, for example by ropes coming down from large distances.  The black hole
need not be close to Schwarzschild; it could be strongly distorted from
sphericity.  This, of course, does not violate the no-hair principle because
the black hole is not an isolated black hole. 

Now suppose the charges, initially at rest, are set into slow motion, for
instance by being lowered slowly with help of the ropes.   Let $\upsilon$ be
the {\it signed scale\/} of the velocity involved.  For example, this could
be the typical proper radial velocity of one of the charges.  The sign of 
$\upsilon$ distinguishes one slow motion from its exact reversal, all
starting from the same configuration.

Let $\xi^\alpha$ be the Killing vector of the background geometry before
motion sets in and let $T^\alpha{}_\beta$ denote the exact energy-momentum
tensor, at all times, of the sources and their fields.  The spacelike
components of $T^\alpha{}_\beta\, \xi^\beta$, the energy flux components
defined with respect to the background metric, $g^{(0)}_{\mu\nu}$, obviously
switch sign together with $\upsilon$.  If we assume that $T^\alpha{}_\beta\,
\xi^\beta$ can be expanded in a series in $\upsilon$, that series must thus
start with ${\cal O}(\upsilon)$.  By the Einstein equations linearized about
$g^{(0)}_{\mu\nu}$, some of the components of the metric perturbations coming
from the motion, $\delta g_{\mu\nu}$,  must be of   ${\cal O}(\upsilon)$.  
This means that the black hole is generically distorted to ${\cal
O}(\upsilon)$.  However, it does {\it not\/} follow that the horizon's area
changes to ${\cal O}(\upsilon)$. 
 
We saw in Sec.~\ref{areatheorem} that for a static situation
$\rho=\sigma= C_{\alpha\beta\gamma\delta}\, l^\alpha m^\beta l^\gamma
m^\delta=0$; thus $g^{(0)}_{\mu\nu}$ by itself must give $\rho=\sigma=
C_{\alpha\beta\gamma\delta}\, l^\alpha m^\beta l^\gamma m^\delta=0$ while
$\delta g_{\mu\nu}$ [which has some components of ${\cal O}(\upsilon)$] will
generate corrections $\delta\rho$, $\delta\sigma$ and
$\delta(C_{\alpha\beta\gamma\delta}\, l^\alpha m^\beta l^\gamma m^\delta)$ of
${\cal O}(\upsilon)$ {\it or higher\/}.  Now look at
Eqs.~(\ref{eq:changesigma}).   Its right hand side is of ${\cal O}(\upsilon)$
so $d\delta\sigma/d\lambda ={\cal O}(\upsilon)$.  This is consistent
with $\delta\sigma$ being of ${\cal O}(\upsilon)$ because
$\delta\sigma\rightarrow 0$ in the far future when things settle down to
staticity.

Eq.~(\ref{eq:changerho}) now tells us that the question of
whether $d\delta\rho/d\lambda$ is of ${\cal O}(\upsilon)$ or ${\cal
O}(\upsilon^2)$ is determined by the order of $T_{\alpha\beta} l^\alpha
l^\beta$, the other terms in the right hand side being necessarily of ${\cal
O}(\upsilon^2)$.  We showed in Sec.~\ref{areatheorem} that $T_{\alpha\beta}
l^\alpha l^\beta=0$ for the background situation involving no motion.  Can
$T_{\alpha\beta} l^\alpha l^\beta$ be of ${\cal O}(\upsilon)$  for the
dynamic situation in question ?  No !  That eventuality would allow it to
switch sign with $\upsilon$, but this would contravene the weak energy
condition Eq.~(\ref{eq:energycondition}).  We conclude that $T_{\alpha\beta}
l^\alpha l^\beta={\cal O}(\upsilon^2)$, or higher.  Then
Eq.~(\ref{eq:changerho}), together with the requirement that
$\delta\rho\rightarrow 0$ in the future when all changes die out, tells us
that $\delta\rho = {\cal O}(\upsilon^2)$.  It finally follows from
Eq.~(\ref{eq:changeA}) that the overall change in horizon area is of ${\cal
O}(\upsilon^2)$.  This shows that the change in horizon area is of
higher order of smallness than those of the changes undergone by the
perturbation of the hole; but this is precisely what the adiabatic theorem
would claim.

\section{Black hole quantization}  \lb{Heuristic}

\setcounter{equation}{0}
\renewcommand{\theequation}{4.\arabic{equation}}

Quantum gravity effects supposedly become important only at the Planck
scale, variously stated as ${\cal M}_{P}= (\hbar/G)^{1/2} \approx 1.2 \times
10^{20}\, {\rm MeV}$ or ${\cal L}_P= (\hbar G)^{1/2} \approx 1.6 \times
10^{-33}\,{\rm cm}$.  Now this scale is so extreme by laboratory standards
that it would seem one shall never be able to put quantum gravity to the test
in the laboratory.  Is this really so or is it possible that by some
recondite effect quantum gravity may make itself felt well below the Planck
energy (well above the Planck length) ?  The Hawking radiance, it is true, is
expected also well away from the Planck scale.  However, it is generally
acknowledged that derivations of it (at least those not based on superstring
theory) are semiclassical in nature (no quantum gravity), and cannot tell us
what would really happen at the Planck scale.  In this lecture I show how one
can use a mixture of classical hints and quantum ideas to guess what the
departure from Hawking's simple spectrum should be.  The surprise is that
there are serious departures expected well away from the Planck regime.  

I stated the basic idea~\cite{BekNC} immediately after the appearance of the
Hawking radiance paper.  It was taken up later by Mukhanov~\cite{Mukhanov},
and we eventually synthesized our ideas~\cite{BekMukh,QG}.  
Other references are my Marcel Grossman VIII talk~\cite{MG8} and my talk at
the XVII Brazilian Meeting on Particles and Fields~\cite{BekBrazil}.

Henceforth in this section I use units with $G=c=1$ and denote the charge of
the electron by $-e$.

\subsection{Quantum numbers of a black hole}  \lb{Quantum_numbers}

In setting out to give a quantum description of black holes, a primary
question (first asked by Wheeler in the late 1960's) is what is the complete
set of quantum numbers required to describe a black hole in a stationary
quantum state.  Quantum numbers are first and foremost attributes of
elementary particles.  Now an elementary object with mass below ${\cal M}_P$
has its gravitational radius tucked below its Compton wavelength; it is thus
properly termed ``elementary particle''.  By contrast an elementary object
with mass above ${\cal M}_P$ has its Compton wavelength submerged under
the gravitational radius; it is best called a black hole.  The discontinuity
between the two occasioned by the emergence of the horizon is illusory
because at the Planck scale the spacetime geometry should be quite fuzzy.  So
there is no in between regime here, and by continuity the smallest black
holes should be quite like elementary particles, and should merit description
by a few quantum numbers like mass, charge, spin,  etc. 

As the black hole gets larger, it should become more classical and thus
come into the province of Wheeler's no-hair principle (see
Secs.~\ref{updown}-\ref{hairy}):  a black hole is parametrized only by mass,
spin angular momentum, electric and magnetic charge.  Of course there are the
nonabelian generalizations of the Kerr-Newman solutions.  But as we saw, with
the exception of the Skyrmionic black hole, these are all unstable.  I now
argue, by analogy with field theory, that we need not promote the parameters
of these unstable solutions to the status of quantum numbers.

Recall the Higgs field with Mexican hat potential in flat spacetime.  A
homogeneous configuration of Higgs field taking on a value on the slope of
the potential is not a stationary classical solution.  No stationary quantum
state corresponds to it.  A configuration with the field at a minimum of the
potential is a classical stationary stable solution. Small perturbations away
from it, which classically oscillate around it, are interpreted in the
quantum theory as excitations of the field above the minimum state.  By
contrast, a configuration with the field at a maximum of the potential is a
classical stationary but unstable solution. A small perturbations away from
it runs away.  In the quantum theory such perturbations are reinterpreted as
tachyonic excitations.  To us this really means that the underlying
stationary configuration are pathological.

By analogy we may conclude that to each {\it stable stationary\/}
classical black hole solution corresponds a stationary quantum state which is
capable of excitation.  Again by analogy, the excited state can be
interpreted as the base black hole state  plus quanta of various fields
propagating on its background.  By contrast, an unstable stationary classical
black hole solution cannot be associated with a stationary quantum state
because excitations of the later would be tachyonic in nature.   Thus, the
unstable nonabelian hair black holes and the BBM black hole {\it do not\/}
furnish classical analogues of quantum stationary states.  

Of course the above argument cannot rule out quantum stationary black hole
states without classical analogs.  But it does suggest that, as far as
present evidence requires, the only quantum numbers of a stationary black hole
state are mass, spin angular momentum, electric and magnetic charge and
Skyrmionic topological number.  As mentioned in Sec.~\ref{hairy}, this last
is a kind of winding number, and as such not obviously additive.  For this
reason I strike it from the list.   

\subsection{Mass spectrum of a black hole}\lb{mass_spectrum}

I thus focus on black hole eigenstates of the operators mass $\hat M$,
angular momentum $\hat {\bf J}^2$ and $\hat J_z$, electric charge $\hat Q$,
magnetic charge $\hat {\cal G}$ and, of course, linear momentum
${\hat {\bf P}}$. This last can be set to zero if we agree to work in the black
hole's center of mass.  The eigenvalues of $\hat Q, \hat {\cal G}, \hat
{\bf J}^2, {\hat J}_z$ are well known.  By making the standard assumption that
these operators are mutually commuting, we may immediately establish the
spectrum of the mass for the extremal black holes~\cite{Mazur}.

The classical {\it extremal\/} Kerr-Newman black hole is defined by the 
vanishing of the square root in the expression for the Boyer-Lindquist radius
of the horizon:
\be
r_{\cal H} = M + \sqrt{M^2-Q^2-{\cal G}^2-{\bf J}^2/M^2}
\lb{eq:horizon_radius}
\ee
This means
\be
M^2=Q^2+{\cal G}^2 +{\bf J}^2/M^2
\lb{eq:constraint}
\ee
Now solve for $M$ and discard the negative root solution (it gives
imaginary $M$).  One enforces the quantization of charge, magnetic charge and
spin angular momentum by replacing in this expression 
$Q\rightarrow qe$, ${\cal G}\rightarrow g\hbar/2e$ and $J^2\rightarrow 
j(j+1)\hbar^2$ with $q, g$ integers and $j$ a nonnegative integer or
half-integer.  One thus obtains the mass eigenvalues first found by
Mazur~\cite{Mazur}
\bea
M_{qgj}&=& {\cal M}_P\left[\beta_{qg} + \sqrt{\beta_{qg}^2 +
 j(j+1)}\right]^{1/2}
\lb{eq:mass_extreme}
\\
\beta_{qg} &\equiv& q^2 e^2/2\hbar + g^2 \hbar/8 e^2
\lb{eq:betaqg} 
\eea
What the above manipulations really mean is the following.  The
classical constraint (\ref{eq:constraint}) is replaced by the quantum
statement 
\be (\hat M^2-\hat Q^2 - \hat{\cal G}^2 -\hat {\bf J}^2/\hat M^2)\,|
qgj\rangle=0
\lb{eq:quantum_constraint}
\ee
which picks out the extreme black hole states $\,| qgj\rangle$ whose mass
eigenvalues are given by Eq.~(\ref{eq:mass_extreme}).  Any black hole state
not anhilated by the shown operator is just not the quantum analog of an
extreme black hole.  For the moment I sidestep the question of factor
ordering ($\hat M^2$ and $\hat {\bf J}^2$ may not commute).

For nonextremal black holes one does not have a constraint like
Eq.~(\ref{eq:constraint}).  One can, however, proceed from the
Christodoulou-Ruffini formula for the mass of the Kerr-Newman black hole in
terms of its area (irreducible squared mass):~\cite{CR}  
\be
M^2 = {A\over 16\pi}\left(1+{4\pi (Q^2 + {\cal G}^2)\over A}\right)^2 + {4\pi
{\bf J}^2\over A}
\lb{eq:CR}
\ee
This can be obtained by substituting Eq.~(\ref{eq:horizon_radius}) into the
generalization of Eqs.~(\ref{eq:AreaK}) and (\ref{eq:AreaRN}), namely
\be
A = 4\pi (r_{\cal H}{}^2 +{\bf J}^2/M^2),
\lb{eq:AreaKN}
\ee
and solving for $M^2$.  One should note that only the parameter domain 
\be
A^2 \ge 16\pi^2[ (Q^2 + {\cal G}^2)^2  + 4 {\bf J}^2]
\lb{eq:restriction}
\ee
of Eq.~(\ref{eq:CR}) is physical.  For smaller $A$ the Ruffini-Christodoulou
formula has $M^2$ {\it decreasing\/} with increasing $A$ (for fixed $Q$,
${\cal G}$ and ${\bf J}^2$), a trend which contradicts Eq.~(\ref{eq:AreaKN})
with Eq.~(\ref{eq:horizon_radius}) substituted in.  Obviously when
restriction~(\ref{eq:restriction}) does not hold, formula~(\ref{eq:CR}) is an
extraneous root.

In converting Eq.~(\ref{eq:CR}) to a quantum relation between the
operators $\hat M$, $\hat Q$, $\hat {\cal G}$ and $\hat {\bf J}$, one faces
the problem of factor ordering.  Now the area of a black hole should be
invariant under rotations of its spin; since ${\bf \hat J}$ is the generator
of such rotations, one sees that $[\hat A, \hat{\bf J}] = 0$.  Similarly,
area should remain invariant under gauge transformation whose generator is,
as usual, the charge $\hat Q$.  Hence  $[\hat A, \hat Q] = 0$.  Duality
invariance of the Einstein-Maxwell equations would then suggest that $[\hat
A, \hat {\cal G}] =0$. Hence one may merely replace the parameters in
Eq.~(\ref{eq:CR}) by the corresponding operators:
\be
\hat M^2 = \left[{\hat A\over 16\pi}\left(1+{4\pi (\hat Q^2 + \hat
{\cal G}^2)\over
\hat A}\right)^2 + {4\pi \hat {\bf J}^2\over \hat A}\right] \Theta\left
(\hat A^2 -16\pi^2[(\hat Q^2 + \hat {\cal G}^2)^2  + 4 \hat {\bf J}^2]\right)
\lb{eq:spectrum}
\ee
The Heavyside $\Theta$ (step) function  enforces the physical restriction
Eq.~(\ref{eq:restriction}); when this last is violated, a zero mass
eigenvalue is predicted, which means there is no such black hole.  One may
thus read off the mass eigenvalues of the Kerr-Newman black hole; this
approach was first used in Ref.~\cite{BekNC}.

Two comments are in order.  One might object that it is not obvious that the
operators $\hat A$, $\hat M$, $\hat Q$, $\hat {\cal G}$ and $\hat {\bf J}^2$
are related in exactly the same way as the classical quantities.  Might not
the classical relation (\ref{eq:CR}) arise as an expectation value of
some more complicated looking quantum relation ?   This is possible, but the
available evidence does not seem to require any such complication.  If one
can neglect fluctuations of the various observables, the expectation value
of formula ({\ref{eq:spectrum}) will reproduce the Christodoulou-Ruffini
formula.  The second comment is that it seems nothing was gained in putting
(\ref{eq:spectrum}) forward.  To judge from the classical situation, $\hat A$
would seem to have a continuous spectrum, and so all that (\ref{eq:spectrum})
tells us is that there are several continuum mass sectors, one for each set
of eigenvalues of \{$\hat Q, \hat {\cal G}, \hat {\bf J}^2\}$.  The next
section shows the evidence pointing to a discrete spectrum for $\hat A$.    

\subsection{Discrete spectrum for horizon area}\lb{discrete}

As we saw in Sec.~\ref{adiabatic}, the horizon's area of a nonextremal
black hole is an analog of an adiabatic invariant in mechanics.   This is
interesting to us because one can often understand classical
adiabatic invariance in simple quantum terms.  As an example consider the
plain harmonic oscillator.  When it is in a stationary state (labeled by
quantum number $n$), $E/\omega = (n+{\scriptstyle 1\over \scriptstyle
2})\hbar$.  One expects $n$ to remain constant during an adiabatic change
(changing the spring constant or the length of a pendulum)  because the
perturbations imposed on the system have frequencies $\ll \omega$, so that by
perturbation theory, quantum transitions between states of different
$n$ are strongly suppressed.  Therefore, the ratio $E/\omega$ should be
preserved. Now for the harmonic oscillator the Jacobi action is $\oint p\,dq
= 2\pi E/\omega$ so it should be preserved.  Thus a quantum insight here gives
us an easy understanding of the classical adiabatic invariance of the Jacobi
action involved.

Ehrenfest generalized this insight into a principle~\cite{Ehrenfest}: any
classical adiabatic invariant (action integral or not) corresponds to a
quantum entity with discrete spectrum.  Again, the {\it rationale\/} is that
an adiabatic change, by virtue of its slowness, is expected to lead only to
continuous changes in the system, not to jumps that change a discrete quantum
number.  The preservation of the value of the quantum entity then explains
the classical invariance.   Ehrenfest's idea was embodied in the
Bohr-Wilson-Sommerfeld quantization rules of the old quantum theory:
\be
\,\oint p\, dq=2\pi \hbar n
\lb{BWS}
\ee 

Ehrenfest's hypothesis can be used profitably in many problems. 
A not too well known example concerns a relativistic particle of rest mass
$m$ and charge $e$ spiralling in a magnetic field ${\bf B}$. One knows that
the Larmor spiralling frequency is
\be
\Omega={e|{\bf B}|\over \gamma m}={e|{\bf B}|\over E}
\lb{eq:Larmor}
\ee
where $\gamma$ is Lorentz's factor $(1-\upsilon^2)^{-1/2}$, and $E$ the total
energy.  When ${\bf B}$ varies (in space or in time) slowly over one Larmor
radius $r$ or over one Larmor period $2\pi/\Omega$, there exists, by
Ehrenfest's theorem, an adiabatic invariant of the form 
\be
\oint p\, dq =\,\oint m\gamma\Omega r\, d\ell =2\pi e|{\bf B}|r^2 =
2e\varphi,
\lb{eq:flux}
\ee
namely, the magnetic flux $\varphi$ through one loop of orbit~\cite{Jackson}. 
Now rewrite the energy
\be
E=m\left(1-\dot r^2-\dot z^2-r^2\Omega^2\right)^{-1/2}
\lb{eq:energy}
\ee
by replacing $\dot z\rightarrow p_z/m\gamma$, taking into account that
$\dot r$ is nearly vanishing, and replacing $\Omega$ and $r^2$ by means of
Eq.~(\ref{eq:Larmor}) and Eq.~(\ref{eq:flux}) to get
\be
 E^2=m^2+p_z^2+e^2 r^2 B^2 = m^2+p_z^2+e^2 \varphi B/\pi
\lb{eq:temporary}
\ee

By Ehrenfest's principle, in the quantum problem $\varphi$ should have a
discrete spectrum. One is thus led to expect that for fixed $p_z$, $E^2$
should be quantized, possibly with uniformly spaced eigenvalues.  And indeed,
the {\it exact\/} solution of the relativistic Landau problem with the
Klein-Gordon equation~\cite{LLQE} leads to the spectrum
\be
E^2=m^2+p_z^2+e\hbar B (2n+1); \qquad n=0, 1, \cdots
\lb{eq:Landau levels}
\ee
which justifies the prediction from the Ehrenfest principle.

I now take seriously the analogy between horizon area and adiabatic
invariants to conjecture, in harmony with Ehrenfest's principle, that the
area of an equilibrium black hole has a discrete spectrum.  We do not have
any evidence that one can express horizon area in the form $\oint p\, dq$. 
Therefore, one should not immediately jump to the conclusion that the area
eigenvalues are equally spaced.  After all, what if horizon area corresponded
to $(\oint p\, dq)^2$ rather than to $\oint p\, dq$ ?   Thus at first I only
write the area eigenvalues as
\be
a_n = f(n);\qquad n=1, 2, 3, \cdots
\lb{eq:eigenareas}
\ee
The function $f$ must clearly be positive and monotonically increasing (this
last just reflects the ordering of eigenvalues by magnitude). In light of
Eq.~(\ref{eq:spectrum}) and the quantization of charge, magnetic charge, and
angular momentum, this conjecture implies that the nonextremal Kerr-Newman
black hole also has a discrete {\it mass\/} spectrum.  Its form will be
elucidated in Sec.~\ref{spacing}.

How are the area eigenvalues really spaced ? One can obtain a hint
by elaborating on Christodoulou's reversible processes,  a special case of
which  (for a Reissner-Nordstr\"om black hole) was discussed
Sec.~\ref{particleabs}.  More generally Christodoulou and Ruffini~\cite{CR}
showed that the assimilation of a {\it point\/} classical particle by a
Kerr-Newman black hole can be made reversibly if the particle,  which may be
electrically charged and carry angular momentum,  is injected at the horizon
from a radial turning point in its orbit.  In this case the horizon area (or
equivalently the irreducible mass) is left unchanged, so that the effects on
the black hole can be undone by a second reversible process which adds
charges and angular momentum opposite in sign to those added by the first. 
One can check that Christodoulou and Ruffini's calculation establishes
reversibility only for {\it nonextremal\/} black holes.

\subsection{Quantum Christodoulou processes}\lb{quantum}  

In the Christodoulou-Ruffini process the particle follows a bound classical
orbit, and must be a point particle in order for its absorption to leave the
area unchanged.  Particularly the first requirement clashes with quantum
theory.  The particle cannot both be {\it at the horizon\/} and be {\it at a
turning point\/}; this contradicts the uncertainty principle because
``turning point'' means the radial momentum is exactly zero and this is
incompatible with being precisely ``at the horizon''.  How then do we
formulate the Christodoulou-Ruffini process while taking cognizance of
quantum mechanics for the particle ?  

The first thing to settle is the condition under which one can work
with classical bound orbits at all.  According to the tenets of quantum
mechanics this requires that the particle be in a state with large quantum
number.  Thus let us imagine a particle of mass $\mu<{\cal M}_P$ and charge
$e$ in a quantum stationary state in the spherically symmetric field of a
Reissner-Nordstr\"om black hole with mass $M\gg {\cal M}_P$ and charge
$Q<M$.  In this preliminary investigation we ignore relativistic effects, and
focus on Schr\"odinger's problem in the attractive potential $V=(-M\mu+Qe)/r$
(hence $M\mu>Qe$). The  radii of Bohr orbits of order $n$
are~\cite{Merzbacher}
\be
 R_n={n^2\hbar^2\over \mu (M\mu-Qe)}
\lb{eq:Bohr_radii}
\ee
We require $R_n \approx M$ for a fairly large $n$ so that only
semiclassical states are involved in describing the particle near the black
hole.  Thus
\be
{\hbar\over\mu M}\ll \left(1-{eQ\over \mu M}\right)^{1/2}
\lb{eq:Compton}
\ee
An additional requirement is that the particle's Compton length $\hbar/\mu$
be much smaller than $r_{\cal H}\sim M$, so that one can speak of the
particle localized near ${\cal H}$.  For $Qe$ not approximately equal to $\mu
M$, the second requirement guarantees that restriction (\ref{eq:Compton}) is
satisfied; for  $Qe\approx \mu M$, restriction (\ref{eq:Compton}) already
takes care of making the Compton length small on scale $M$.  In any case,
consideration of semiclassical orbits near ${\cal H}$ requires $\hbar/(\mu M)
\ll 1$.

We now turn to the general relativistic problem.  We generalize the black
hole to a Kerr-Newman one (mass $M$, charge  $Q$ and spin parameter $a\equiv
J/M$).   Boyer-Lindquist coordinates $r$ and $\theta$ are used; the following
abbreviations are useful:
\bea
\Delta &\equiv& r^2-2Mr+a^2+Q^2
\lb{eq:Delta}
\\
\rho^2 &\equiv& r^2+a^2 \cos^2\theta
\lb{eq:rho2}
\eea
It may be noted that $\Delta = 0$ at ${\cal H}$.

The condition  $\hbar/(\mu M) \ll 1$ granted, we should be able to describe
the motion of the particle away from a turning point by applying WKB
approximation~\cite{Merzbacher} to a wave packet representing the particle. 
This means the packet's center of mass will move classically on a geodesic
or---if the particle is charged---on the appropriate solution of the Lorentz
equation in Kerr-Newman spacetime.  At the turning point the WKB
approximation breaks down~\cite{Merzbacher}, so we expect the above
description to fail.  However, since the gist of the quantum description is
the existence of uncertainty relations, it should be possible to obtain
correct relations between the various parameters of the orbit if we replace
the physical radial momentum at the turning point by the radial momentum
uncertainty $\delta P$, and the proper radial distance of the turning point
from the horizon by the radial proper distance uncertainty $\hbar/\delta P$. 
This has to be done in the integrated classical orbits.

Carter~\cite{Carter3} was first to find the first integrals for the meridional
($\theta$) and radial ($r$) motions in the Kerr-Newman background.  The first
integrals can be combined as in Misner, Thorne and Wheeler~\cite{MTW}:
\bea
0 &=& \tilde\alpha E^2 -2\tilde\beta E + \tilde\gamma 
\\
\tilde\alpha &\equiv& (r^2+a^2)^2 - a^2 \sin^2\theta \Delta
\lb{eq:alpha}
\\
\tilde\beta &\equiv& (r^2+a^2) (a L_z + eQr)  - a L_z \Delta
\lb{eq:beta}
\\
\tilde\gamma &\equiv&  (a L_z + eQr)^2 -  (L_z^2/\sin^2\theta+
\mu^2\rho^2) \Delta -\rho^4\left[(p^r)^2+ (p^\theta)^2 \Delta\right]
\lb{eq:gamma}
\eea
Here $E$ and $L_z$ denote the total energy and angular momentum about the
symmetry axis of the particle, while $p^r$ and $p^\theta$ denote the
appropriate contravariant momentum components.  It proves useful to express
these last in terms of the physical components (in an orthonormal
tetrad) $P\equiv\Delta^{-1/2}\rho\, p^r$ and $\Pi \equiv\rho\, p^\theta$. 
Note that $\Pi$ bears dimension of a linear momentum.

Following Christodoulou and Ruffini (see also  Appendix A
in Ref.~\cite{BekEntropy}), one solves the quadratic for $E$ taking care to
select the root which would give positive energy far from the hole.  We are
only interested in the expression near the horizon where $\Delta$ is small,
so one may replace $r\rightarrow r_{\cal H}$ and $\rho^2\rightarrow
\rho_{\cal H}{}^2$ everywhere except in $\Delta$ itself.  Pulling a factor
$\Delta$ out of the root we get after some rearrangement
\bea
E-\Omega L_z-e\Phi &=& {\Delta^{1/2}\over
r_{\cal H}{}^2+a^2}\sqrt{{B^2\over \sin^2\theta} +\rho_{\cal
H}{}^2(\mu^2+P^2+\Pi^2) }+ {\cal O}(\Delta^{3/2})
\lb{eq:integral_of_motion}
\\
\Omega &\equiv& a(r_{\cal H}{}^2+a^2)^{-1} \rightarrow {\rm rotational\
angular\ frequency}
\\
\Phi &\equiv& Q\,r_{\cal H}(r_{\cal H}{}^2+a^2)^{-1} \rightarrow 
{\rm electrical\  potential}
\lb{eq:Phi}
\\
B &\equiv & L_z-\Omega(a L_z+r_{\cal H}\,Qe)\sin^2\theta
\eea
$\Omega$ is identified with the angular frequency of the hole because it
turns out to coincide with the dragging angular frequency at the
horizon~\cite{MTW}.  Likewise $\Phi$ is interpreted as the hole's electrical
potential because it equals the component $A_t$ of the vector potential
evaluated at ${\cal H}$.

As mentioned, one may not set $\Delta =0$ in
Eq.~(\ref{eq:integral_of_motion}) even if one is interested in capture by the
black hole because of the uncertainty principle.  It thus pays to reexpress
the prefactor of the square root in Eq.~(\ref{eq:integral_of_motion}) in
terms of the (small) {\it proper\/} radial distance $\ell$ of $r$ from
$r_{\cal H}$.  One easily calculates that
\be
\ell \equiv \int_{r_{\cal H}}^r \sqrt{g_{rr}}\, dr \approx {2\rho_{\cal H}
(r-r_{\cal H})^{1/2}\over (r_{\cal H}-r_{\cal C})^{1/2}} 
\approx {2\rho_{\cal H} \Delta^{1/2}\over (r_{\cal H}-r_{\cal C})}
\ee
where $r_{\cal C}$ is the radius of the inner (Cauchy) horizon [negative
square root in Eq.~(\ref{eq:horizon_radius})].  We may thus rewrite
Eq.~(\ref{eq:integral_of_motion}) as
\be
E-\Omega L_z-e\Phi = { (r_{\cal H}-r_{\cal C})\ell\over 2\rho_{\cal
H}(r_{\cal H}{}^2+a^2)}\sqrt{{B^2\over \sin^2\theta} +\rho_{\cal
H}{}^2(\mu^2+P^2+\Pi^2) }
\lb{eq:new_integral}
\ee

Let us now calculate the change in horizon area occasioned by the particle's
capture. The differential of Eq.~(\ref{eq:AreaKN}) is
\bea
d A &=& \Theta_{KN}^{-1} (d M-\Omega d J -\Phi d Q)
\lb{eq:dAKN}
\\
\Theta_{KN} &\equiv& (r_{\cal H}{}^2+a^2)^{-1}(r_{\cal H} - r_{\cal C})/16\pi
\eea
Not coincidentally, $\Omega$ and $\Phi$ appear here.  Their physical
identifications are again clear from analogy with thermodynamic formulae. We
must of course substitute $d M = E$, $d Q = e$ and $d J = L_z$ in accordance
with energy, charge and angular momentum conservation.  In view of the result
(\ref{eq:new_integral}) we have
\be
d A = 8\pi\ell \sqrt{{B^2\over\rho_{\cal H}{}^2\sin^2\theta} +\mu^2+P^2+\Pi^2}
\lb{eq:final_dA}
\ee

In this equation $\theta$ represent the meridional angle at which the capture
takes place while $\ell$ is a measure of the radial proper distance from the
horizon at which the particle can be said to merge with the black hole.  In
the {\it classical\/} case the  limit $\ell \rightarrow 0$ recovers for us
Christodoulou's reversible process for the nonextremal black holes (the
turning point condition is $p^r\rightarrow 0$ or $\ell P\rightarrow 0$).  But
$d A$ cannot be zero in the quantum case.  We are interested in the minimum
possible value for $d A$ required by quantum mechanics.  Actually quantum
mechanics places no onerous limits on $\Pi$; since we are not terribly
interested on precisely where on the horizon the particle hits, one can
tolerate substantial uncertainty in angle $\theta$, or equivalently in
the linear coordinate $r_{\cal H}\theta$.   As its canonically conjugate
momentum, $\Pi$ is allowed to have a small uncertainty. More precisely, we
could have $\Pi\sim \delta
\Pi \sim \hbar/r_{\cal H}$ so that the contribution to $d A$ is $\sim
\hbar(\ell/r_{\cal H})$.  Of course our whole treatment presupposes that the
particle can approach the horizon close compared to the latter's radius, so
the contribution to $d A$ can be made negligible compared to $\hbar ={\cal
L}_P{}^2$.

Now the sign of $L_z$, the angular momentum along the symmetry axis, is
free.  One can classically arrange for $L_z$ to be such as to nullify the
quantity $B$.  In the quantum theory  $L_z$, is quantized, as usual, with the
spectrum (no spin) $\hbar \times \{\cdots\, -2, -1, 0, 1, 2,\,\cdots\}$.  But
this occasions no special problem.   Suppose first that $Qe=0$.  Then we can
certainly pick $L_z=0$ so that $B$ makes no contribution to $dA$.  Now
suppose $Qe$ is nonzero.  Our whole treatment presupposes that absorption of
the particle is a small perturbation on the hole, so that $|e|\ll |Q|$. 
Since charge is quantized in of $\approx (\hbar/137)^{1/2}$ we see that
$|eQ|\gg \hbar/137$.  Now for the  Kerr-Newman black hole $\Omega r_{\cal H}
\leq 1/2$ and $\Omega a \leq 1/2$.  Thus unless the black hole is nearly
nonrotating, it is possible to select a nonzero eigenvalue of $L_z$ which
nullifies $B$ with an error no larger than about $\hbar$.  For a hole very
close to nonrotating or an accretion point near the hole's pole (so that
$|a\Omega e Q|\sin^2\theta < \hbar$), this can be accomplished with $L_z=0$. 
Hence the term $B^2$ under the root in Eq.~(\ref{eq:final_dA}) contributes to
$d A$ at most a term of
${\cal O}(\hbar\ell/\rho_{\cal H})$.  Again the contribution to $d A$ can be
made small compared to ${\cal L}_P{}^2$.

The contribution to $d A$ of the $P^2$ term under the square root cannot be
made so small.  At the turning point $P$ cannot be said to vanish, but
must be replaced by its uncertainty $\delta P$.  And the center of the
particle cannot be placed at the horizon with accuracy better than the radial
position uncertainty $\hbar/\delta P$; thus $\ell^2 P^2 > \hbar^2$.  
Likewise, the particle cannot be localized to better than a Compton
wavelength $\hbar/\mu$ so that $\ell^2\mu^2 > \hbar^2$.  It follows that
there must exist a quantum lower bound on $d A$:
\be
 (d A)_{\rm min} =  8\pi\xi \hbar = 
\alpha\,{\cal L}_P{}^2
\lb{eq:universal}
\ee
where the numerical coefficient $\xi$ takes into account the inherent
fuziness of the uncertainty relation. Incidentally, this conclusion  fails for
{\it extremal\/} black holes because $\Theta_{KN}$ in Eq.~(\ref{eq:dAKN})
diverges in that case.  The minimal increase in area is then not
Eq.~(\ref{eq:universal}), but a quantity dependent on $M$, $Q$ and $J$, just
as in the example discussed at the end of Sec.~\ref{particleabs}.   But,
surprisingly, for nonextremal black holes $(d A)_{\rm min}$ turns out to be
independent of the black hole parameters $M$, $Q$ and $J$.  

It is in order to emphasize the approximations made in obtaining
Eq.~(\ref{eq:universal}).  We assumed the particle only slightly perturbs the
black hole.  Thus if it is charged, $Q\gg (\hbar/137)^{1/2}$ and in any case
$M\gg \mu$.  We also assumed the particle can get close to the horizon which
means $M\gg \ell > \hbar/\mu$.  Of course, the last two inequalities are
consistent by our original assumption that $M\gg\surd\hbar = {\cal M}_P$.

\subsection{Spacing and multiplicity of the area eigenvalues}\lb{spacing} 
   
The fact that, as soon as one allows quantum nuances to the problem, there
is, for nonextremal black holes, a minimum horizon area increase  suggests
that this  $(d A)_{\rm min}$ corresponds to the spacing between eigenvalues
of  $\hat A$ in the quantum theory.  And the fact that $(d A)_{\rm min}$ is a
universal constant suggests that the spacing between eigenvalues is a uniform
spacing.  For it would be strange indeed if that spacing were to vary, say, as
mass of the black hole, and yet the increment in area resulting from the best
approximation to a reversible process would contrive to come out universal, as
in Eq.~(\ref{eq:universal}), by involving a number of quantum steps inversely
proportional to the eigenvalue spacing.  I thus conclude that for nonextremal
black holes the spectrum of
$\hat A$ is
\be
a_n = \alpha\, {\cal L}_P{}^2\, (n+\eta);\quad\ \eta>-1; \quad n = 1,
2,\cdots
\lb{eq:areaspectrum} 
\ee
where the condition on $\eta$ excludes nonpositive area eigenvalues.  Since
Eq.~(\ref{eq:universal}) fails for an extremal Kerr-Newman black hole, one
cannot deduce as above that its area eigenvalues are evenly spaced.  This is
entirely consistent with Eq.~(\ref{eq:mass_extreme}) according to which the
area spectrum is then very complicated.

For nonextremal black holes the evidence of Sec.~\ref{quantum} only suggests
a uniformly spaced spectrum well above the Planck scale.  Thus
Eq.~(\ref{eq:areaspectrum}) is supported for large $n$, or for any $n$ if
$\eta\gg 1$.  However, I shall go beyond the concrete evidence and assume
that the formula is valid also at low quantum numbers even if $\eta={\cal
O}(1)$.  Some support for this comes from the heuristic picture of a
patchwork horizon discussed below.

Thus far I have said nothing about entropy; the discussion has been at the
level of mechanics, not statistical physics. But Eq.~(\ref{eq:areaspectrum})
allows us to understand, in a pleasant and intuitive way, the mysterious
proportionality between black hole entropy and horizon area. 

The quantization of horizon area in {\it equal\/} steps brings to mind
an horizon formed by patches of equal area $\alpha\,{\cal L}_P{}^2$ which get
added one at a time.  There is no need to think of a specific shape or
localization of these patches.  It is their standard size which is important,
and which makes them all equivalent.   This patchwork horizon can be regarded
as having many degrees of freedom, one for each patch.  After all, the
concept ``degree of freedom'' emerges for systems whose parts can act
independently, and here the patches can be added to the patchwork one at a
time.  In quantum theory degrees of freedom independently manifest distinct
states.  Since the patches are all equivalent, each will have the same number
of quantum states, say, $k$.  Therefore, the total number of quantum states
of the horizon is
\be
N = k^{A/(\alpha {\cal L}_P{}^2)}
\lb{eq:numberstates}
\ee
where $k$ is a positive  integer and the effects of the $\eta$ zero point in
Eq.~(\ref{eq:areaspectrum}) are glossed over in this, heuristic, argument.

The $N$ states may not all be equally probable.  But if the $k$ states of
each patch are all equally likely, then all $N$ states are
equally probable.  In that case the statistical (Boltzmann) entropy
associated with the horizon is $\ln N$ or
\be
S_{BH} = {\ln k\over \alpha}{A\over {\cal L}_P{}^2}
\lb{eq:BHentropy}
\ee
Thus is the proportionality between black hole entropy and horizon
area justified in simple terms. Even if not all $k$
states are equally probable, one can still use Eq.~(\ref{eq:BHentropy})
provided $k$ is regarded as an effective number of equally probable states.
Only at this point thus one compare Eq.~(\ref{eq:BHentropy}) with Hawking's
formula for $S_{BH}$ to calibrate the constant $\alpha$: 
\be
\alpha=4\ln k
\lb{eq:calibration}
\ee

The above argument depends crucially on the uniformly spaced area spectrum.
 The logic leading to the number of states Eq.~(\ref{eq:numberstates}) was
used in the early days of black hole thermodynamics by me~\cite{BekinWheeler}
and by Sorkin~\cite{Sorkin} without regard to any particular area spectrum,
but these early arguments are not really convincing because their partition
of the horizon into equal area cells would be without basis if the desired
result, entropy $\propto$ area, were not known.

Mukhanov's~\cite{Mukhanov,BekMukh} alternate route to 
Eqs.~(\ref{eq:numberstates}) and (\ref{eq:calibration}) starts from the
accepted formula relating black hole area and entropy. In the spirit of the
Boltzmann-Einstein formula, he views
$\exp(S_{BH})$ as the degeneracy of the particular area  eigenvalue because
$\exp(S_{BH})$ quantifies the number of microstates of the black hole
that correspond to a particular macrostate (a black hole with definite $M$,
$Q$ and ${\bf J}$).  Since black hole entropy is determined by thermodynamic
arguments only up to an additive constant, one writes, in this approach,
$S_{BH}= A/4{\cal L}_P{}^2+$ const. Substitution of the area eigenvalues 
from Eq.~(\ref{eq:areaspectrum}) gives the degeneracy corresponding to
the $n$-th area eigenvalue:
\be
g_n = \exp\left({a_n\over 4{\cal L}_P{}^2} + {\rm const.}\right) = 
g_1\ e^{\alpha
(n-1)/4}
\lb{eq:exp}
\ee

As stressed by Mukhanov, since $g_n$ has to be integer for every $n$,
this is only possible when~\cite{BekMukh,QG}
\be
g_1 = 1, 2, \cdots\quad{\rm and}\quad \alpha = 4\times\left\{\ln 2,\ln 3, 
\cdots\right\} 
\lb{eq:options}
\ee
The simplest option would seem to be $g_1 = 1$ (nondegenerate
black hole ground state).  Here the additive constant in
Eq.~(\ref{eq:exp}) must be negative: were it zero, the area $a_1$ would also
vanish which seems an odd thing for a black hole.  Just this case was studied
in Ref.~\cite{BekMukh}; it is a bit ugly in that the eigenvalue law
Eq.~(\ref{eq:areaspectrum}) and the black hole entropy include related but
undetermined additive constants.  

The next simplest case, $g_1=2$ (doubly degenerate black hole ground state),
no longer requires the ugly additive constant in the black hole entropy to
keep $a_1$ from vanishing.  With this constant set to zero and the choice
$\alpha=4\ln 2$ corresponding to $k=2$, Eqs.~(\ref{eq:areaspectrum})  and
(\ref{eq:exp}) require that  $\eta=0$ so that one is rid of the second ugly
constant as well.  The area spectrum is
\be
a_n = 4{\cal L}_P{}^2\ln 2\cdot n;   \quad  n=1,2,\cdots
\lb{eq:truespectrum}
\ee
't Hooft has independently found evidence for a fundamental unit of area on
the horizon of size $ 4{\cal L}_P{}^2\ln 2$~\cite{tHooft}.   

Spectrum (\ref{eq:truespectrum}), which I shall adopt henceforth, is good for
nonextremal Kerr-Newman black holes.   The corresponding degeneracy of area
eigenvalues
\be
g_n = 2^n
\lb{eq:degeneracy}
\ee
corresponds to a doubling of the degeneracy as one passes from one area
eigenvalue to the next largest.  Mukhanov~\cite{Mukhanov} thought of this
multiplicity as the number of ways in which a black hole in the $n$-th area
level can be made by first making a black hole in the ground state,
and then proceeding to ``excite it'' up the ladder of area levels in
all possible ways.  Danielsson and Schiffer~\cite{DanSch} considered
this multiplicity as representing rather the number of ways the black hole
with area $a_n$ can ``decay'' down the staircase of levels to the ground
state. In either case there are $2^{n-1}$ ways.  The extra factor of two in
the scheme here adopted comes from the double degeneracy of the ground state. 

To what extent do these intuitively physical predictions correspond to
results from more formal quantum gravity schemes ?   Mention should be made
of Kogan's string theoretic argument~\cite{Kogan}, and the quantum membrane
approaches of Maggiore~\cite{Maggiore} and Lousto~\cite{Lousto} which
establish the uniformly spaced area eigenstates as the base for excitations
of the black hole.  The efforts of D-brane aficionados (for a review see
Ref.~\cite{Peet}) have rather concentrated on the question of degeneracy {\it
qua\/} entropy, and it is not clear that they have anything to say about a
discrete mass spectrum.   

There are also several canonical quantum gravity treatments of a shell or 
ball of dust collapsing on its way to black hole formation.  Those by
Schiffer~\cite{Schiffer} and Peleg~\cite{Peleg} obtain a uniformly spaced
area spectrum.  But Berezin~\cite{Berezin}, as well as Dolgov and
Khriplovich~\cite{DolgKhrip}, obtain mass spectra for the ensuing black hole
which correspond to discrete area spectra with {\it nonuniform\/} spacing
(and in Berezin's approach the levels are infinitely degenerate).  Other 
canonical quantum gravity approaches by  Louko and M\"akel\"a~\cite{LoukMak},
Barvinskii and Kunstatter~\cite{BarvKunst}, M\"akel\"a~\cite{Makela} and
Kastrup~\cite{Kastrup} treat rather a spherically symmetric {\it vacuum\/}
spacetime that gets endowed with dynamics by some subtlety; they also come up
with a uniformly spaced area spectrum.  There is, however, no general
agreement on the spacing of the levels.  The analogous treatment of the
charged black hole by M\"akel\"a and Repo~\cite{MakelaRepo} gets a nonuniform
area spectrum.     

In the loop quantum gravity approach (for a review see Ref.~\cite{Ashtekhar})
the black hole area spectrum is discrete  but with a spacing  which narrows
with increasing area, becoming virtually continuous in the infinite area
limit.  This is completely at variance with the uniformly spaced spectrum. 
However, as noticed by I. Khriplovich~\cite{Khriplovich}, the area spectrum
for the extremal neutral Kerr black hole according to Mazur
[Eq.~(\ref{eq:mass_extreme}] coincides in part with the loop gravity horizon
area spectrum.

The contradictory conclusions mentioned support the view that none of the
existing formal schemes of quantum gravity is as yet a quantum theory of
gravity. Clearly a role exists for the heuristic approach.

\section{Black hole spectroscopy}  \lb{Consequences}

\setcounter{equation}{0}
\renewcommand{\theequation}{5.\arabic{equation}}

Of the ramifications of the discrete area spectrum, the most surprising is
the prediction of quasidiscrete spectral lines from a black hole, even one
well away from the Planck scale.  In this last lecture I explore this
aspect.  I continue to use units with $G=c=1$.

\subsection{The mass levels and a paradox}\lb{paradox}

In the operator relation (\ref{eq:spectrum}) we substitute the area
spectrum (\ref{eq:truespectrum}).  The mass eigenvalues of the Kerr-Newman
black hole are thus~\cite{BekNC}
\bea
M_{nqgj} &=& {\cal M}_P\left[{n\ln 2\over 4\pi}\left(1+{2\pi\beta_{qg}\over
n \ln 2}\right)^2 + {\pi j(j+1)\over n \ln 2}\right]^{1/2}
\lb{eq:mass_spectrum}
\\
n &>& {2\pi\over \ln 2}\sqrt{\beta_{qg}^2 + j(j+1)}
\lb{eq:inequality}
\eea
where $\beta_{qg}$ is the same as in Eq.~(\ref{eq:betaqg}) and the constraint
on
$n$ comes from the Heavyside function in Eq.~(\ref{eq:spectrum}); we have
written a strict inequality because we know that formula
(\ref{eq:mass_spectrum}) applies to nonextremal black holes only.  

For zero charges and spin the mass spectrum is of the form
\be
M\propto \surd n; \qquad n=1, 2, \, \cdots
\lb{eq:massspectrum}
\ee
implying the $n\Longrightarrow n-1$ transition frequency
\be
\omega_0\equiv d M/\hbar = ( 8\pi M)^{-1}\ln 2
\lb{eq:Bohr}
\ee
This simple result is in agreement with Bohr's correspondence principle:
``transition frequencies at large quantum numbers should equal classical
oscillation frequencies'', because a classical Schwarzschild black hole
displays `ringing frequencies' which scale as $M^{-1}$, just as
Eq.~(\ref{eq:Bohr}) would predict.  This agreement would be destroyed if the
area eigenvalues were unevenly spaced.  Indeed, the loop gravity spectrum
mentioned in Sec.~\ref{spacing} fails this correspondence principle test
(practitioners of loop gravity are content with trying to recover the Hawking
semiclassical spectrum in some limit---see review in Ref.~\cite{Ashtekhar}.

It follows from the discussion in Sec.~\ref{quantum} that absorption of a
{\it massive\/} particle by the hole always causes a jump in black hole mass
of at least $d M$ (this corresponds to $d A = 4 {\cal L}_P\ln 2$).  What if
the particle is very light ($\mu\ll \hbar/M$), or if we replace it by a
photon ($\mu=0$).  The discussion in Sec.~\ref{quantum} is no longer relevant
since we cannot follow the localized particle to near the horizon: we have to
treat the particle as a wave.  Then a paradox---the treshold
paradox---arises.  Scatter off a Schwarzschild black hole an electromagnetic
wave whose frequency $\omega$ is below $\omega_0$, or is not a precise
multiple of
$\omega_0$.  Photons in the wave do not have the right frequency to cause
a transition between two mass levels.  It would seem that none of the wave
can be absorbed.  Admittedly, the transmissivity is small at small
frequencies, but the quantum prediction of no absorption contrasts starkly
with the classical picture of some absorption.  And when $\omega\gg
\omega_0$, the classical transmission coefficient for electromagnetic waves
is close to unity, so one sees no correspondence between the quantum picture
espoused here and the accepted classical picture,  even in the limit of large
black holes.  Does all this mean the area spacing we have postulated is not
really there ?  

One should not confuse the question of the classical transmissivity with the
question of quantum absorptivity.  The transmissivity is determined by the
potential barrier around the black hole that shows up in the electromagnetic
wave equation.  By contrast, the statement that a photon cannot get absorbed
unless its frequency is $\omega_0$ or a multiple thereof is a quantum gravity
statement.  A single photon with $\omega< \omega_0$ should never be absorbed
(modulus the question of line broadening and splitting to be discussed below)
even though it has some probability of penetrating the potential
barrier~\cite{Starobinskii}.  But if we are dealing with a macroscopic
wave with $\omega< \omega_0$, multiple photon absorption may help to achieve
the treshold; re-emission of photons with frequency $\neq \omega$ is then
possible.   This would be analogous to multiphoton processes in nonlinear
optical media where the incident frequencies are shifted.  This anomalous
absorption would be interpreted in classical theory as the expected
absorption of subtreshold frequencies.  Now consider a photon with
$\omega=100.3\,
\omega_0$.  It also is not in resonance with the black hole levels.  But
after negotiating the potential barrier, which it does easily because of its
high frequency~\cite{Starobinskii}, it may get absorbed with re-emission of a
quantum with frequency $0.3\,\omega_0$, or $1.3 \,\omega_0$,  etc.  One would
thus expect that a macroscopic wave with $\omega=100.3\,\omega_0$ can get
partially absorbed with accompanying re-emission of lower frequency radiation.

Admittedly this absorbing behavior of black holes is at variance with what
one is accustomed to expect from quantum field theory on a fixed background,
where a wave's frequency is not shifted while scattering off a
stationary object.  But such shifts are seen in nonlinear optics, and
gravitation is a nonlinear phenomenon.  

\subsection{The black hole line emission spectrum}\lb{lines}

By analogy with atomic transitions, a black hole at some particular
mass level would be expected to make a transition to some lower level with
emission of one or more quanta of any of the fields in nature.  In the sequel
I call these photons for short.  The corresponding line spectrum---very
different from the Hawking semiclassical continuum---was first discussed in
Ref.~\cite{BekNC} and further analyzed much later~\cite{BekMukh,QG}. 
According to Eq.~(\ref{eq:Bohr}) the spacing between mass levels is uniform
over a small range of $M$.  Thus quantum jumps larger than the minimal
produce emission at all frequencies which are integral multiples of
$\omega_0$: $\omega =
\omega_0\, \delta n$ with $\delta n =1, 2, \,\cdots\,$.  

As Mukhanov was first to remark~\cite{BekMukh,QG}, this simple spectrum
provides a way to make quantum gravity effects detectable even for black holes
well above the Planck mass:  the uniform frequency spacing of the black hole
lines occurs at all mass scales, and the unit of spacing is inversely
proportional to the black hole mass over all scales.  Of course, for very
massive black holes, one would expect all the lines to become dim and 
unobservable (just as in the semiclassical description the Hawking radiance
intensity goes down as
$1/M^2$), but there should be a mass regime (primordial mini-black holes ?)
well above Planck's for which the first few uniformly spaced lines should be
detectable under optimum circumstances.  It is thus important to
understand clearly the nature of the line spectrum.

First we must know the ratio of line intensities.  Again proceeding by
analogy with the perturbation theory of atomic line transitions, each line
intensity should be proportional to the square of a matrix element, to the
photon energy
$\hbar \omega_0 \delta n$, to the photon phase space factor, and to the
degeneracy of the {\it final\/} black hole state.  We  do not know anything
about the matrix element, or even what the relevant operator is.  Thus it
seems wisest to assume that the matrix element does not vary much as one goes
from a nearest neighbor transition (frequency $\omega=\omega_0$) to one 
between somewhat farther neighbors ($\omega=\omega_0\, \delta n$).  Thus,
aside from the question of normalization of the spectrum, the matrix element
does not enter into our simple estimate.   

The phase space factor is, as usual, $\omega^2=(\omega_0\, \delta n)^2$. 
The final black hole state's degeneracy factor is $2^{n-\delta n}$ where $n$
refers to the initial state.  Thus all possible transitions from the
state $n$ will give lines with frequencies $\omega=\omega_0\, \delta n$ and
intensities proportional to $ (\omega_0\delta n)^3 \exp(-\delta n \ln
2)= \omega^3 \exp(-\delta n \ln 2)$.  The same result can be had by relying on
the relation between the Einstein coefficient of spontaneous emission
$A_\downarrow$ and that for absorption $B_\uparrow$ (this last equivalent to
the squared matrix element):
\be
A_\downarrow = B_\uparrow{\hbar\omega^3\over 4\pi^2}{g_{\rm down}\over
g_{\rm up}}= B_\uparrow{\hbar\omega^3\over 4\pi^2} {2^{n-\delta n}\over 2^n}
\lb{eq:Einstein}
\ee
With $\omega\rightarrow \omega_0\,  \delta n$ this gives the same result
stated earlier.  Therefore, the line spectrum emitted by the black hole is
expected to be
\be
I(\omega) \propto  \sum _{\delta n =1}^\infty
\Gamma(\omega)\,\hbar\omega^3\exp(-\omega
\ln 2/\omega_0) \,\delta(\omega-\omega_0\,\delta n)
\lb{eq:lines}  
\ee
where $\Gamma(\omega)$ is the transmission coefficient through the potential
barrier surrounding the black hole averaged over angular momenta of the
quanta. 

This result should be compared with Hawking's semiclassical spectrum
\be
I(\omega) \propto {\Gamma(\omega)\hbar\omega^3\over \exp(\hbar\omega/T_{\rm
BH})-1} = {\Gamma(\omega)\hbar\omega^3\over \exp(\omega
\ln 2/\omega_0)-1}
\lb{eq:Hawking_spectrum}
\ee
where we have used Eq.~(\ref{eq:Bohr}) and the standard expression for
the Hawking temperature $T_{\rm H} = \hbar/8\pi M$.  It may be seen that,
apart from the question of normalization, Hawking's spectrum becomes the
envelope of the line spectrum for $\omega \gg \omega_0$ while ``overshooting''
slightly the first few lines.  Both the existence of lines and the
``deficiency'' in the first few as compared to the thermal spectrum are
predictions of the heuristc approach.

The above is not to say that the emission spectrum should be a pure line
spectrum. Multiple photon emission in one transition will also contribute a
continuum.  To go back to atomic analogies, the transition from the 2s to the
1s states of atomic hydrogen, being absolutely forbidden by one-photon
emission, occurs with the long lifetime of 8 s by two-photon emission (photon
splitting in the jargon).  The hydrogenic spectrum is thus a continuum over
the relevant frequency range.  We have already mentioned multiphoton {\it
absorption\/} as a possible resolution of the ``treshold paradox''.  By
detailed balance some multiple photon emission should accompany decay of the
black hole from higher to lower mass levels which should generate a continuum
that would compete with the line spectrum~\cite{BekMukh,QG}.  However, for
the black hole no reason is known why one-photon transition would be
forbidden.  Thus my expectation, again based on the atomic analogy, is that
most of the energy will get radiated in one-photon transitions which give
lines.  The spectrum, in first approximation, should be made up of lines
sticking quite clearly out of a lowly continuum.

\subsection{Broadening and splitting of black hole lines} \lb{broadening} 

Another question is whether natural broadening of the lines
will not smear the spectrum into a continuum.  First explored by
Mukhanov~\cite{Mukhanov}, this issue has been revisited recently by both of
us~\cite{BekMukh,QG}.  By the usual argument the reciprocal broadening of a
line, $(\delta\omega)^{-1}$, should be of order $\tau$, the typical time (as
measured at  infinity) between transitions of the black hole from level to
level.  One may estimate the rate of loss of black hole mass as
\be
{dM\over dt} \approx -{\hbar\omega_0\over \tau}=- {\hbar\,\ln 2\over 8\pi M\,
\tau }
\lb{eq:rate}
\ee
Alternatively, one can estimate $dM/dt$ by assuming, in accordance with 
Hawking's  semiclassical result, that the radiation is black body radiation,
at least in its intensity.  Taking the radiating area as 
$4\pi (2M)^2$ and the temperature as  $\hbar/8\pi M$ one gets
\be
{dM\over dt} = - {\gamma\hbar\over 15360\pi M^2}
\lb{eq:blackbody}
\ee
where $\gamma$ is a fudge factor that summarizes the grossness of our
approximation.  By comparing Eq.~(\ref{eq:blackbody}) with Eq.~(\ref{eq:rate})
one infers $\tau$ which then gives
\be
{\delta\omega\over \omega_o} \sim 0.019\,\gamma
\lb{eq:relative}
\ee

Mukhanov and I regard $\gamma$ to be of order unity, which would make
the natural broadening weak and the line spectrum sharp.  More recently
M\"akel\"a~\cite{Makela2} has estimated a much larger value, and claimed
that the line  spectrum effectively washes out into a continuum.  He
views this as a welcome development because it brings the ideas about black
hole quantization, as here described, into consonance with Hawking's
smooth semiclassical spectrum.  

M\"akel\"a uses Page's~\cite{Page} estimate of black hole luminosity which
takes into account the emission of several species of quanta, whereas our
value $\gamma \sim 1$ is based on one species.  It is, of course, true that a
black hole will radiate all possible species, not just one.  This is
expected to enhance $\gamma$ by an order or two over the naive value.   But it
is also true that because the emission is, in the first instance, in lines,
part of the frequency spectrum is thus blocked, which should lead to a
reduced value for $\gamma$ in Eq.~(\ref{eq:blackbody}).  Mukhanov and I
consider the two tendencies to partly compensate, and expect $\gamma$ to
exceed its putative value of unity by no more than an order of magnitude. 
According to Eq.~(\ref{eq:relative}) this should leave the emission lines
unblended.

Anyway, the most important thing to get out of Eq.~(\ref{eq:relative}), of
which only the value of $\gamma$ is in contention, is that the  natural
broadening scales in proportion to the line spacing.  Thus natural broadening
is not the way to get a spectrum which gradually becomes a continuum for more
massive black holes.  If the lines are smeared into a continuum by natural
broadening, then this is true even at the Planck scale.

We now come to line splitting.  In atomic physics emission spectra display
a hierarchy of splittings which can be viewed as reflecting the hierarchical
breaking of the various symmetries.  Thus in atomic hydrogen the $O(4)$
symmetry of the Coulomb problem, which is reflected in the Rydberg-Bohr
spectrum, is broken by relativistic effects (spin-orbit interaction and
Thomas precession) thus giving rise to fine structure splitting of lines. 
But even an exact relativistic treatment in the framework of Dirac's equation
leaves the 2s and the 2p levels perfectly degenerate.  They are split by a
minute energy by vacuum polarization effects and the Lamb shift. In addition,
the rather weak interaction of the electron with the nuclear proton's
magnetic moment leads to a small hyperfine splitting of members of some of
the other fine structure multiplets.  The very simple spectrum in
Eq.~(\ref{eq:massspectrum}) is analogous to the hydrogenic Rydberg-Bohr
spectrum.  Are there any splittings of the lines here discussed ?

There is certainly room for splitting because of the $2^n$-fold degeneracy 
of the levels, particularly well above the Planck scale where $2^n$ is
large.  And we must remember that the higher the mass level, the smaller the
mass spacing between adjacent levels.  Thus, contrary to what happens with
natural broadening, degeneracy splitting could give a spectrum which becomes
quasicontinuous at some mass well above the Planck mass.  To answer
the question of whether there is level splitting and how much, we obviously
need a more formal derivation of the black hole mass spectrum which could
take into account the lifting of symmetries.  This is the purpose of the
algebraic approach to be described in Sec.~\ref{algebraic}

\subsection{Algebraic approach to the quantum black hole}\lb{algebraic}

In quantum theory one usually obtains spectra of operators from the algebra 
they obey.  For instance, Pauli~\cite{Pauli} obtained the complete
spectrum of hydrogen in nonrelativistic theory from the $O(4)$ algebra
of the relevant operators.  This approach sidesteps the question of
constructing the wavefunctions for the states.  I will now describe an
axiomatic algebraic approach, whose genesis goes back to joint work with
Mukhanov, and which  gives an area spectrum identical to the one found above.
It thus supports the results obtained previously, and illuminates the
question of level splitting~\cite{BekBrazil}.  

In Sec.~\ref{mass_spectrum} I introduced some of the relevant operators for
a black hole: mass $\hat M$, charge $\hat Q$, magnetic charge $\hat {\cal G}$
and spin $\hat {\bf J}$.  The spectrum of $\hat Q$ is $\{qe| q\in {\bf Z}\}$,
that of $\hat {\bf J}^2$ is $\{j(j+1)\hbar^2| j=0,{\scriptstyle 1\over
\scriptstyle 2}, 1, \cdots\,\}$, while that of $\hat J_z$ is $ \{-j\hbar,\
-(j-1)\hbar,\ \cdots,\ (j-1)\hbar,\  j\hbar\}$, where
${\bf Z}$ denotes the set of integers.  For brevity I shall ignore $\hat {\cal
G}$ henceforth.  Our first axiom expands the algebra to include horizon area:

\medskip
\noindent {\bf Axiom 1}: Horizon area is represented by a {\it positive 
semi-definite\/} operator $\hat A$ with a {\it discrete\/} spectrum $\{a_n;
\ n=0, 1, 2\cdots\ \}$.  The degeneracy of the eigenvalue $a_n$, denoted
$g(n)$, is independent of the $j, m$ and $q$.
\medskip

I do {\it not\/} prove discreteness of the area spectrum.  It is here
an assumption justified by the adiabatic invariant character of horizon area. 
One imagines the eigenvalues to be arranged so that $a_0=0$, $g(0)=1$
corresponds to the vacuum ${|{\rm vac}\rangle}$ (state devoid of any black
holes)  while the rest of the $a_n$ are arranged in  order of increasing
value. (Since I do not refer to $\hat {\cal G}$ in what follows, no confusion
will arise with the use of $g$ for degeneracy.)   The independence of $g(n)$
from, say $j$, is here an assumption. 

As argued in Sec.~\ref{mass_spectrum}, $\hat A$, $\hat Q$,  ${\bf \hat
J}^2$ and  $\hat J_z$ mutually commute.  We have as yet said nothing about
mass $\hat M$.  It is premature to think of it as the Hamiltonian because in
relativity the last can vanish.  Thus, rather than introducing $\hat M$ into
the algebra, we assume it can be gotten from $\hat A$, $\hat Q$ and  ${\bf
\hat J}^2$ by the usual relation from classical black holes:

\medskip
\noindent {\bf Axiom 2}: The Christodoulou-Ruffini formula
Eq.~(\ref{eq:spectrum}) is valid as a relation between operators.
\medskip

As mentioned, the commutativity of $\hat A$, $\hat Q$ and  ${\bf \hat J}^2$
makes this formula immune to factor ordering problems.  Thus, as already done 
in Secs.~\ref{mass_spectrum} and \ref{paradox}, one can infer the spectrum of
$\hat M$  directly from those of $\hat A$, $\hat Q$ and  ${\bf \hat J}^2$. 

The algebra so far is too trivial to tell us anything about the spectrum of
$\hat A$.  Of course we do not want to assume the uniformly spaced spectrum. 
That is a {\it desideratum\/}.   Recall now the discussion in 
Sec.~\ref{spacing}: the horizon is envisaged as being built one patch at a
time.  There is a temptation here to introduce a ``patch creation operator''
which makes one new patch each time it is applied to the black hole quantum
state.  But if we assume that, then we are prejudicing the formalism in favor
of equally spaced area eigenvalues, since the patches would then be
equivalent.  Or in other words, a single ``area raising operator'' can give
nothing but an equally spaced spectrum.  So let us be more general.

\medskip
\noindent {\bf Axiom 3}: There exist operators $\hat R_{njmqs}$  with the
property that  $\hat R_{njmqs}{|{\rm vac}\rangle}$ is a {\it one\/}
black hole state with horizon area $a_n$, squared spin
$j(j+1)\,\hbar^2$, $z$-component of spin $m\,\hbar$, charge $qe$ and
internal quantum number $s$.  All one-black hole states are spanned by the
basis $\{\hat R_{njmqs}{|{\rm vac}\rangle}\}$.
\medskip

The stress here is on creation operators for single black holes, rather than
on raising operators that convert one black hole into another with different
quantum numbers because,  as mentioned, introducing raising operators runs
the risk of assuming what we would like to establish.  Of course, assuming
that each basis black hole state is created by its own operator is a very
mild assumption.  It amounts to {\it defining\/} the operators by their
simple action.  Introduction of the internal quantum number $s$ is necessary
because from the black hole entropy one knows that each state seen by an
external observer, even that of an uncharged nonrotating black hole,
corresponds to many internal states; these need to be distinguished by an
additional quantum number (below called variously $s, t$ or $r$). When no
misunderstanding can arise, I write $\hat R_{\kappa\, s}$ or plain  $\hat
R_{\kappa}$ for $\hat R_{njmqs}$.

Commutation of the operators now available creates more operators.  If this
process continues indefinitely, no information can be obtained from the
algebra unless additional assumptions are made.  Faith that it is possible to
elucidate the physics from the algebra leads me to require closure of the
algebra at an early stage.  I suppose the algebra to be linear in analogy
with many physically successful algebras.  All these assumptions are
formalized in

\medskip
\noindent {\bf Axiom 4}: The operators $\hat A,\ \hat {\bf J},\ \hat Q$, and
$ \hat R_{\kappa\, s}$ form a closed, linear, infinite dimensional nonabelian
algebra.
\medskip

This assumption has two different parts: the closure at some low level of
commutation (simplicity), and the linear character of the algebra
when formulated in terms of $\hat A$.  As we shall see presently, this last
implies the additivity of horizon area, which is a reasonable property. 
Additivity of mass for several black holes is not reasonable (nonlinearity of
gravity), and this is really the reason why one cannot assume linearity of
the algebra of $\hat M$, $\hat Q$, $\hat {\bf J}$ and $\hat R_{\kappa}$.  In
this sense
$\hat A$ is singled out as special among all functions of the
black hole observables. 

Since $\hat R_{njmqs}|{\rm vac}\rangle$ is defined as a state with spin
quantum numbers $j$ and $m$, the collection of such states with fixed $j$ and
all allowed $m$ must transform among themselves under rotations of the black
hole like the spherical harmonics $Y_{jm}$ (or the corresponding spinorial
harmonic when $j$ is half-integer).  Since  ${|{\rm vac}\rangle}$ must
obviously be invariant under rotation, one learns that the
$\hat R_{njmqs}$ may be taken to behave like an irreducible spherical tensor
operator of rank $j$ with the usual $2j+1$ components labeled by
$m$~\cite{Merzbacher}.  This means that
\be
[\hat J_z, \hat R_{\kappa}] = m_\kappa\,\hbar\, \hat R_{\kappa}
\lb{eq:commuteJz}
\ee
and
\be
[\hat J_\pm, \hat R_{\kappa}] =
\sqrt{j_\kappa(j_\kappa+1)-m_\kappa(m_\kappa\pm 1)}\,\hbar\,\hat R_{\kappa\,
m_\kappa\pm 1}
\lb{eq:commuteJ+}
\ee
where $\hat J_\pm$ are the well known raising and lowering operators for
the $z$-component of spin.  To check these commutators I first operate with
Eq.~(\ref{eq:commuteJz}) on  ${|{\rm vac}\rangle}$ and take into account that
$\hat {\bf J} {|{\rm vac}\rangle}\ = 0$ (the vacuum has zero spin) to get
\be
\hat J_z\,\hat R_{\kappa\, s}{|{\rm vac}\rangle} = m_\kappa\hbar \hat
R_{\kappa\, s}{|{\rm vac}\rangle}
\lb{eq:eigenvalueJz}
\ee
Also from the relation~\cite{Merzbacher} $\hat {\bf J}^2 = (\hat J_+\hat J_-
+ \hat J_-\hat J_+)/2 +\hat J_z^2$, one can work out $[\hat {\bf J}^2, \hat
R_{\kappa\, s}]$ and operate with it on ${|{\rm vac}\rangle}$; after double
use of Eqs.~(\ref{eq:commuteJz}) and (\ref{eq:commuteJ+}) one gets
\be
\hat {\bf J}^2\,\hat R_{\kappa\, s}{|{\rm vac}\rangle}\, =
j_\kappa(j_\kappa+1)\hbar^2 \hat R_{\kappa\, s}{|{\rm vac}\rangle}
\lb{eq:eigenvalueJ^2}
\ee
Of course both of these results were required by the definition of 
$\hat R_{njmqs}{|{\rm vac}\rangle}$.

Moving on one recalls that $\hat Q$ is the generator of (global) gauge
transformations of the black hole, which means that for an arbitrary real
number $\chi$, $\exp(\imath\chi\hat Q)$ elicits a phase change of the
black hole state:
\be
\exp(\imath\chi \hat Q)\, \hat R_{\kappa\, s}{|{\rm vac}\rangle}\, =
\exp(\imath\chi q_\kappa e)\,\hat R_{\kappa\,  s}{|{\rm vac}\rangle}:
\lb{eq:gauge}
\ee
 This equations parallels
\be
\exp(\imath\phi \hat J_z/\hbar)\, \hat R_{\kappa
s}{|{\rm vac}\rangle}\, = \exp(\imath\phi m_\kappa)\,\hat R_{\kappa
s}{|{\rm vac}\rangle},
\lb{eq:rotation}
\ee
which expresses the fact that $\hat J_z$ is the generator of rotations of
the spin about the $z$ axis.  Thus by analogy with Eq.~(\ref{eq:commuteJz}) 
one may settle on the commutation relation
\be
[\hat Q, \hat R_{\kappa\, s}] = q_\kappa e \hat R_{\kappa\, s}.
\lb{eq:commuteQ}
\ee
Operating with this on the vacuum (recall that $\hat Q{|{\rm vac}\rangle} =
0$) gives
\be
\hat Q\, \hat R_{\kappa\, s}{|{\rm vac}\rangle} = q_\kappa e
\hat R_{\kappa\, s} {|{\rm vac}\rangle}
\lb{eq:charge}
\ee
so that $\hat R_{\kappa\, s}{|{\rm vac}\rangle}$ is indeed a one-black hole
state with definite charge $q_\kappa e$, as required.

In addition to Eqs.~(\ref{eq:commuteJz}-\ref{eq:commuteJ+}) and
(\ref{eq:commuteQ}), one would like to determine $[\hat A,
\hat R_{\kappa\, s}]$, but since it is unclear what kind of symmetry
transformation
$\hat A$ generates, a roundabout route is indicated.

\subsection{Algebra of the area observable}\lb{area_algebra}

Consider the Jacobi {\it identity\/}
\be
[\hat B, [\hat V, \hat C]] + [\hat V, [\hat C, \hat B]] + [\hat C, [\hat B,
\hat V]]=0
\lb{eq:Jacobi}
\ee
valid for three {\it arbitrary\/} operators $\hat B$, $\hat V$ and
$\hat C$.  Substitute $\hat B\rightarrow \hat A$, $\hat C\rightarrow \hat
R_{\kappa\, s}$, replace $\hat V$ in turn by $\hat J_z$,  $\hat J_\pm$ and
$\hat Q$, and then make use of Eqs.~(\ref{eq:commuteJz}-\ref{eq:commuteJ+})
and (\ref{eq:commuteQ}) as well as the mutual commutativity of $\hat J_z,
\hat J_\pm,$ $\hat Q,$ and $\hat A$ to obtain the three commutators
\bea
& & [\hat J_z, [\hat A, \hat R_{\kappa\, s}]]  =  m_\kappa\,\hbar\,
[\hat A, \hat R_{\kappa\, s}], \nonumber \\
 & & [\hat J_\pm, [\hat A, \hat R_{\kappa\, m_\kappa\, s}]]  = 
\sqrt{j_\kappa(j_\kappa+1)-m_\kappa(m_\kappa\pm 1)}\,\hbar\,[\hat A,\hat
R_{\kappa\, m_\kappa\pm 1\, s}], \nonumber \\
 & & [\hat Q, [\hat A, \hat R_{\kappa\, s}]]  =  q_\kappa e\, [\hat A,
\hat R_{\kappa\, s}].
\label{eq:three}
\eea
Now compare these equations with
Eqs.~(\ref{eq:commuteJz}-\ref{eq:commuteJ+}) and (\ref{eq:commuteQ}). 
Obviously, for fixed $\{jmq\}$, a particular $[\hat A,
\hat R_{njmqs}]$  has commutators with $\hat J_z, \hat J_\pm$ and $\hat Q$ 
of the same form as would all the $\hat R_{njmqs}$ with the same $\{jmq\}$.  
This means $[\hat A, \hat R_{njmqs}]$ transforms under rotations and gauge
transformations just like a $R_{njmqs}$ with the same $\{jmq\}$.  Thus
\be
[\hat A, \hat R_{\kappa\, s}]=\sum_{n_\lambda t} h_{\kappa\, s}{}^{\lambda
t}\, \hat R_{\lambda t} + \hat T_{\kappa\, s}
\lb{eq:commuteA} 
\ee
where $n_\lambda$ belongs to the set $\lambda$, the
$h_{\kappa\, s}{}^{\lambda\, t}$ are structure constants, and $\hat
T_{\kappa\, s}$ are operators not in the class of $\hat R_{\kappa\, s}$ which
are defined in such a way as to make Eq.~(\ref{eq:commuteA}) true. 

By Axiom 4 the $\hat T_{\kappa\, s}$ can only include $\hat A,\ \hat {\bf J}$ 
and $\hat Q$. But $\hat A$ and $\hat Q$ are both gauge and rotationally
invariant, so they can appear in Eq.~(\ref{eq:commuteA}) only for the case
$\{nqjms\} =
\{n000s\}$.  Further, $\hat {\bf J}$ constitutes a gauge invariant {\it
vector\/} operator, namely its (spherical) components $\hat J_-, J_z$ and
$J_+$ correspond to $\{nqjms\} = \{n01ms\}$.  Because the algebra is to be
linear we can thus rewrite Eq.~(\ref{eq:commuteA}) as
\be
[\hat A, \hat R_{\kappa\, s}]=\sum_{n_\lambda t} h_{\kappa\, s}{}^{\lambda
t}\, \hat R_{\lambda t} +\delta_{q_\kappa}{}^0\big[\delta_{j_\kappa}{}^0\,
(D\hat Q + E\hat A) + \delta_{j_\kappa}{}^1\, F\hat J_{m_\kappa}\big]
\lb{eq:commuteAnew} 
\ee
where $D, E$ and $F$ are numbers depending only on $n_\kappa$ and $s$.

Operating with Eq.~(\ref{eq:commuteAnew}) on the vacuum, and remembering
that $\hat A, \hat Q$ and ${\bf \hat J}$ all anhilate it (because it is
gauge invariant, rotationally invariant and contains no horizons), one gets
\be
a_\kappa\,\hat R_{\kappa\, s}\,{|{\rm vac}\rangle}\, = \sum_{n_\lambda t}
h_{\kappa s}{}^{\lambda t}\, \hat R_{\lambda t}{|{\rm vac}\rangle}
\lb{eq:operate} 
\ee
Now because the $\hat R_{\lambda t}{|{\rm vac}\rangle}$ with various 
$n_\lambda$ and $t$ are independent, one must set
\be
h_{\kappa\, s}{}^{\lambda t}\,  =a_\kappa\, \delta_{n_\kappa}{}^{n_\lambda}
\,\delta_s{}^t 
\lb{eq:h=}
\ee
so that the final form of Eq.~(\ref{eq:commuteA}) is
\be
[\hat A, \hat R_{\kappa\, s}]=a_\kappa\, \hat R_{\kappa\, s} 
+\delta_{q_\kappa}{}^0\big[\delta_{j_\kappa}{}^0\,
(D\hat Q + E\hat A) + \delta_{j_\kappa}{}^1\, F\hat J_{m_\kappa}\big]
\lb{eq:newcommuteA} 
\ee

Let us now define a new creation operator
\be
\hat R_{\kappa\, s}^{\rm new}\equiv  \hat
R_{\kappa\, s} + 
(a_\kappa)^{-1} \delta_{q_\kappa}{}^0\big[\delta_{j_\kappa}{}^0\,
(D\hat Q + E\hat A) + \delta_{j_\kappa}{}^1\, F\hat J_{m_\kappa}\big]
\lb{eq:newR}
\ee
Since $\hat A, \hat J_m$ and $\hat Q$ all anhilate ${|{\rm vac}\rangle}$, it
is seen that $\hat R_{\kappa\, s}^{\rm new}$ creates the same one-black hole
state as $\hat R_{\kappa\, s}$.  But the $\hat R_{\kappa\, s}^{\rm new}$ turn
out to satisfy simpler commutation relations.  Substituting in $[\hat A, \hat
R_{\kappa\, s}^{\rm new}]$ from Eq.~(\ref{eq:newcommuteA}),
(\ref{eq:commuteJz}-\ref{eq:commuteJ+}) and (\ref{eq:commuteQ}) one gets the
commutator
\be
[\hat A, \hat R_{\kappa\, s}^{\rm new}]=a_\kappa\, \hat R_{\kappa\, s}^{\rm
new}
\lb{eq:newnewcommuteA} 
\ee
which supplements Eqs.~(\ref{eq:commuteJz}-\ref{eq:commuteJ+}) and 
(\ref{eq:commuteQ}) and completes the algebra.  Henceforth I use only $\hat
R_{\kappa\, s}^{\rm new}$ but drop the ``new''.

\subsection{Algebraic derivation of the area spectrum}\lb{last} 

Now that we have the full algebra, we can get on with the job of elucidating
the spectrum of $\hat A$. Operating with $\hat R_{\kappa\, s}\hat R_{\lambda
t}$ on ${|{\rm vac}\rangle}$ and simplifying the result with
Eq.~(\ref{eq:newnewcommuteA}) gives
\be 
{\hat A\hat R_{\kappa\, s}\hat R_{\lambda t}{|{\rm
vac}\rangle} =\hat R_{\kappa\, s}(\hat A+a_\kappa)\hat R_{\lambda t}{|{\rm
vac}\rangle} =(a_\kappa+a_{\lambda}) \hat R_{\kappa s}\hat R_{\lambda
t}{|{\rm vac}\rangle} }
\lb{eq:commute} 
\ee
so that the state $\hat R_{\kappa\, s}\hat R_{\lambda t}{|{\rm vac}\rangle}$ 
has horizon area equal to the  sum of the areas of the states $\hat
R_{\kappa\, s}{|{\rm vac}\rangle}$ and $\hat R_{\lambda t}{|{\rm
vac}\rangle}$.  Analogy with field theory might lead one to believe that
$\hat R_{\kappa\, s}\hat R_{\lambda t}{|{\rm vac}\rangle}$ is just a two-black
hole state, in which case the result just obtained would be trivial. 
But in fact, the axiomatic approach allows other possibilities.

Recall Eqs.~(\ref{eq:commuteJz}), (\ref{eq:commuteQ}) and
(\ref{eq:newnewcommuteA}), namely 
\be
[\hat X, \hat R_\kappa] = x_\kappa \hat R_\kappa\qquad {\rm for}\qquad \hat X
=\{\hat A, \hat Q, \hat J_z\}
\lb{eq:commutators}
\ee
The Jacobi identity, Eq.~(\ref{eq:Jacobi}), can then be used to infer that
\be
[\hat X, [\hat R_{\kappa}, \hat R_{\lambda}]] = (x_\kappa+x_\lambda) [\hat
R_{\kappa},\hat R_{\lambda}]
\lb{eq:commutesummary}
\ee
which makes it clear that $[\hat R_{\kappa}, \hat R_{\lambda}]$ has the same
transformations under rotations and gauge transformations as a single $\hat
R_{\mu}$ with the index $\mu\equiv \{nqjms\}$ defined by the condition
\be
 x_\mu\equiv x_\kappa+x_\lambda 
\lb{eq:additivity}
\ee
Axiom 4 then allows one to conclude that ($\varepsilon_{\kappa\lambda}{}^\mu$ 
are structure constants)
\be
[\hat R_{\kappa}, \hat R_{\lambda}]
 =\sum_{\mu} \varepsilon_{\kappa\lambda}{}^{\mu}\hat
R_{\mu} + \delta_{q_\mu}{}^0\big[\delta_{j_\mu}{}^0\,
(\tilde D\hat Q + \tilde E\hat A) + \delta_{j_\mu}{}^1\, \tilde F\hat
J_{m_\mu}\big]
\lb{eq:twoR}
\ee
where $\tilde D, \tilde E$ and $\tilde F$ are numbers depending only on
$n_\mu$, $s_\kappa$ and $s_\lambda$. Although closure was postulated with
respect to the old $\hat R$'s, we use the new $\hat R$'s here.  This causes
no difficulty because the two differ only by a superposition of $\hat A$,
$\hat Q$ and  $\hat J_z$, and these have been added anyway. 

When one operates with Eq.~(\ref{eq:twoR}) on ${|{\rm vac}\rangle}$ one gets
\be
[\hat R_{\kappa}, \hat R_{\lambda}]{|{\rm vac}\rangle}\, = \,|\bullet\rangle
\lb{eq:one}
\ee
where $|\bullet\rangle$ stands for a {\it one\/}-black
hole state, a superposition of states with various $\mu$.  Were $\hat
R_{\kappa\, s}\hat R_{\lambda t}{|{\rm vac}\rangle}$ purely a two-black hole
state, as suggested by the field-theoretic analogy, one could not get
Eq.~(\ref{eq:one}).  Inevitably
\be
\hat R_{\kappa\, s}\hat R_{\lambda t}|{\rm
vac}\rangle\,=\,|\bullet\bullet\,\rangle\, +
\,|\bullet\rangle
\lb{eq:two}
\ee
with $|\bullet\bullet\,\rangle$ a two-black hole
state, symmetric under exchange of the $\kappa s$ and $\lambda t$ pairs.  The
superposition of one and two-black hole states means that the rule of
additivity of eigenvalues, Eq.~(\ref{eq:additivity}), applies to one black
hole as well as two: {\it the sum of two eigenvalues of $\hat Q$,  $\hat J_z$
or $\hat A$ of a single black hole is also a possible eigenvalue of a single
black hole.\/}   For charge or $z$-spin component this rule is consistent with
experience with quantum systems whose charges are always integer multiples of
the fundamental charge (which might be a third of the electron's), and whose 
$z$-spins are integer or half integer multiples of $\hbar$.  This agreement
serves as a partial check of our line of reasoning.

In accordance with Axiom 1, let $a_1$ be the smallest nonvanishing eigenvalue
of $\hat A$. Then Eq.~(\ref{eq:additivity}) says that any positive
integral multiple $na_1$ (which can be obtained by repeatedly adding $a_1$ to
itself) is also an eigenvalue.  This spectrum of $\hat A$ agrees with that
found in Sec.~\ref{spacing} by heuristic arguments.  But the question
is, are there any other area eigenvalues in between the integral ones (this
has a bearing on the question of whether splitting of the area eigenvalues of
Sec.~\ref{spacing} is at all possible) ?

To answer this query, I write down the hermitian conjugate of
Eq.~(\ref{eq:newnewcommuteA}):
\be
[\hat A, \hat R_{\kappa}^\dagger]=-a_\kappa \hat R_{\kappa}^\dagger
\lb{eq:conjugate}
\ee
Then
\be
\hat A\, \hat R_{\kappa}^\dagger \hat R_{\lambda}\,
{|{\rm vac}\rangle} =\, \left(\hat R_{\kappa}^\dagger\hat A - a_\kappa
\hat R_{\kappa}^\dagger
\right)\hat R_{\lambda} {|{\rm vac}\rangle}\, =
\left(a_{\lambda}-a_\kappa\right) \hat R_{\kappa}^\dagger 
\hat R_{\lambda}{|{\rm vac}\rangle}
\lb{eq:commuteconj}
\ee
Thus differences of area eigenvalues are area eigenvalues in their own
right.  Since $\hat A$ has no negative eigenvalues, if $n_\lambda\,\leq
\,n_\kappa$, the operator $\hat R_{\kappa}^\dagger$ must anhilate the
one-black hole state $\hat R_{\lambda}\, {|{\rm vac}\rangle}$ and there is no
black hole state  $\hat R_{\kappa}^\dagger \hat R_{\lambda}\,
{|{\rm vac}\rangle}$.  By contrast, if $n_\kappa\,<\,n_\lambda$,
$R_{\kappa}^\dagger$ obviously lowers the area eigenvalue of $\hat
R_{\lambda}$.  There is thus no doubt that $\hat R_{\kappa}^\dagger \hat
R_{\lambda}\, {|{\rm vac}\rangle}$ is a purely one-black hole state (a
``lowering'' operator cannot create an extra black hole:
Eq.~(\ref{eq:commuteconj}) shows that
$\hat R_{\kappa}^\dagger$ anhilates the vacuum). In conclusion, {\it positive
differences of one-black hole area eigenvalues are also allowed area
eigenvalues of a single black hole.\/}  

If there were fractional eigenvalues of $\hat A$, one could, by substracting
a suitable integral eigenvalue, get a positive eigenvalue below $a_1$, in
contradiction with $a_1$'s definition as lowest positive area eigenvalue. 
Thus the set $\{na_1;\ n=1, 2, \cdots\}$ comprises the totality of
$\hat A$ eigenvalues for one black hole, in complete agreement with the
heuristic arguments of Sec.~\ref{spacing} (but the algebra by itself
cannot set the area scale $a_1$).

What about the degeneracy of area eigenvalues ? According to Axiom 1, $g(n)$,
the degeneracy of the area eigenvalue $na_1$, is independent of $j, m$ and
$q$.  Thus for fixed $\{n_\kappa, j_\kappa, m_\kappa, q_\kappa\}$ where not
all of $j_\kappa, m_\kappa$ and $q_\kappa$ vanish, there are $g(n_\kappa)$
independent one-black hole states $\hat R_{\kappa\,s}{|{\rm vac}\rangle}$
distinguished by the values of $s$.  Analogously, the set $\{n_\lambda=1,
j_\lambda=0, m_\lambda=0, q_\lambda=0\}$ specifies $g(1)$ independent states
$\hat R_{\lambda\,t}{|{\rm vac}\rangle}$, all different from the previous
ones because not all quantum numbers agree.  One can thus form $g(1)\cdot
g(n_\kappa)$ one-black hole states, $[\hat R_{\kappa\,s}, \hat
R_{\lambda\,t}]{|{\rm vac}\rangle}$, with area eigenvalues $(n_\kappa +
1)a_1$ and charge and spin just like the states $\hat R_{\kappa\,s}{|{\rm
vac}\rangle}$.
 {\it If\/} these new states are independent, their number cannot exceed the
total number of states with area $(n_\kappa + 1)a_1$, namely $g(n_\kappa +1)
\geq g(1)\cdot g(n_\kappa)$.  Iterating this inequality
starting from
$n_\kappa=1$ one gets
\be
g(n) \geq  g(1)^n
\lb{eq:degeneracy2}
\ee
The value $g(1)=1$ is excluded because one knows that there
is some degeneracy.  Thus the result here is consistent with the law
(\ref{eq:degeneracy}) which we obtained heuristically.  In particular, it 
supports the idea that the degeneracy grows exponentially with area.  The
specific value $g(1)=2$ used in Sec.~\ref{spacing}  requires further input.

\acknowledgements{I thank A. Mayo, M. Milgrom, V. Mukhanov, M. Schiffer and 
L. Sriramkumar for many remarks, and Mayo for help with the graphics.  The
research on which these lectures are based was supported in part by a grant
from the Israel Science Foundation which was established by the Israel
Academy of Sciences.}

{}


\begin{iapbib}{}

\bibitem{Achu}  Achucarro A.,  Gregory R. and Kuijken K., 1995, Phys. Rev. D
52, 5729

\bibitem{Adler} Adler S. A.  and Pearson R. P., 1978, Phys.
Rev. D 18, 2798

\bibitem{Arnold}  Arnold V. I., 1989, {\it Mathematical Methods of Classical
Mechanics\/}. Springer, New York, 2nd edition

\bibitem{Ashtekhar}  Ashtekhar A. and Krasnov K., 1998, preprint
gr-qc/9804039

\bibitem{Ayon1} Ayon E., 1997, preprint gr-qc/9606081

\bibitem{Ayon2} Ayon E., 1997, preprint gr-qc/9606082

\bibitem{Ayon-Beato} Ayon-Beato E., 1997, preprint gr-qc/9611069

\bibitem{BarvKunst} Barvinskii A. and  Kunstatter G., 1996, Phys. Lett. B 289,
231

\bibitem{Bekthesis} Bekenstein J. D., 1972, Dissertation, Princeton 
University 

\bibitem{BekinWheeler} Bekenstein J. D., 1972, see Wheeler J. A., 1990,
{\it A Journey into Gravitation and Spacetime\/}. Freeman, San Francisco 

\bibitem{Beknohair} Bekenstein J. D., 1972, Phys. Rev. Lett.  28, 452;
Bekenstein J. D., 1972, Phys. Rev. D 5, 1239 and 2403

\bibitem{Bek_super} Bekenstein J. D., 1973, Phys. Rev. D 7, 949

\bibitem{BekEntropy} Bekenstein J. D., 1973, Phys. Rev. D 7, 2333

\bibitem{Bekscalar}  Bekenstein J. D., 1974, Ann. Phys. (NY)   82, 535

\bibitem{BekNC} Bekenstein J. D., 1974,  Lett. Nuovo Cimento 11, 467

\bibitem{Bekbh}  Bekenstein J. D., 1975, Ann. Phys. (NY)  91, 72

\bibitem{BekPT} Bekenstein J. D., 1980, Physics Today  33, 24 

\bibitem{Bek95} Bekenstein J. D., 1995, Phys. Rev. D 51, R6608

\bibitem{BekMukh} Bekenstein J. D. and  Mukhanov V. F., 1995, Phys. Lett. B 
360, 7

\bibitem{BekBrazil} Bekenstein J. D., 1996, eds  da Silva A. J. {et. al\/}, 
in {\it XVII Brazilian National Meeting on Particles and Fields\/}. Brazilian
Physical Society 

\bibitem{Sakharov} Bekenstein J. D., 1997, eds Dremin I. and Semikhatov A.,
in {\it The Second Andrei D. Sakharov Conference in Physics\/}.  World
Scientific, Singapore

\bibitem{QG} Bekenstein J. D. and  Mukhanov V. F., 1997, eds  Berezin V. A., 
Rubakov V. A. and  Semikoz  D.  V., in {\it Sixth Moscow
Quantum  Gravity Seminar\/}.  World Publishing,  Singapore

\bibitem{BekSchiff} Bekenstein J. D. and Schiffer M., 1998, Physical
Review D 58, 64014.  Preprint gr-qc/9803033

\bibitem{BHTrail}   Bekenstein J. D., 1998, eds   Iyer B. R. and  Bhawal B., 
in {\it Black Holes, Gravitational  Radiation and the Universe\/}.  Kluwer,
Dordrecht

\bibitem{MG8} Bekenstein J. D., 1998, eds Piran T. and  Ruffini R., in {\it
Proceedings of the VIII Marcel Grossmann Meeting on General Relativity\/}.  
World Scientific, Singapore, preprint gr-qc/9710076

\bibitem{Berezin}  Berezin V., 1997, Phys. Rev. D 55, 2139

\bibitem{Bizon1} Bizon P., 1990, Phys. Rev. Lett. 64, 2844

\bibitem{BBM} Bocharova N., Bronnikov K. and Melnikov  V., 1970, Vestn. Mosk.
Univ. Fiz. Astron.  6, 706

\bibitem{Brodbeck} Brodbeck O. and Straumann N., 1994, Phys. Lett. B 324 
309; 1996, J. Math. Phys. 37 1414

\bibitem{BroKir}   Bronnikov K. A. and  Kireyev Yu. N., 1978, Phys. Lett. A 
67, 95 

\bibitem{Carter2} Carter B., 1968, Commun. Math. Phys. 10, 280

\bibitem{Carter3} Carter B., 1968, Phys. Rev. D 174, 1559
 
\bibitem{Carter} Carter B., 1971, Phys. Rev. Lett.  26, 331 

\bibitem{Chase} Chase J. E., 1970, Commun. Math. Phys.  19, 276

\bibitem{Christodoulou} Christodoulou D., 1970, Phys. Rev. Lett.  25, 1596 

\bibitem{CR} Christodoulou D. and  Ruffini R., 1971,  Phys. Rev. D 4, 3552

\bibitem{DanSch}  Danielsson U. H. and Schiffer M., 1993, Phys. Rev. D
48, 4779

\bibitem{DolgKhrip} Dolgov A. D. and  Khriplovich I. B., 1997, Phys. Lett. B
400, 12

\bibitem{Droz}  Droz S.,  Heusler M. and  Straumann N., Phys. Lett. B,
1991, 268, 371

\bibitem{Ehrenfest} Ehrenfest, P., 1914, work mentioned in Born
M., 1969, {\it Atomic Physics\/}. Blackie, London, 8th edition 

\bibitem{Gibbons} Gibbons G. W., 1991, ed  Barrow J. D., in {\it The Physical
World: The Interface Between Cosmology, Astrophysics and Particle Physics.
Lecture Notes in Physics  383\/}. Springer, New York

\bibitem{GinzFr} Ginzburg  V. L. and Frank I. M., 1947, Dokl. Akad.
Nauk. SSSR  56, 583

\bibitem{Ginz} Ginzburg  V. L., 1993, ed Wolf E., in {\it Progress in Optics
XXXII\/}. Elsevier, Amsterdam

\bibitem{Greene}  Greene B. R.,  Mathur S. D. and  O'Neill C. M., 1993,
Phys. Rev. D 47, 2242 

\bibitem{Hartle}  Hartle J., 1971, Phys. Rev. D 3, 2938

\bibitem{Hawking_area} Hawking S. W., 1971, Phys. Rev. Lett.  26, 1344

\bibitem{Hawk_Ellis} Hawking S. W. and G. F. R. Ellis, 1973, {\it The Large
Scale Structure of Spacetime\/}. Cambridge Univ. Press, Cambridge

\bibitem{HawkingHartle}  Hawking S. W. and  Hartle J. B., 1972, Commun.
Math. Phys.  27, 283

\bibitem{Heu} Heusler M., 1992, J. Math. Phys. 33, 3497; 1995, Class. Quant. 
Grav. 12, 779

\bibitem{Skyrmion_stable} Heusler M., Straumann N. and  Zhou Z. H., Helv. 
Phys. Acta, 1993, 66, 614  

\bibitem{Heusler} Heusler M., 1996, {\it Black Hole Uniqueness Theorems.\/}
Cambridge Univ. Press, Cambridge

\bibitem{Israel} Israel W., 1967, Phys. Rev. 164, 1776; 1968, Commun. Math.
Phys.  8, 245

\bibitem{Jackson} Jackson J. D., 1962, {\it Classical Electrodynamics.} 
Wiley, New York

\bibitem{Kastrup} Kastrup H., 1996, Phys. Lett. B 385, 75 

\bibitem{Khriplovich} Khriplovich I. B, 1998, preprint gr-qc/9804004

\bibitem{Kogan}  Kogan Ya. I., 1986,  JETP Lett.  44, 267; 
Kogan I. I., 1994, preprint hep-th/9412232

\bibitem{Lahiri} Lahiri A., 1993, Mod. Phys. Lett. A 8, 1549

\bibitem{LLFields} Landau L. D. and Lifshitz E. M., 1975, {\it The Classical
Theory of Fields\/}. Pergamon, Oxford, 4th edition 

\bibitem{LLMech} Landau L. D. and Lifshitz E. M., 1976, {\it Mechanics\/}.
Pergamon, Oxford, 3rd edition

\bibitem{LLSP1} Landau L. D., Lifshitz E. M. and Pitaevskii L. P., 1980, {\it
Statistical Physics, Part 1\/}. Pergamon, Oxford

\bibitem{LLECM} Landau L. D., Lifshitz E. M. and Pitaevskii L. P., 1984, {\it
Electrodynamics of Continuous Media\/}.  Pergamon, Oxford,  2nd edition

\bibitem{Lee} Lee K-Y., Nair  V. P. and Weinberg E., 1992, Phys. Rev. Lett.
68, 1100
  
 \bibitem{LLSP2} Lifshitz E. M. and Pitaevskii L. P., 1980, {\it Statistical
Physics, Part 2\/}. Pergamon, Oxford, 3rd ed

\bibitem{LLQE} Lifshitz E. M., Pitaevskii L. P. and  Berestetskii V. I.,
1982, {\it Quantum Electrodynamics\/}. Pergamon, Oxford

\bibitem{LoukMak} Louko J. and M\"akel\"a J., 1996, Phys. Rev. D 54,
4982

\bibitem{Lousto} Lousto  C. O., 1995, Phys. Rev. D 51, 1733

\bibitem{Maggiore} Maggiore, M., 1994, Nucl. Phys. B 429, 205

\bibitem{Makela} M\"akel\"a J., 1996, unpublished preprint gr-qc/9602008.

\bibitem{Makela2} M\"akel\"a J., 1997, Phys. Lett. B 390, 115

\bibitem{MakelaRepo}  M\"akel\"a J. and  Repo P., 1998, Phys. Rev. D 57,
4899

\bibitem{Matzner} Matzner R. A.,  1968, Phys. Rev. D 9, 163

\bibitem{Mavromatos} Mavromatos N. E. and Winstanley,  E., 1996,  Phys. Rev.
D 53 3190

\bibitem{MB} Mayo A. E. and Bekenstein J. D., 1996, Phys. Rev. D 54, 5059

\bibitem{Mayo}  Mayo A. E., 1998, Phys. Rev. D 58, in press,
preprint gr-qc/9805047 

\bibitem{Mazur} Mazur P. 1987, Gen. Rel. Grav.  19, 1173

\bibitem{Merzbacher} Merzbacher E., 1970, {\it Quantum Mechanics.\/} Wiley,
New York, 2nd edition

\bibitem{MTW} Misner  C. W.,  Thorne K. S. and  Wheeler J. A., 1973, 
{\it Gravitation\/}. Freeman, San Francisco

\bibitem{Misner} Misner C. W., 1972, unpublished

\bibitem{Mukhanov} Mukhanov V. F., 1986, JETP Lett.  44, 63; 
1990, ed  Zurek W. H., in {\it Complexity, Entropy and the
Physics of Information: SFI Studies in the Sciences of Complexity.\/}
Addison-Wesley, New York, Vol 3

\bibitem{NP} Newman E. and Penrose R., 1962, J. Math. Phys. 3, 566

\bibitem{Nunez} N\'u\~nez, D., Quevedo H. and Sudarsky D., 1996 ,Phys. Rev.
Lett. 76, 571 

\bibitem{Page} Page D., 1976, Phys. Rev. D 13, 198 

\bibitem{Pauli} Pauli W., 1926, Z. Phys. 36, 336

\bibitem{Peet}  Peet A., 1997, preprint hep-th/9712253

\bibitem{Peleg} Peleg Y., 1995, Phys. Lett. B 356, 462

\bibitem{Penrose} Penrose R. and  Floyd R. M., 1971, Nature  229, 177

\bibitem{Pirani}  Pirani F. A. E., 1965, eds Deser S. and  Ford K. W., in 
{\it Lectures on General Relativity: Brandeis Summer Institute in Theoretical
Physics\/}. Prentice-Hall, Englewood Cliffs, Vol 1

\bibitem{Wheeler} Ruffini R. and Wheeler J. A., 1971, Physics Today 24, 30

\bibitem{Saa1} Saa A., 1996, J. Math. Phys.  37, 2346

\bibitem{Saa2} Saa A., 1996, Phys. Rev. D 53, 7377

\bibitem{SaaSchiffer} Saa A. and Schiffer M., preprint gr-qc/9805080

\bibitem{Schiffer} Schiffer M., 1989, unpublished preprint IFT/P-38/89
``Black hole spectroscopy'', S\~ao Paulo

\bibitem{Sorkin} Sorkin R. D., 1998, ed  Wald R. M., in {\it Black Holes and
Relativistic Stars\/}.  University of Chicago Press, Chicago

\bibitem{Starobinskii} Starobinskii A. A., 1973, Sov. Phys. JETP  37, 28;
Starobinskii A. A. and  Churilov S. M., 1973, Sov. Phys. JETP   38, 1

\bibitem{Straumann} Straumann N. and Zhou Z. H., 1991, Phys. Lett. B 243,
33; 1991, Nucl. Phys. B 369, 180.

\bibitem{Sud} Sudarsky  D., 1995, Class. Quant. Grav. 12, 579

\bibitem{SudZan} Sudarsky D. and Zannias T., preprint gr-qc/9712083. 

\bibitem{Teitelboim} Teitelboim C., 1972, Lett.  Nuov. Cim.  3, 326 and 397

\bibitem{PressTeuk1} Teukolsky S. A. and  Press W. H., 1974, ApJ 193, 443

\bibitem{tHooft}  't Hooft G., 1996, Int. J. Mod. Phys. A  11, 4623

\bibitem{Unruh} Unruh W. G., 1974, Phys. Rev. D 10, 3194

\bibitem{Vishu}   Vishveshwara C. V., 1968, J. Math. Phys.  9, 1319  

\bibitem{Virb}  Virbhadra K. S. and  Parikh J. C., 1994, Phys. Lett. B 331,
302

\bibitem{Volkov}  Volkov A. M.,  Izmest'ev A. A. and  Skrotskii,  G. V., 1970,
Sov. Phys. JETP  32, 686

\bibitem{VolkovM} Volkov M. S. and  Gal'tsov D. V., 1989, JETP Lett.  50,
312; 1990, JETP Lett. 50, 346

\bibitem{Wald} Wald R. M., 1971, Phys. Rev. Lett.  26, 1653 

\bibitem{Xanthopoulos}  Xanthopoulos B. C. and Zannias T., 1991, J. Math.
Phys. 32, 1875   
 
\bibitem{Yasskin}  Yasskin P., 1972, unpublished

\bibitem{Zannias} Zannias  T., 1995, J. Math. Phys.  36, 6970

\bibitem{Zeld1}  Zel'dovich Ya. B., 1971, JETP Lett.  14, 180

\bibitem{Zeld2}  Zel'dovich Ya. B., 1971, Sov. Phys. JETP  35, 1085

\end{iapbib}
\end{document}